\documentclass[a4paper,11pt]{article}
\pdfoutput=1 

\usepackage{jcappub} 

\usepackage[T1]{fontenc} 
\usepackage{graphicx}
\usepackage{epsfig}
\usepackage{rotate}
\usepackage{amsmath}
\usepackage{amssymb}
\usepackage{amsfonts}
\usepackage{bm}
\usepackage{verbatim}
\usepackage{tablefootnote}
\usepackage{enumerate}
\usepackage{afterpage}
\usepackage{xcolor}
\usepackage{natbib, hyperref}

\usepackage{multicol}
\usepackage{multirow}
\usepackage{booktabs}
\usepackage[section]{placeins}

\UseRawInputEncoding

\allowdisplaybreaks

\newcommand{\ud}{\mathrm{d}}
\newcommand{\p}{\partial}
\newcommand{\cH}{\mathcal{H}}
\newcommand{\Q}{\mathcal{Q}}
\newcommand{\fnl}{f_{\rm NL}}
\newcommand{\E}{\mathcal{E}}
\newcommand{\K}{\mathcal{K}}

\def\be{\begin{equation}}
\def\ee{\end{equation}}
\def\bea{\begin{eqnarray}}
\def\eea{\end{eqnarray}}

\newcommand{\red}[1]{{\color{black}{#1}}}
\newcommand{\teal}[1]{{\color{teal}{#1}}}

\usepackage{pgf,tikz}
\usetikzlibrary{arrows}

\title{Primordial non-Gaussianity -- the effects of 
relativistic and wide-angle corrections to the power spectrum}

\author{S\^ecloka L. Guedezounme$^{1,a}$, Sheean Jolicoeur$^{2,1,b}$, \\ Roy Maartens$^{1,3,4,c}$}
\affiliation{$^{1}$Department of Physics \& Astronomy, University of the Western Cape, Cape Town 7535, South Africa \\
$^{2}$Department of Physics, Stellenbosch University, Matieland 7602, South Africa \\ 
$^{3}$Institute of Cosmology \& Gravitation, University of Portsmouth, Portsmouth PO1 3FX, United Kingdom\\
$^4$National Institute for Theoretical \& Computational Sciences, Cape Town 7535, South Africa}

\abstract{
Wide-angle and relativistic corrections to the Newtonian and flat-sky approximations are important for accurate modeling of the galaxy power spectrum of next-generation  galaxy surveys. 
In addition to Doppler and Sachs-Wolfe relativistic corrections, we include the effects of lensing convergence, time delay and integrated Sachs-Wolfe.
We investigate the impact of these corrections on measurements of the local primordial non-Gaussianity parameter $f_{\mathrm{NL}}$, using two futuristic spectroscopic galaxy surveys, planned for SKAO2 and MegaMapper. In addition to the monopole, we include the quadrupole of the galaxy Fourier power spectrum. The quadrupole is much more sensitive to the corrections than the monopole. The combination with the quadrupole improves the precision on $\fnl$ by 
$\sim {{45}}\%$ and $\sim {{63}}\%$ for SKAO2 and MegaMapper respectively.
{Neglecting the wide-angle and relativistic corrections produces a  shift in $f_{\mathrm{NL}}$ which is very sensitive to the magnification bias and the redshift evolution of the comoving number density. 
In the case of SKAO2, the shift in $\fnl$ is negligible -- since the contributions to the shift from integrated and non-integrated effects nearly cancel.
For MegaMapper, there is only partial cancellation of integrated and non-integrated effects and the shift is  $\sim {0.6} \, \sigma$.} 
We point out that some of the approximations made in the wide-angle and relativistic corrections may artificially suppress the shift in $\fnl$.}

\emailAdd{$^{a}$seclokaguedezounme@gmail.com}
\emailAdd{$^{b}$jolicoeursheean@gmail.com}
\emailAdd{$^{c}$roy.maartens@gmail.com}

\usepackage[toc]{appendix}

\begin{document}

\maketitle
\date{\today}
%\flushbottom

\newpage

\section{Introduction}
\label{sec1}

Primordial non-Gaussianity (PNG) is a key probe of Inflation models that are assumed to generate the primordial perturbations -- which in turn seed the  cosmological fluctuations measured by cosmic microwave background (CMB) and large-scale structure surveys \cite{Dalal:2007cu, Komatsu:2010hc, Chen:2010xka}. 
Among the various forms of PNG, the local type, parametrised by $f_{\rm NL}$, is of primary importance due to its characteristic scale-dependent impact on galaxy clustering. This dependence makes large-scale structure an indispensable tool for constraining inflationary physics through precise measurements of $f_{\rm NL}$. 

The analysis of the galaxy power spectrum often relies on simplifying theoretical assumptions, such as the flat-sky (or plane-parallel)  and Newtonian approximations. While sufficient for small-scale studies, such approximations fail to account for relativistic and wide-angle effects, which become increasingly significant on ultra-large scales -- where the local PNG signal is strongest. The corrections include local effects (wide-angle, Doppler and Sachs-Wolfe)  as well as integrated effects (lensing convergence, integrated Sachs-Wolfe and time-delay)  \cite{Challinor:2011bk,Raccanelli:2015vla,Alonso:2015uua}. Neglecting these effects risks introducing systematic shifts (biases) in estimates of $f_{\rm NL}$ and other key cosmological parameters \cite{Namikawa:2011yr,Bruni:2011ta,Camera:2014sba,Lorenz:2017iez,Viljoen:2021ocx}. 

Incorporating these relativistic and wide-angle corrections is therefore essential to exploit the full potential of the galaxy power spectrum as a cosmological probe. 
Here we treat these effects as leading-order modifications to the standard Fourier galaxy power spectrum, in order to estimate their influence on $f_{\rm NL}$ estimation. 
This is particularly relevant for next-generation surveys, which aim to probe ultra-large scales with high precision. 
We consider two complementary futuristic surveys: the SKAO2 HI galaxy survey which covers  redshift 0 to 2 \cite{Maartens:2021dqy}, and the  MegaMapper LBG survey, covering redshift 2 to 5
\cite{Sailer:2021yzm}.

The bispectrum of CMB temperature anisotropies has delivered the current best measurement and $1\sigma$ constraint on local PNG \cite{Akrami:2019izv}
\begin{align}
    \fnl=-0.9\pm 5.1 \,.
\end{align}
The CMB and dark matter power spectra are not sensitive to local PNG, but  
local PNG changes the power spectrum of biased tracers such as galaxies -- by 
inducing a scale-dependent contribution to the linear clustering bias \red{\cite{Dalal:2007cu,Matarrese:2008nc}}:
\begin{align} \label{sdb}
    b~\to~ b + b_{\mathrm{ng}}~~\mbox{where}~~ {b_{\mathrm{ng}}(z,k) =3 \delta_{c} \big[b(z) - 1\big]\,
    \frac{\Omega_{m0}H_0^2\,(1+z_{\rm in})D_{\rm in}}{D(z)}
    \,\frac{\fnl}{T(k)\,k^2}\,.} 
 \end{align} 
Here $\delta_{c} = 1.686$ is the critical collapse overdensity, $z_{\rm in}$ is a redshift deep in the matter-dominated era, $D$ is the matter growth factor (normalised to 1 at redshift $z=0$)  and $T(k)$ is the matter transfer function. In \eqref{sdb} we have used a simple universality relation. There are serious issues involved in this assumption (see e.g. \cite{Barreira:2022sey,Barreira:2023rxn,Fondi:2023egm,Adame:2023nsx}), but our focus here is on comparing models of the power spectrum and not on realistic modeling of scale-dependent bias.
It is clear that the local PNG signal is non-negligible only on ultra-large scales,  $k \lesssim k_{\rm eq}$, where $T$ asymptotes to 1. These are the scales where the standard galaxy power spectrum acquires wide-angle and relativistic corrections, which therefore need to be incorporated for an accurate measurement of $\fnl$.

\begin{figure}[!htbp]
\centering
\includegraphics[width=0.50\linewidth]{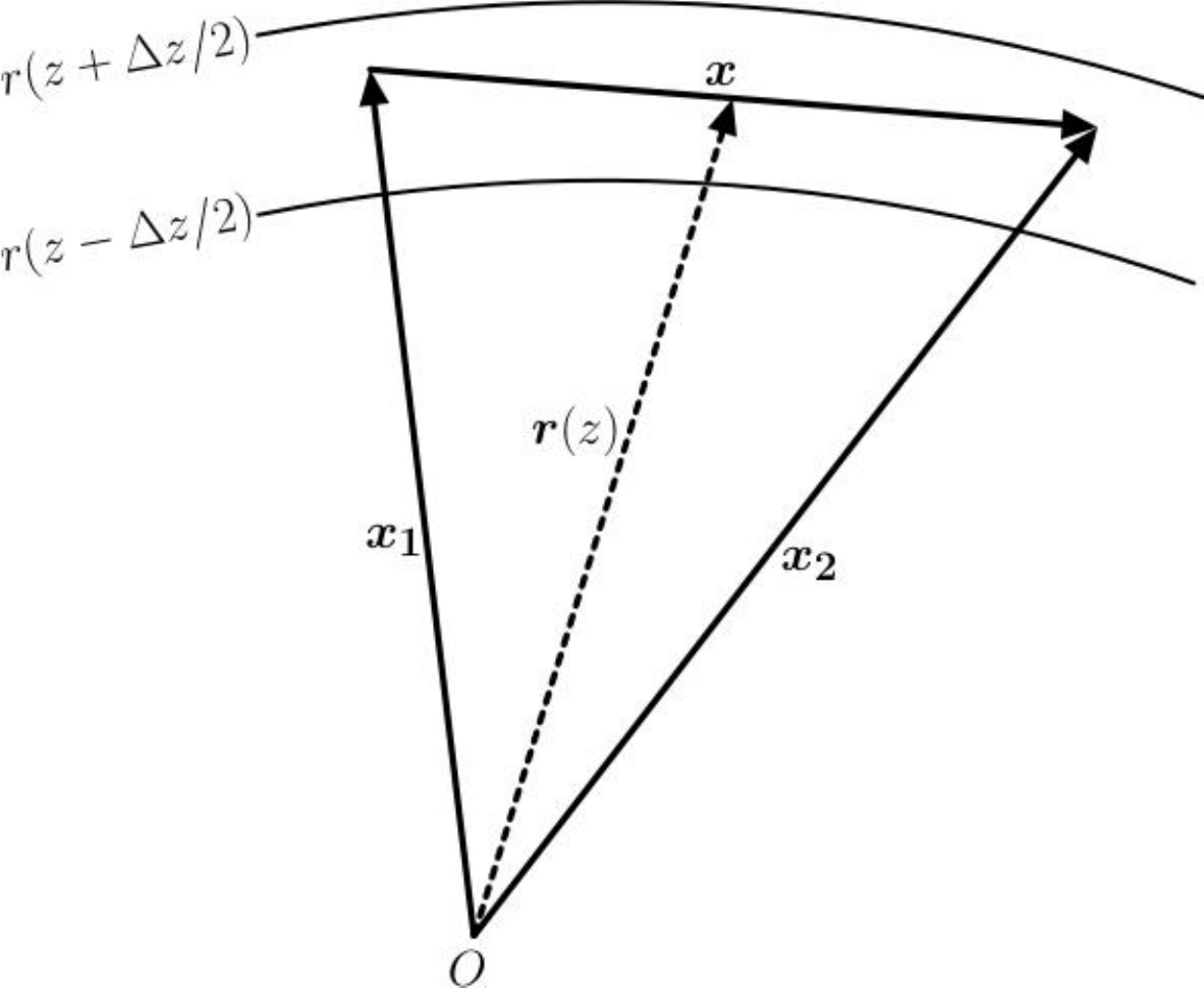}
\caption{Geometry of the correlations.} \label{geom}
\end{figure}

If $n_g$ is the comoving number density of galaxies, then the number density contrast $\delta_g=(n_g-\bar n_g)/\bar n_g$  in configuration space is $\delta_g(\bm x_a)= b_a\,\delta(\bm x_a)$ (see \autoref{geom}), where $b_a$ is the linear Gaussian bias at $x_a$ and $\delta$  is the matter density contrast. The observed number density contrast in redshift space is
\begin{align}
   \Delta_g(\bm x_a)= {b_a}\,\delta(\bm x_a)-\frac{1}{{\cH_a}}\, \hat{\bm x}_a\cdot\bm\nabla \big(\hat{\bm x}_a\cdot \bm v_a \big),
\end{align}
where the redshift dependence is implicit in $x_a(z)$ and $b_a$ and $\cH_a$ (conformal Hubble rate) are evaluated at $x_a$.
The galaxy peculiar velocity is $\bm v_a$ and we have included only the dominant distortion term, in the standard Newtonian approximation. In Fourier space, 
\begin{align} \label{pp}
   \Delta_g({x_a,\bm k_a})= \Big({b_a+ f_a\mu^2_a} \Big)\delta({x_a,\bm k}_a) \equiv \K^{\rm S}({x_a,k_a,\mu_a})\, \delta ({x_a,\bm k}_a) \qquad\mbox{with}~~ {\mu_a}=\hat{\bm{k}}_a\cdot \hat{\bm x}_a \,,
\end{align}
where $\K^{\rm S}$ is the Fourier kernel in the standard (S) approximation.
In this approximation, the line-of-sight direction is fixed, using a flat-sky   assumption,
\begin{align}\label{pp2}
 {\hat{\bm x}_1 = \hat{\bm x}_2 = {\bm n}}\,,   
\end{align}
where {$\bm n$ is a fixed direction.} 
\red{When we use wide-angle corrections (see \autoref{sec2}), $\bm n$ varies across the survey and is the midpoint direction for each pair of galaxies.}
In \autoref{pp}, $f= -\ud \ln D/\ud\ln(1+z)$ is the linear growth rate. 
In order to ensure a physical matter power spectrum $P$ on ultra-large scales, the density contrast should be the one measured in the matter rest-frame -- i.e., $\delta$ should be in comoving gauge. Then the relativistic Poisson equation has the same form as the Newtonian Poisson equation:
\begin{align}
    \nabla^2\Phi = \frac{3}{2}\Omega_m\cH^2\,\delta\,,
    \label{pois}
\end{align}
where $\Phi$ is the metric potential in Newtonian gauge:
\begin{align}
    \ud s^2=a^2\Big[-\big(1+2\Phi \big) \ud\eta^2+ \big(1-2\Phi \big)\ud\bm{x}^2 \big].
    \label{pmet}
\end{align}

The standard power spectrum $P^{\mathrm{S}}_g=\big(\K^{\rm S} \big)^2 P$ implicitly assumes
the flat-sky approximation.
Wide-angle corrections $P^{\mathrm{W}}_g$ arise from removing this assumption.
The second type of correction to  the standard power spectrum is from the relativistic effects introduced by observing on the past lightcone. A theoretically complete treatment of wide-angle + relativistic effects is to use from the start the fully general 2-point correlation function (see e.g. \cite{Matsubara:1999du,Matsubara:2000pr,Bertacca:2012tp,Tansella:2017rpi,Tansella:2018sld, Scaccabarozzi:2018vux}) or its angular harmonic transform (e.g. \cite{Bonvin:2011bg,Challinor:2011bk, Bruni:2011ta,Camera:2014bwa,Raccanelli:2015vla,Alonso:2015uua}). These naturally incorporate all effects from observing the spherical sky on the past lightcone. An alternative, which exploits the computational advantages of the Fourier power spectrum, is to Fourier transform the 2-point correlation function which includes all relativistic and wide-angle effects (e.g. \cite{Grimm:2020ays,Castorina:2021xzs,Foglieni:2023xca}) or transform to spherical Fourier-Bessel space (e.g. \cite{Wen:2024hqj,Semenzato:2024rlc}). 

A simplification of the alternative approach is to include the wide-angle and relativistic effects as leading-order  corrections to the standard Fourier spectrum (e.g. \cite{Noorikuhani:2022bwc,Paul:2022xfx,Jolicoeur:2024oij}):
\begin{align}
P_g = P^{\mathrm{S}}_g + P^{\mathrm{corr}}_g \quad \mbox{with}\quad P^{\mathrm{corr}}_g = P^{\mathrm{NI}}_g + P^{\mathrm{I}}_g 
\quad \mbox{up to}\quad 
\mathcal{O}\Big[\frac{1}{k^2r^2}\,,~  \frac{\cH^2}{k^2}\,,~ \frac{1}{kr} \,\frac{\cH}{k}\Big],
\label{ExpPtotal}
\end{align}
where {the wide-angle expansion parameter is $1/(kr)$.}
Here NI denotes the leading-order non-integrated corrections: wide-angle (W) and relativistic Doppler and potential (R) effects -- including the correlations  R$\,\times\,$S, W$\,\times\,$S and R$\,\times\,$W. This NI correction was derived in \cite{Noorikuhani:2022bwc,Paul:2022xfx,Jolicoeur:2024oij}. 
The integrated relativistic correction $P_g^{\rm I}$ in \autoref{ExpPtotal}  is the leading-order correction to $P^{\mathrm{S}}_g$ arising from lensing convergence (L), time-delay (TD)  and integrated Sachs-Wolfe (ISW) effects.
{We follow \cite{Noorikuhani:2022bwc} and include only I$\,\times\,$S and S$\,\times\,$I in $P_g^{\rm I}$.  We 
note that there is a significant distinction between lensing L, which scales as $(H/k)^0\delta$, and the other integrated terms TD and ISW, which scale as $(H/k)^2\delta$. We are working up to the order $\mathcal{O}\big[r^{-2}k^{-2}\,,{\cH^2}k^{-2}\,, r^{-1}\,{\cH}k^{-2}\big]$, which means that there are further lensing terms that could be included.
These are  L$\,\times\,$L, and the correlation of L with the wide-angle W and {with TD and ISW}. However, these terms
are considerably more difficult to compute and we leave them for future work.}

\section{Relativistic and wide-angle power spectrum}
\label{sec2}

\autoref{geom} shows the configuration for 2-point correlations in a redshift bin, using the midpoint line of sight vector $\bm r$. Then the 2-point correlation function is
$\xi(\bm r, \bm x)= {\big\langle \Delta_g(\bm x_1)\, \Delta_g(\bm x_2) \big\rangle}$,
where $\bm x=\bm x_2-\bm x_1$.
The power spectrum $P_g(\bm{r},{\bm k})$ is a local Fourier transform for each line of sight $\bm r$ \cite{Noorikuhani:2022bwc,Jolicoeur:2024oij}. 
{The monopole and dipole of the non-integrated correction $P_g^{\rm NI}$ are derived in \cite{Jolicoeur:2024oij} and given in \autoref{AppB} for convenience.}

Following \cite{Noorikuhani:2022bwc},
we include in $P_g^{\rm I}$ at leading order only the
correlations of 
{L+TD+ISW} 
with the standard term, i.e., {(L+TD+ISW)}$\,\times\,$S. 
Qualitatively,  in order to produce $P^{\mathrm{I}}_g$, the integrated kernel {$\K^{\rm int}$} is multiplied by the standard kernel $\K^{\rm S}$, and then 
integrated from observer to source, with a radial weighting factor. This kernel is made up of three contributions:
\begin{align}
{\mathcal{K}^{\rm int}}(r,\tilde r,  k, \mu) &= \mathcal{K}^{\rm L}(r,\tilde r, k, \mu) + \mathcal{K}^{\rm TD}(r,\tilde r, k) + \mathcal{K}^{\rm ISW}(r,\tilde r, k)\;, \label{eKI} 
\end{align}
where $0\le \tilde r\le r$ is the integration variable and $r$ (at the source) is fixed for each integration. Note that at linear order, the integration is along a background lightray, so that $\hat{\tilde{\bm r}}= \hat{\bm r}$. 

The relativistic lensing convergence  is defined in terms of the Laplacian on the 2-sphere, $\nabla_{\bm n}^2$
\cite{Challinor:2011bk,Bolejko:2012uj,Bacon:2014uja}:
\begin{align}\label{relk}
    \kappa =\int_0^r\ud\tilde r\, \frac{\tilde r(r-\tilde r)}{r} \,{\tilde\nabla_{\bm n}^2}\tilde \Phi
    \quad \mbox{where}~~ {\tilde\nabla_{\bm n}^2}=\nabla^2-(\bm n\cdot \bm \nabla)^2+\frac{2}{\tilde r}\,\bm n\cdot \bm \nabla {~\mbox{and}~ \bm n=\hat{\tilde{\bm r}}= \hat{\bm r}}\,.
\end{align}
Here and below, a tilde indicates that the quantity is evaluated at the background redshift $\tilde z$ corresponding to $\tilde r$.
The standard Newtonian approximation replaces $\nabla_{\bm n}^2$ by $\nabla^2$ and uses the Poisson \autoref{pois} to eliminate $\Phi$. This gives the dominant contribution to $\kappa$ on scales $k> k_{\rm eq}$:
\begin{align}
    \kappa_{\rm standard}= \frac{3}{2} \int_0^r\ud\tilde r\, \frac{\tilde r(r-\tilde r)}{r}\, \tilde \Omega_{m}\tilde \cH^2 \,\tilde\delta\,.
\end{align}
{Here we focus on ultra-large scales, $k < k_{\rm eq}$, so that we need to use the full expression \autoref{relk}.}
The lensing effect on galaxy number density contrast is given by the term 
\begin{align}
    \Delta_g^{\rm L}= 2\big(\Q-1\big)\kappa\,,
\end{align}
where $\Q$ is the magnification bias of the galaxy survey, {which determines whether a galaxy is brightened or dimmed sufficiently by lensing to  move above or below the flux cut}.
Using the relativistic $\kappa$ in \autoref{relk}, this leads to the lensing kernel:
\begin{align}
\mathcal{K}^{\rm L}(r,\tilde r, k, \mu) = 3  \big(\Q-1\big)
\tilde\Omega_{m}\,\tilde{\cH}^{2}\, \frac{\tilde{r}\big(r - \tilde{r}\big)}{r}  \bigg[1 - \mu^{2} + 2\,\mathrm{i}\,\frac{\mu}{\tilde r}\,\frac{G(r,\tilde{r})}{k}\bigg]\;, \label{eKL}
\end{align}
where {$\mu=\hat{\bm k} \cdot \bm n$} and the geometric weight factor is
\begin{align}\label{eG}
    G(r,\tilde r)=\frac{r+\tilde r}{2r}\,.
\end{align}
The remaining terms in
\autoref{eKI} follow from the time-delay $\Delta_g^{\rm TD}$ and integrated Sachs-Wolfe $\Delta_g^{\rm ISW}$ contributions to the number counts \cite{Challinor:2011bk,Alonso:2015uua}. These lead to the kernels \cite{Noorikuhani:2022bwc}: 
\begin{align}
\mathcal{K}^{\rm TD}(r,\tilde r,  k) &=  6\big(\Q-1\big)  
\frac{\tilde{\Omega}_{m}\, \tilde{\mathcal{H}}^{2}}{r} \left[\frac{G(r,\tilde{r})}{k}\right]^{2} \;, \label{eKTD} \\
\mathcal{K}^{\rm ISW}(r,\tilde r,  k) &= 3\left[\mathcal{E} - 2\mathcal{Q} + \frac{2(\mathcal{Q}-1)}{r \mathcal{H}} - \frac{\mathcal{H}^{\prime}}{\mathcal{H}^{2}}\right] 
\tilde{\Omega}_{m}\, \tilde{\mathcal{H}}^{3} \big(\tilde{f}-1\big) \left[\frac{G(r,\tilde{r})}{k}\right]^{2}\;, \label{eKISW}
\end{align}
where  $\E$ is the evolution bias of the galaxy survey, defined below.

The derivation of $P^{\mathrm{I}}_g$ is given in \cite{Noorikuhani:2022bwc}. For convenience we present our version of the derivation in \autoref{AppA}. This leads to
\begin{align}
P^{\mathrm{I}}_g(r,k,\mu) &= \int^{r}_{0} \ud\tilde r  \, \big[\mathcal{J}(r,\tilde r, k, \mu) + \mathcal{J}^{*}(r,\tilde r, k, \mu)\big]\; , \label{EqPSxI}
\end{align}
where 
\begin{align}
\mathcal{J}(r,\tilde r, k, \mu) = \frac{{D(r)} {D(\tilde r)}}{G(r,\tilde r)^{3}}   \exp\!\left[\! -{\rm i} \frac{\mu\, k}{G(r,\tilde r)} (r - \tilde r)\!\right] 
\!\mathcal{K}^{\rm S}\! \!\left(\!r, \frac{k}{G(r,\tilde r)}, \mu\!\right)\! {\mathcal{K}^{\rm int\,*}}(r, \tilde r, k, \mu) P_0\!\left(\!\frac{k}{G(r,\tilde r)}\!\right)\!.~~ \label{eJ}
\end{align}
Here $P_0$ is the matter power spectrum at $z=0$, and 
\begin{equation}
\mathcal{K}^{\rm S} \left(r, \frac{k}{G(r,\tilde r)}, \mu \right)  = b(r)+f(r)\mu^2 + b_{\rm ng} \left(r,\frac{k}{G(r,\tilde r)}\right)  
\; . \label{EqkernelS}
\end{equation} 
In order to compute the integral \eqref{EqPSxI}, we use the Gauss-Legendre quadrature method and approximate the result as:
\begin{align}
    \int_{y_{\rm min}}^{y_{\rm max}}\ud y\, f(y)  \approx \frac{y_{\rm max} - y_{\rm min}}{2} \sum_{i=1}^{n} w_i f\left( \frac{y_{\rm max} - y_{\rm min}}{2} \eta_i + \frac{y_{\rm max} + y_{\rm min}}{2} \right),
\end{align}
where $\eta_i$ are the roots of the Legendre polynomial $\mathcal{L}_n(\eta)$ of degree $n$, which lie in the interval $[-1, 1]$, and $w_i$ are the corresponding weights associated with each $\eta_i$:
\begin{align}
    \mathcal{L}_n(\eta_i) = 0, \qquad w_i = \frac{2}{\left( 1 - \eta_i^2 \right) \left[ \mathcal{L}^{\prime}_n(\eta_i) \right]^2}\,.
\end{align}
The monopole and quadrupole of  
$P^{\rm I}_g$ involve a further integral; see  \autoref{AppA} for some numerical details.

We consider two futuristic surveys: the Square Kilometer Array Phase 2 (SKAO2) HI galaxy survey \cite{Maartens:2021dqy} and the MegaMapper Lyman-Break Galaxy (LBG) \cite{Sailer:2021yzm} survey. The specifications of these surveys are summarised in \autoref{tab:specifications}.
\begin{table}[h]
\centering
\caption{Specifications of SKAO2 HI galaxy  and MegaMapper LBG surveys.}
\vspace{0.3cm}
\label{tab:specifications}
\begin{tabular}{|c|c|c|c|}
\hline
~~~~ Survey ~~~~ & ~~~~ Sky area [$\mathrm{deg}^{2}$] ~~~~ & ~~~~ Redshift Range ~~~~ \\
\hline
\hline
SKAO2 HI galaxy & 30,000 & $0.1 \leq z \leq 2.0$ \\
\hline
MegaMapper LBG & 20,000 & $2.0 \leq z \leq 5.0$ \\
\hline
\end{tabular}
\end{table}
\begin{figure}[!htbp]
\centering
\includegraphics[width=0.48\linewidth]{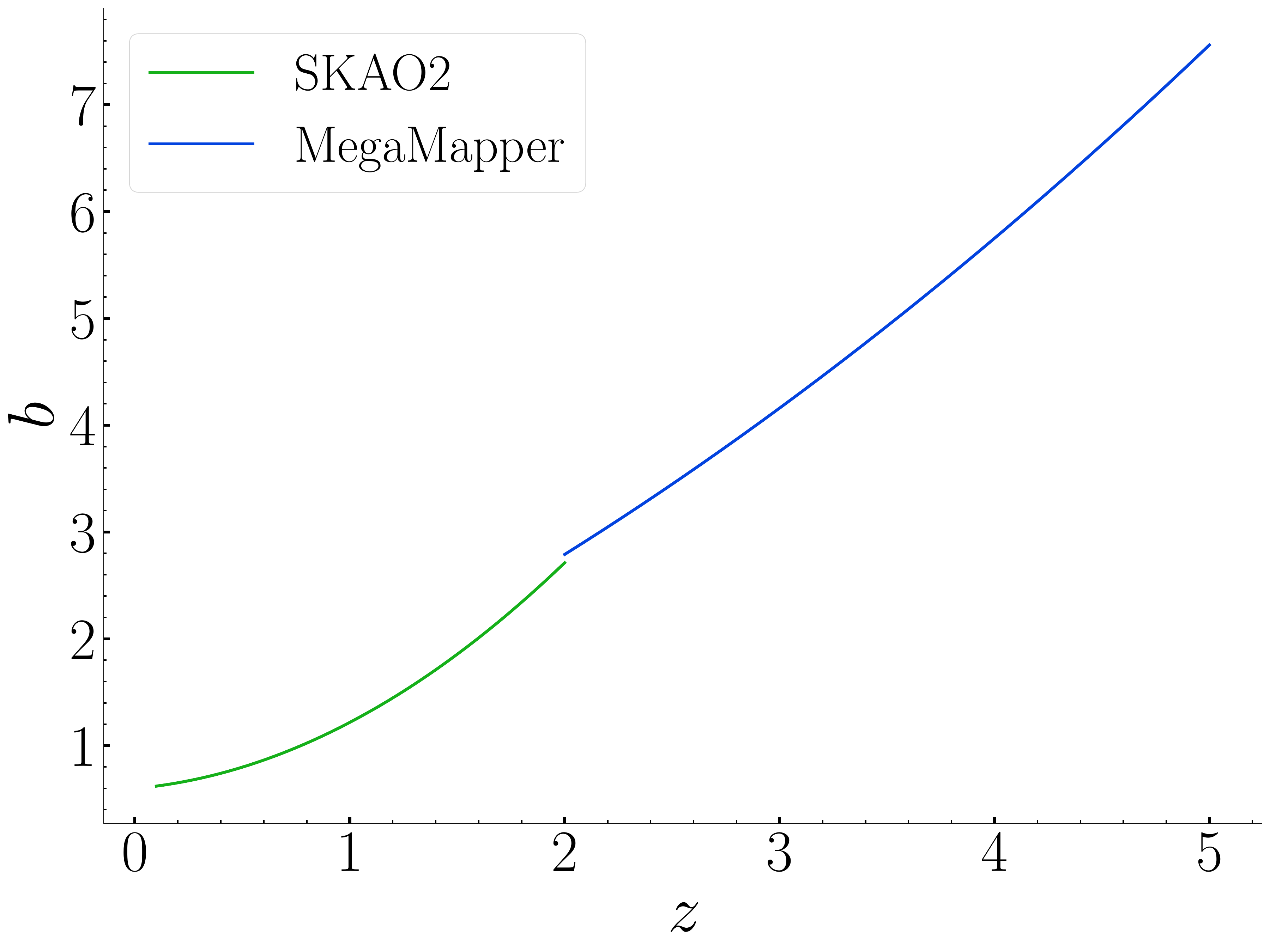}
\includegraphics[width=0.48\linewidth]{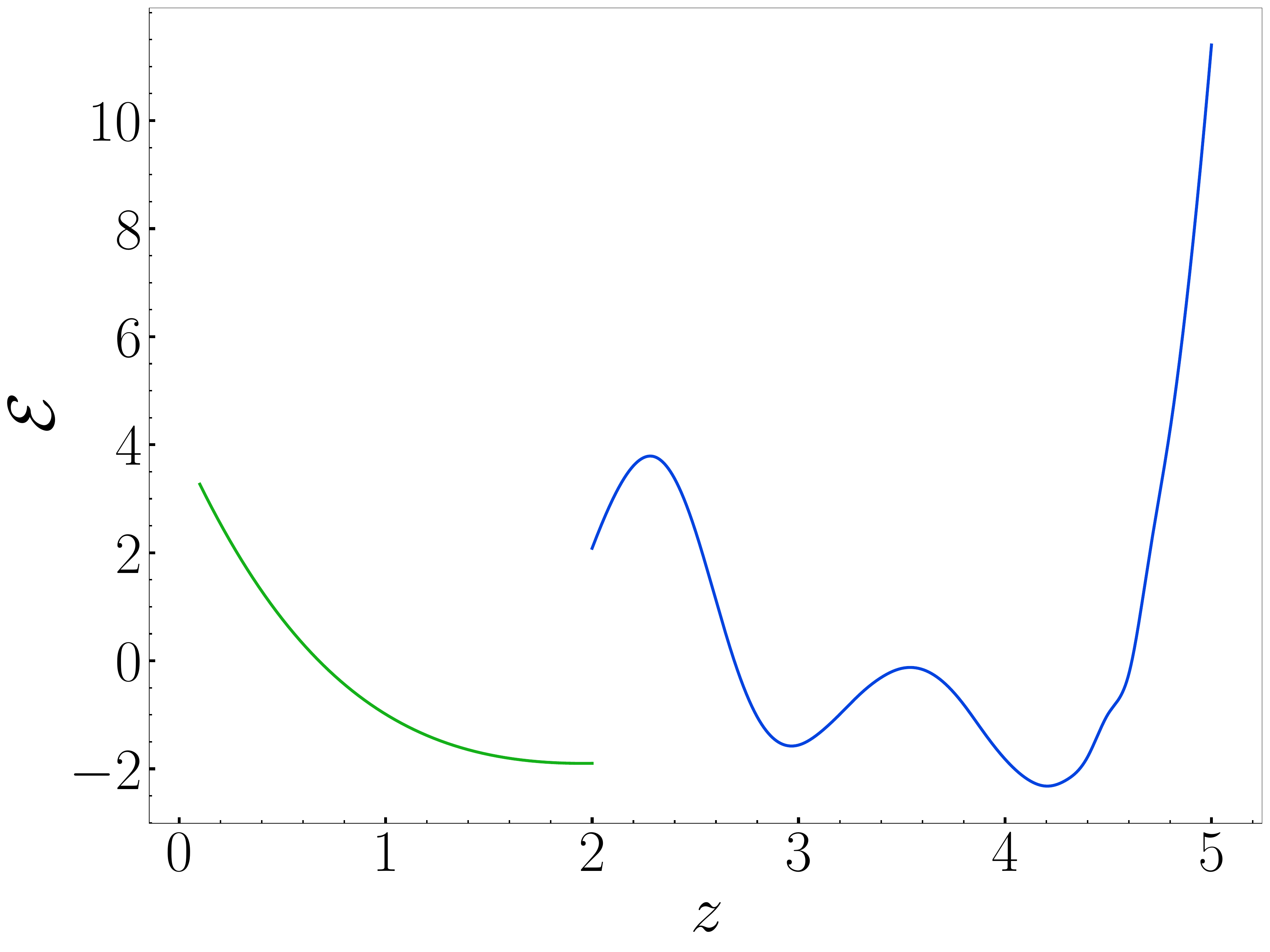}
\includegraphics[width=0.48\linewidth]{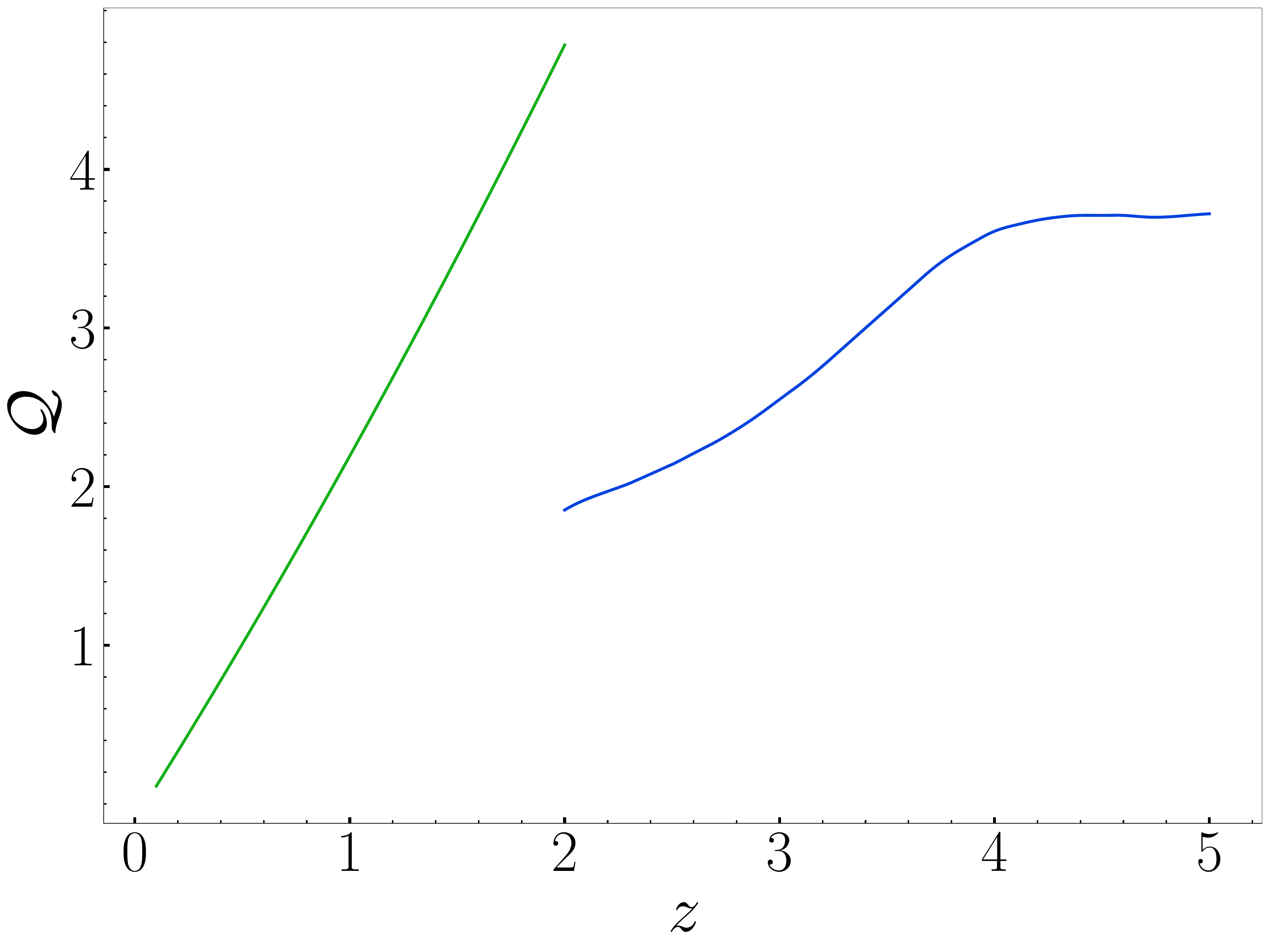} \;
\includegraphics[width=0.48\linewidth]{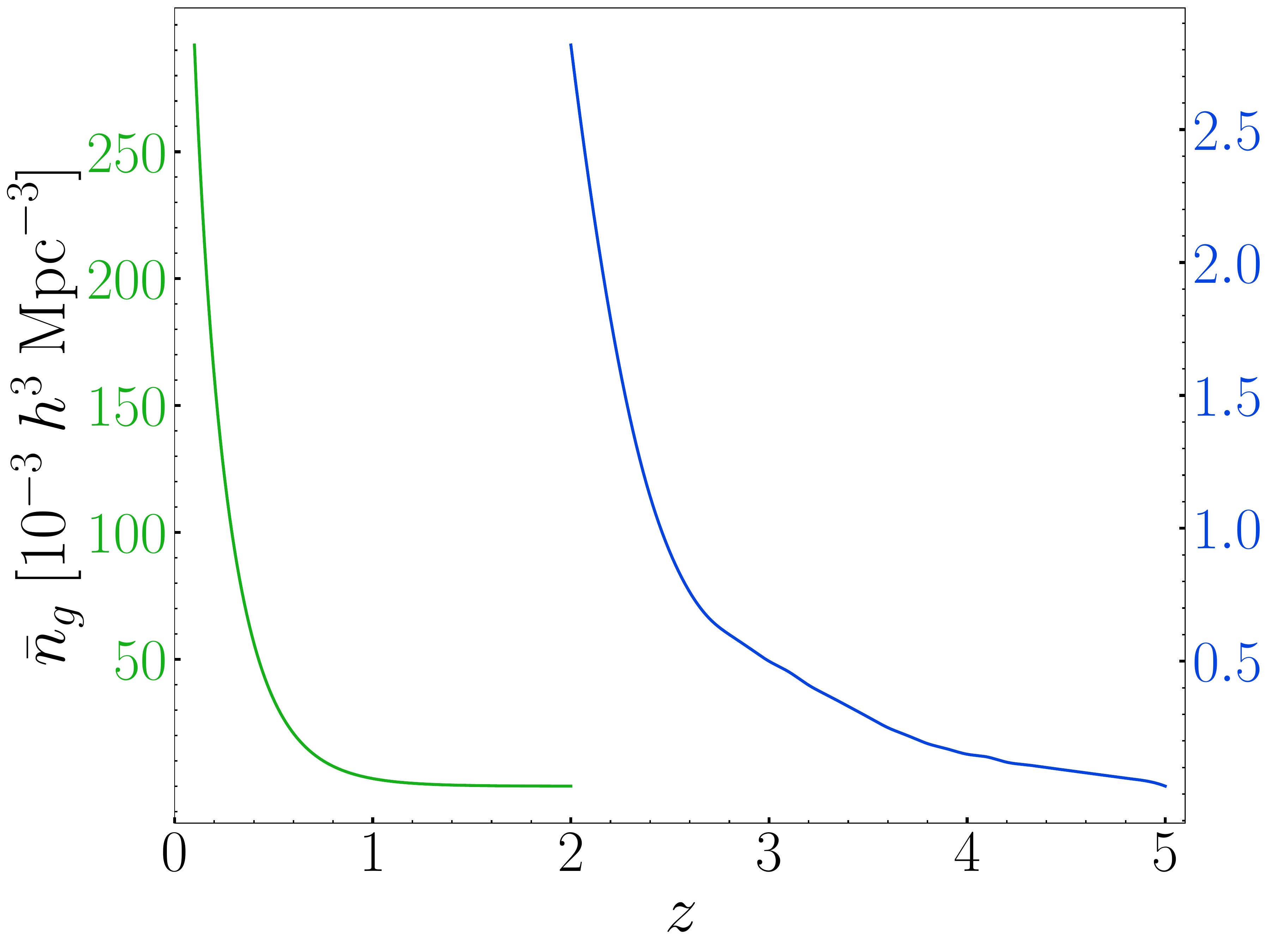}
\caption{Galaxy clustering bias (\emph{top left}), evolution bias (\emph{top right}), magnification bias (\emph{bottom left}) and number density (\emph{bottom right}) for SKAO2 and MegaMapper (with $b_0,\E_0,\Q_0=1$).
} \label{fig:biases}
\end{figure}

\autoref{AppB} shows how the magnification bias $\Q$ and evolution bias $\E$ enter the non-integrated relativistic corrections to the power spectrum monopole and quadrupole.
It is clear from \autoref{eKL}--\autoref{eKISW} that the integrated relativistic corrections also involve these two additional astrophysical parameters, which are defined as  \cite{Maartens:2021dqy}:
\begin{align}
    \Q =-\frac{\p\ln \bar n_g}{\p\ln L_c}\,, \quad \E = -\frac{\p\ln \bar n_g}{\p\ln (1+z)}\,,
\end{align}
where $L_c$ is the luminosity cut corresponding to the survey flux cut.
Then  the Gaussian clustering bias $b$, the background number density $\bar n_{g}$, and $\Q$, $\E$  for each survey are modelled as follows {(note that all fitting functions apply only over the redshift range of the survey)}.
\begin{itemize}
\item SKAO2 HI galaxy survey:
    \begin{align}
    b &= b_{0} \, (0.598 + 0.181 z + 0.438 z^{2}) \, , \qquad &b_{0} = 1.0  \;, \label{eq:bSKA2}
    \\
    {\bar n_g} &= 0.298 z^{-0.191} \exp{(-4.599 z)} ~~~~ h^{3}{\rm Mpc}^{-3} \;, & \label{eq:ngSKA2}
    \\
    \mathcal{Q} &= \mathcal{Q}_{0} \, (-0.104 + 2.150 z + 0.147 z^{2}) \, , \qquad &\mathcal{Q}_{0} = 1.0  \;,  \label{eq:QSKA2}
    \\ 
    \mathcal{E} &= \mathcal{E}_{0} \, (4.085 - 4.491 z - 2.282 z^{2}) \exp{(-z)} \, , \qquad  &\mathcal{E}_{0} = 1.0  \;.  \label{eq:beSKA2}
    \end{align}
The functions  $\bar n_g, \mathcal{Q}, \mathcal{E}$ are understood to be evaluated at the luminosity cut. 
Here the clustering bias $b$ is a fit function derived from simulation data, based on a flux sensitivity threshold of $S_{\rm rms} = 5 \, \mu\rm{Jy}$ over the redshift range $0.1 \leq z \leq 2.0$, as described in Table $\rm A1$ from \cite{Yahya:2014yva}. The functions $\bar n_{g}$, $\mathcal{Q}$ and $\mathcal{E}$ are modelled as fits to simulation data, given in Table~2 of \cite{Maartens:2021dqy}.
Uncertainties in the three biases are accounted for by the amplitude parameters $b_0, \Q_0, \E_0$, each with fiducial value 1.

\item MegaMapper LBG survey:
    \begin{align}
    b &= b_{0} \, (0.710 + 0.820 z + 0.110 z^{2}) \, ,  \qquad \qquad \qquad    &b_{0} = 1.0  \;, \label{eq:bLBG}
    \end{align}
is a fit from \cite{Kopana:2023uew} to the values in \cite{Sailer:2021yzm}. The values of $\bar n_{g}$, $\mathcal{Q}$ and $\mathcal{E}$ in each redshift bin are given in \autoref{appc}, \autoref{tab:LBGParameters}.
As in the case of SKAO2, we multiply the tabulated values $\E,\Q$ by an amplitude factor with fiducial value 1: $\E \to \E_0\,\E$, $\Q \to \Q_0\,\Q$.
\end{itemize}

\noindent
The three bias functions and the number densities for the two surveys  are displayed in \autoref{fig:biases}.\\

\begin{figure}[!htbp]
\centering
\vspace*{-1cm}
\includegraphics[width=0.485\linewidth]{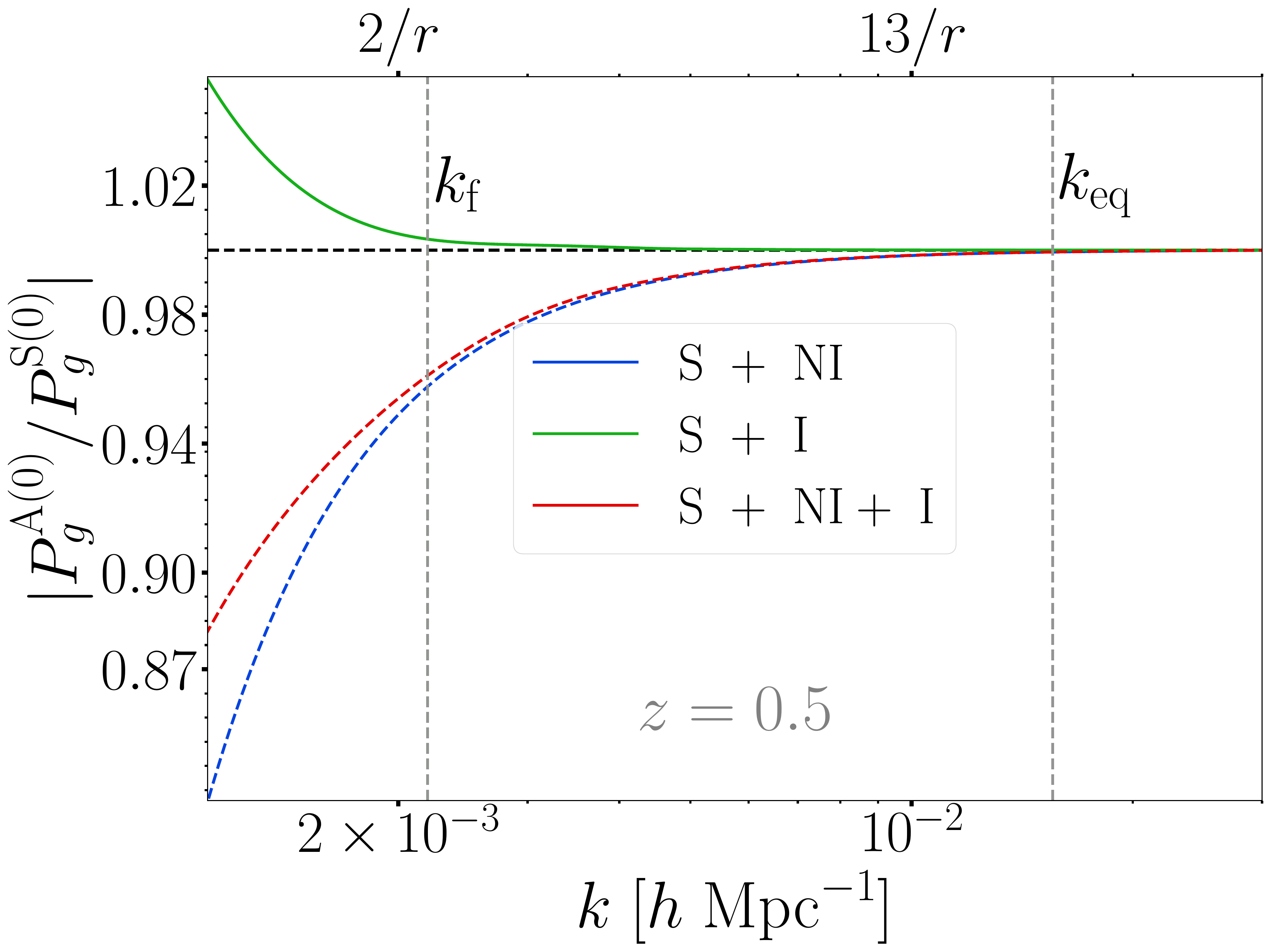}
\includegraphics[width=0.485\linewidth]{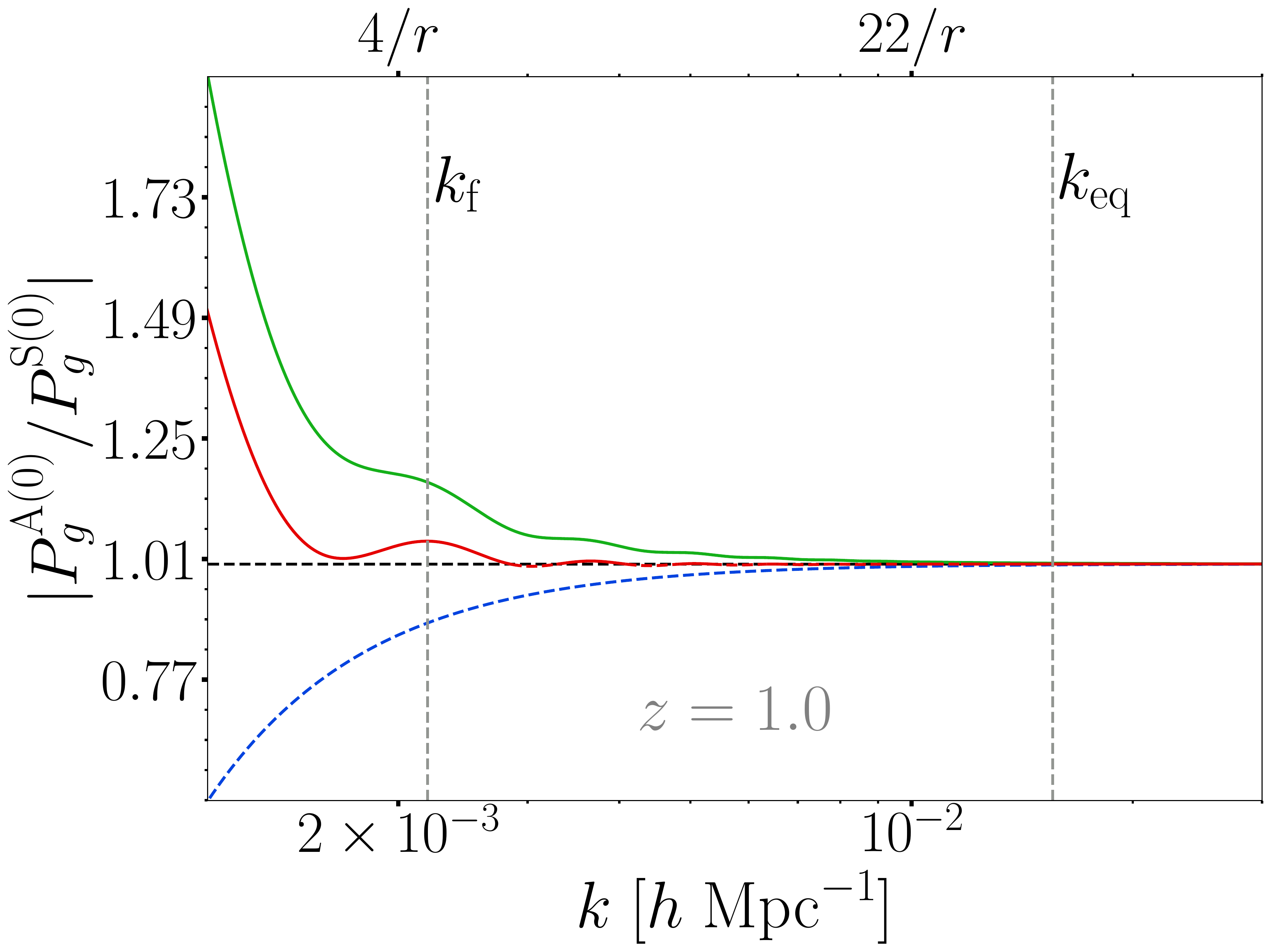}
\includegraphics[width=0.485\linewidth]{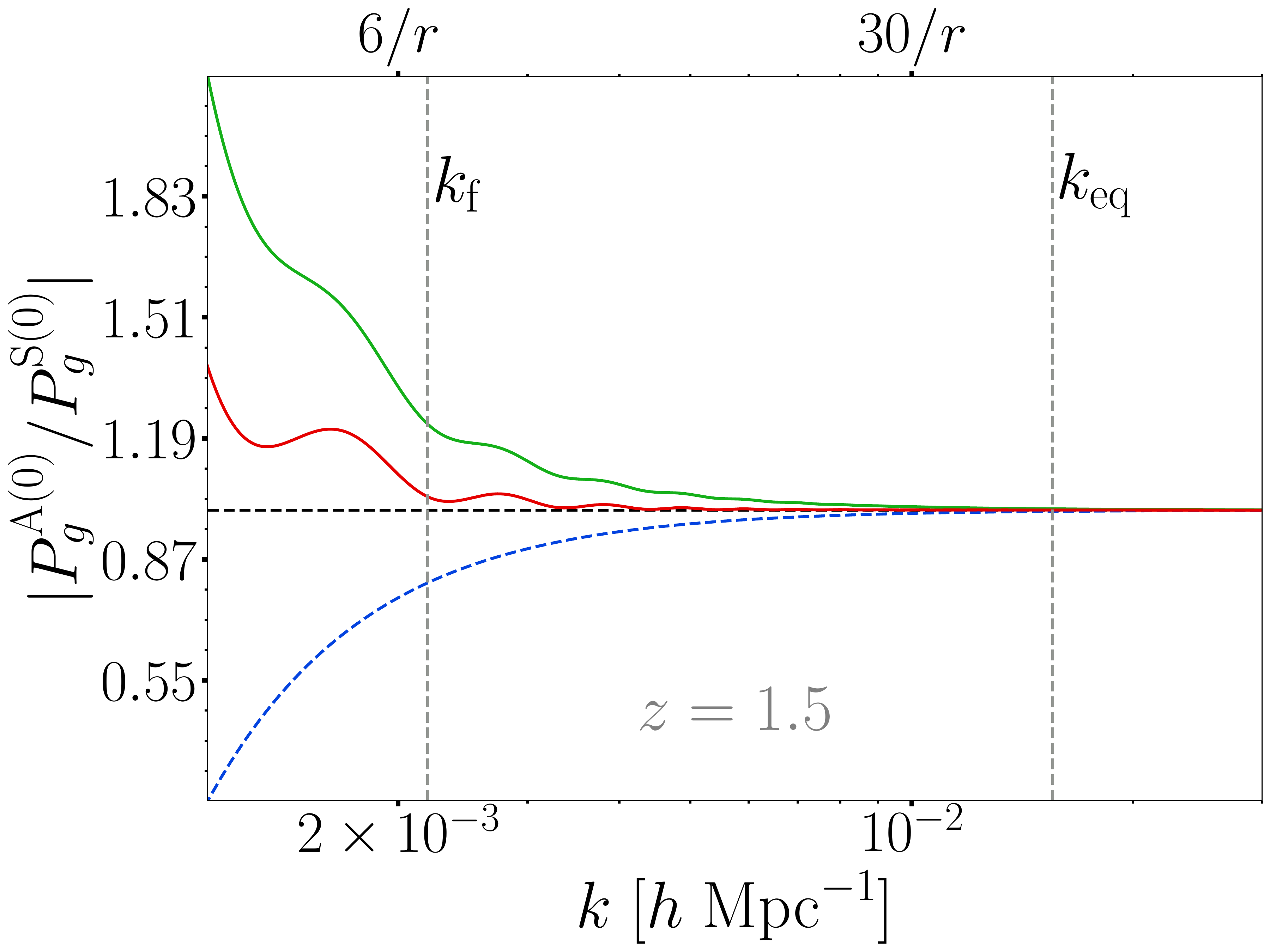}
\includegraphics[width=0.485\linewidth]{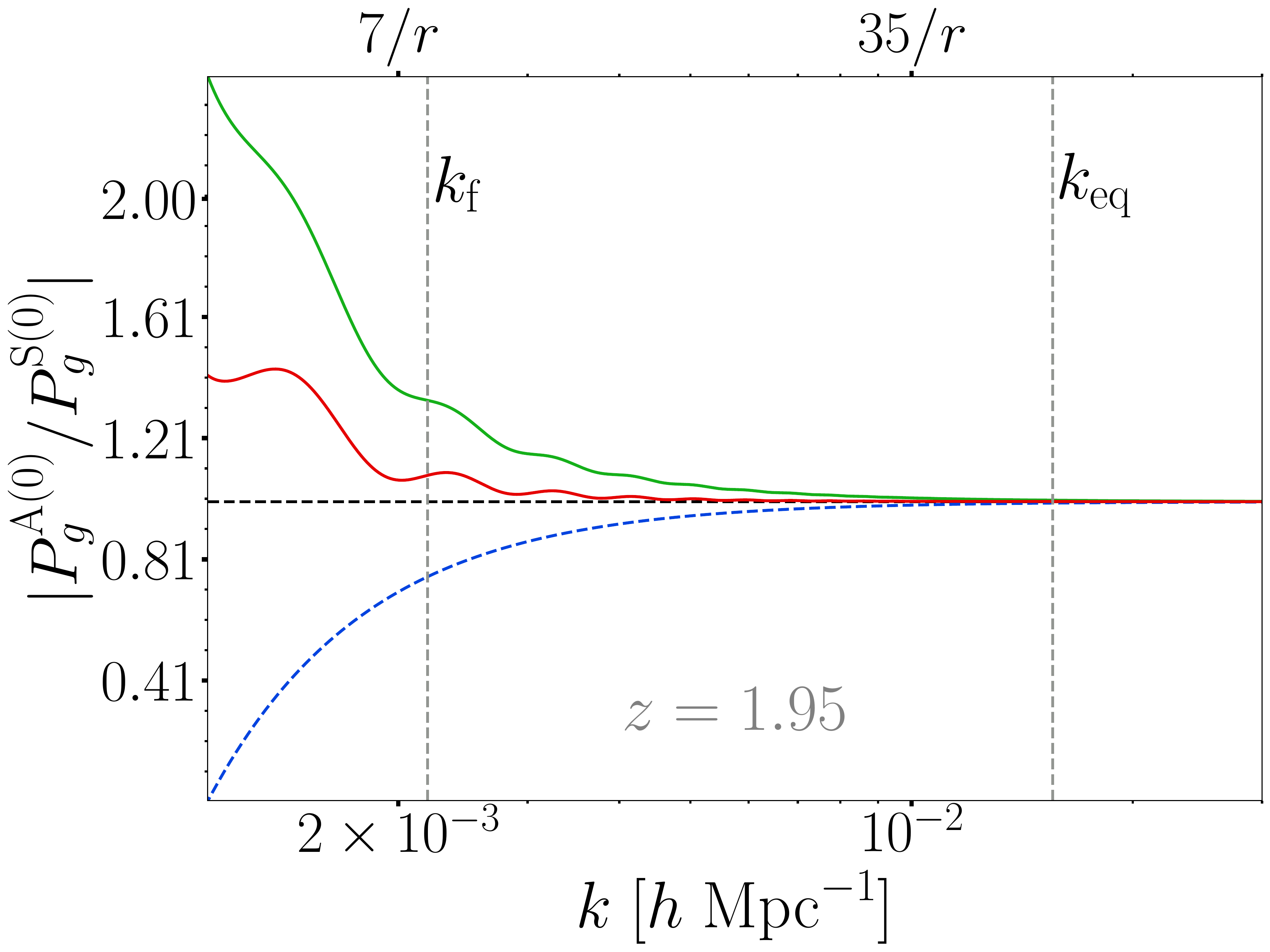}
\includegraphics[width=0.485\linewidth]{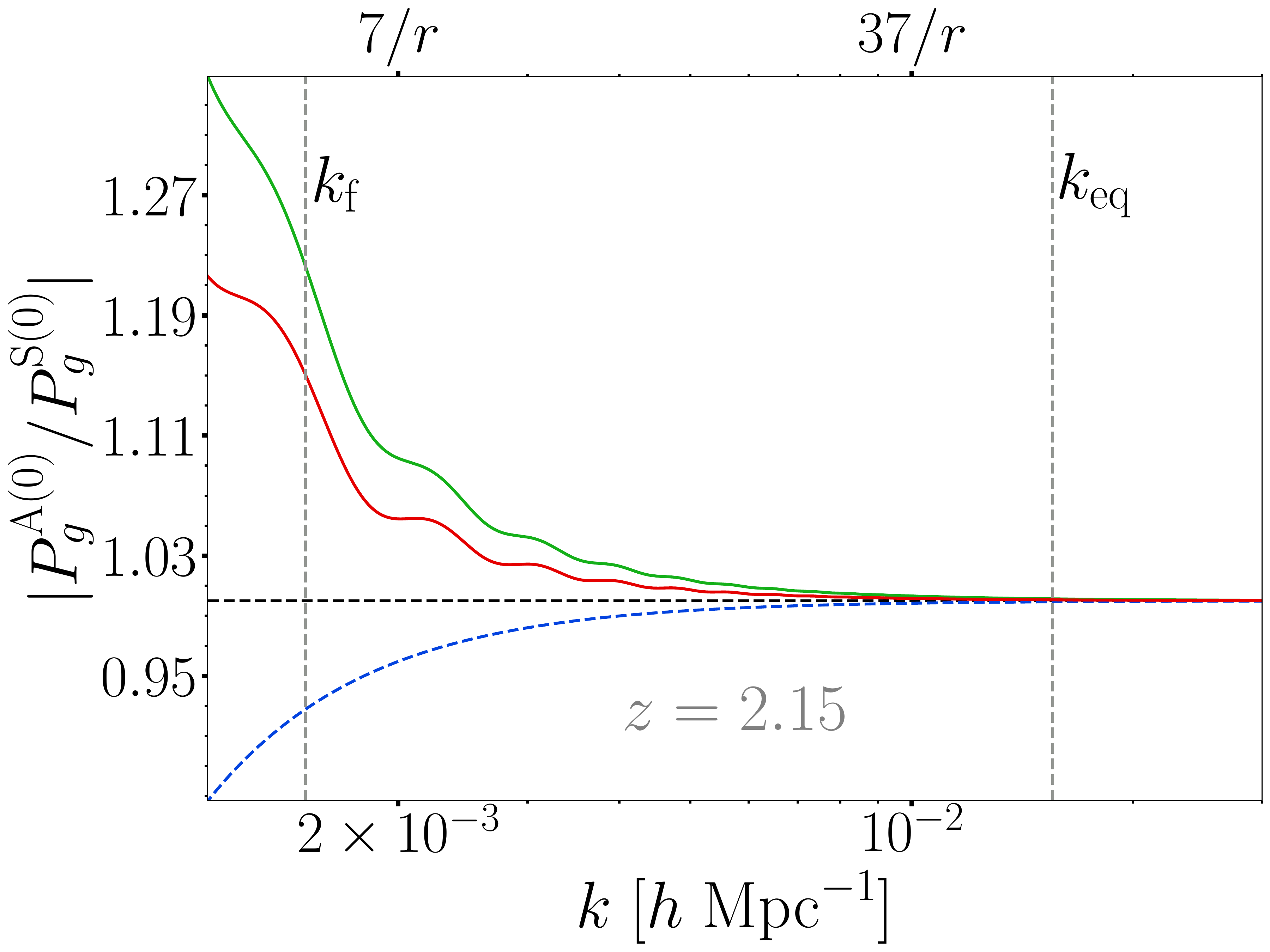}
\includegraphics[width=0.485\linewidth]{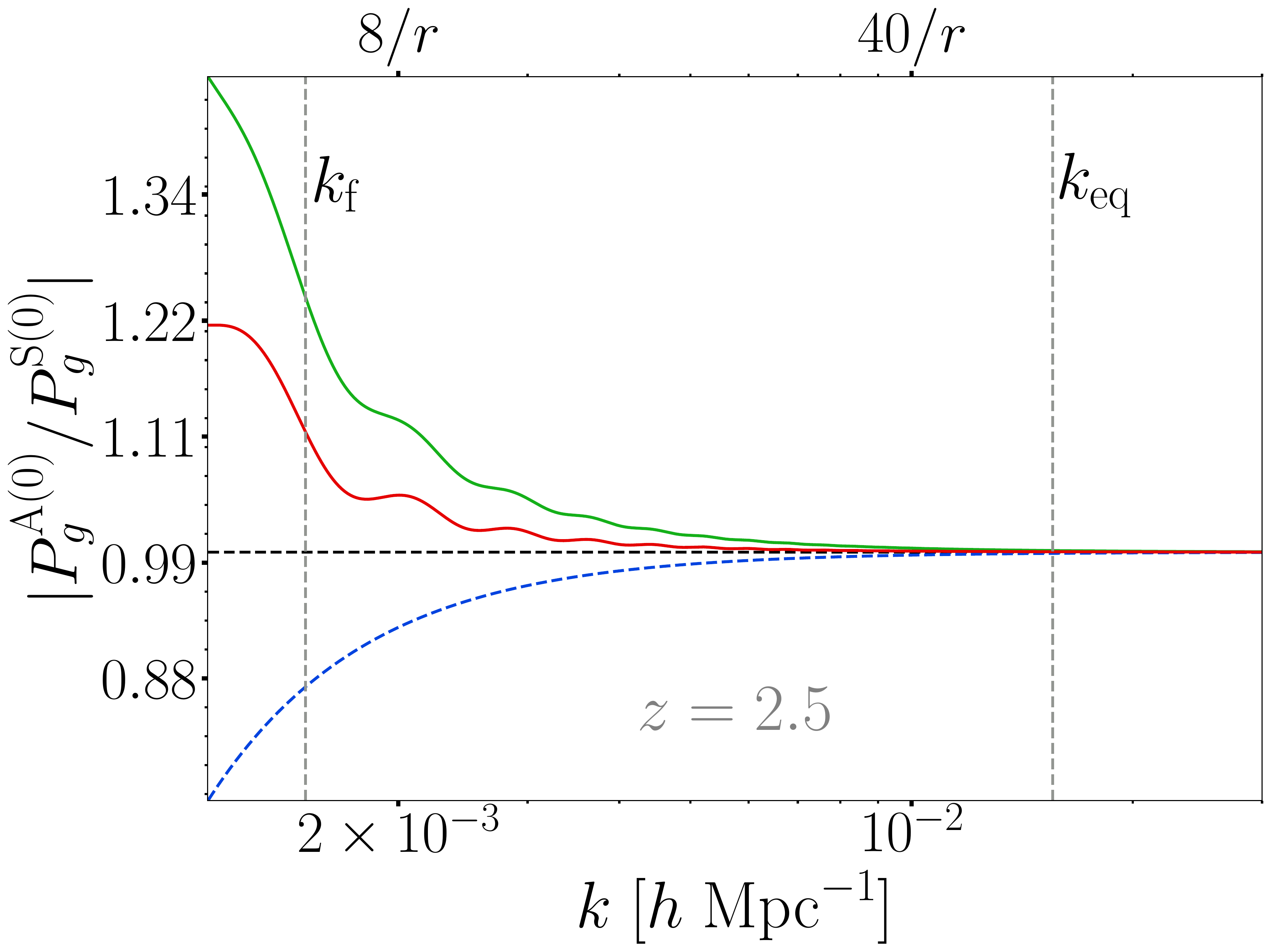}
\includegraphics[width=0.485\linewidth]{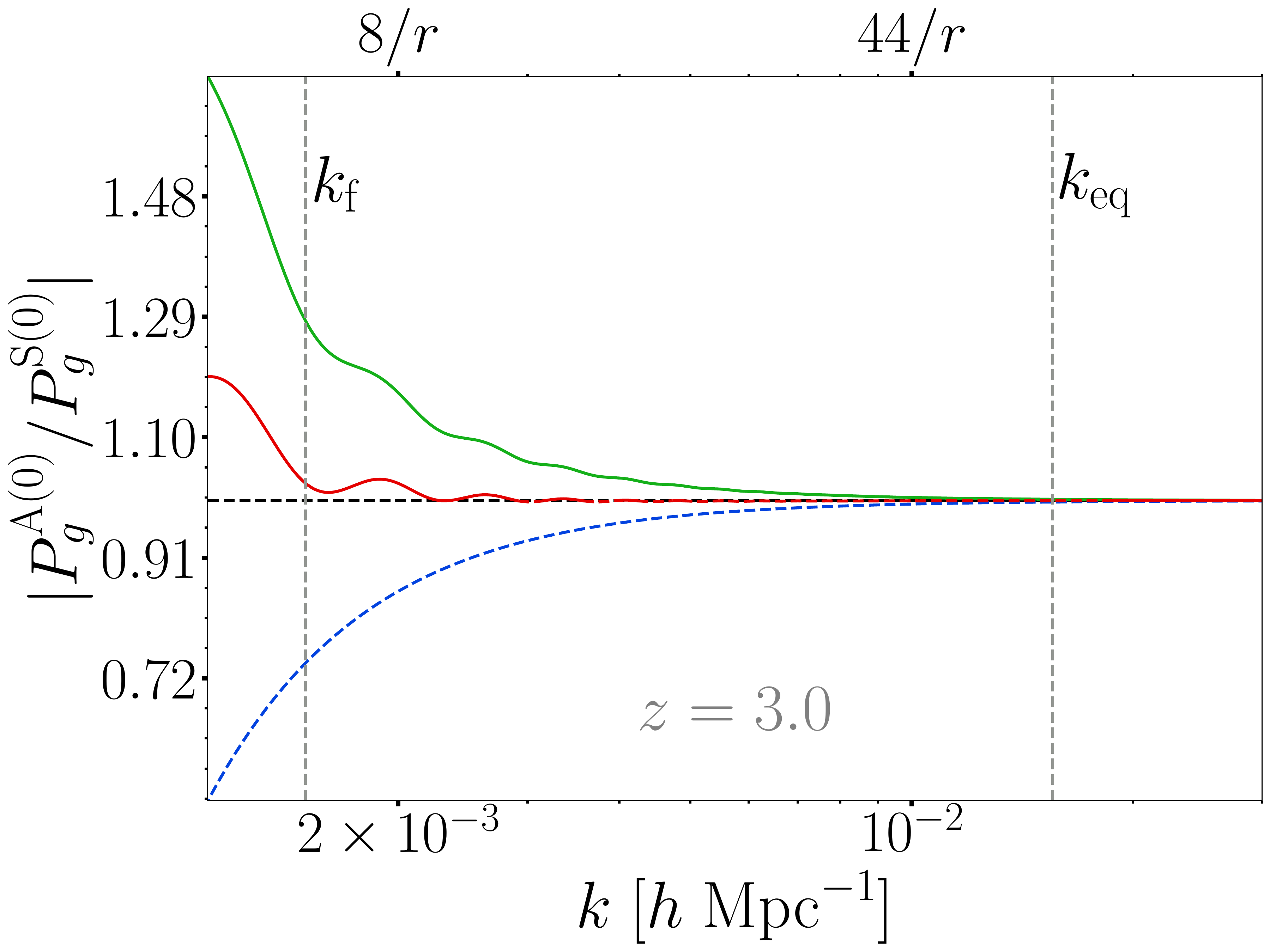}
\includegraphics[width=0.485\linewidth]{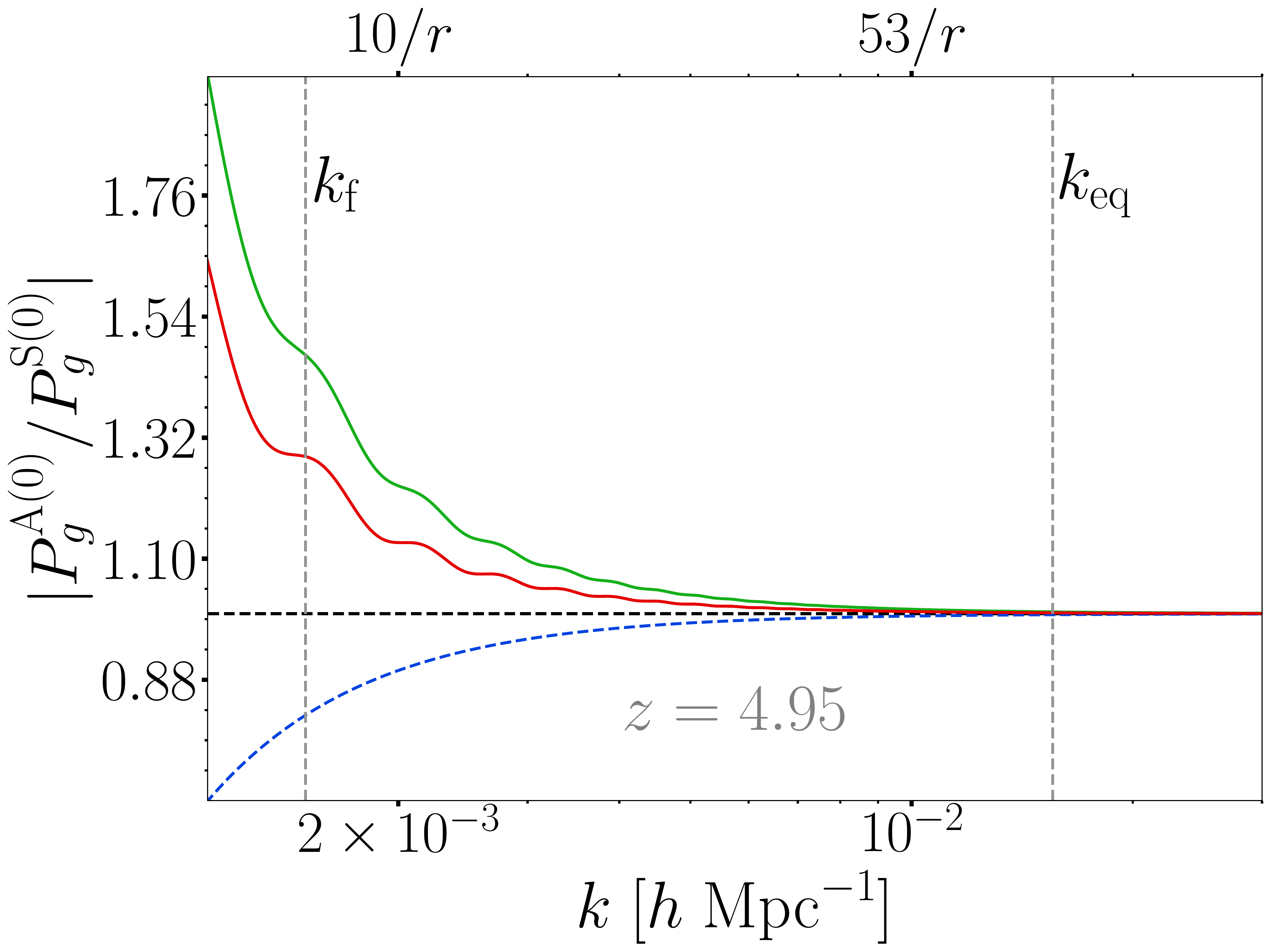}
\caption{Relative contributions
 of non-integrated (NI), integrated (I) and total (NI + I) corrections to the standard monopole -- i.e., $P_g^{{\rm A}(0)}/P_g^{{\rm S}(0)}$, where
 A = S+NI (blue), S+I (green), S+NI+I (red).
{\em Top two rows} SKAO2, {\em bottom two rows} MegaMapper. 
Dashed lines indicate negative values.
\red{Top horizontal axes show multiples of $1/r$ corresponding to $k$ on the bottom axes (wide-angle expansions require $k>1/r$).}
}
\label{fig:monopolesSKA2_LBG_SxI_SxNI}
\end{figure}

\begin{figure}%[!htbp]
\centering
\vspace*{-1.0cm}
\includegraphics[width=0.495\linewidth]{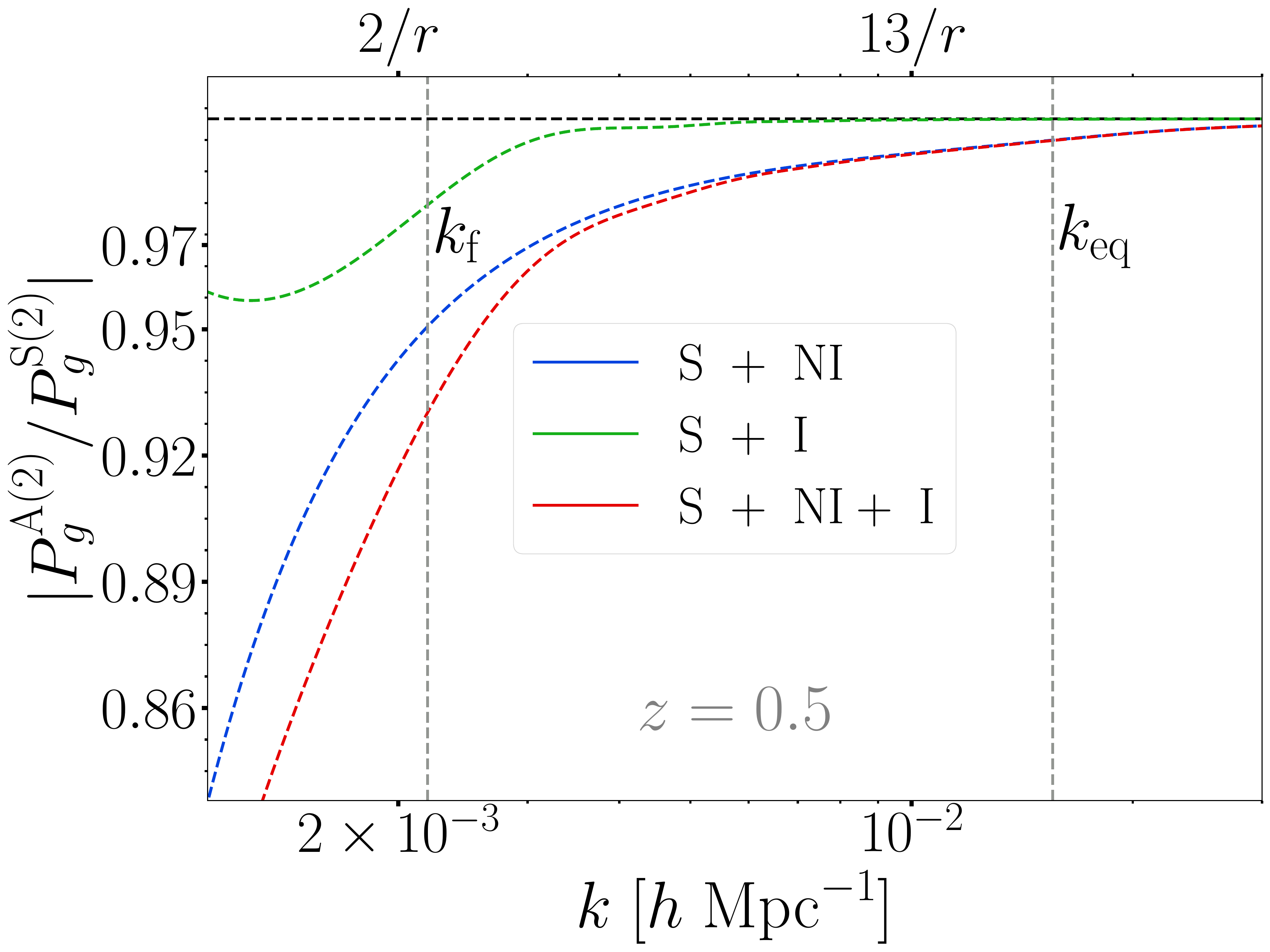}
\includegraphics[width=0.495\linewidth]{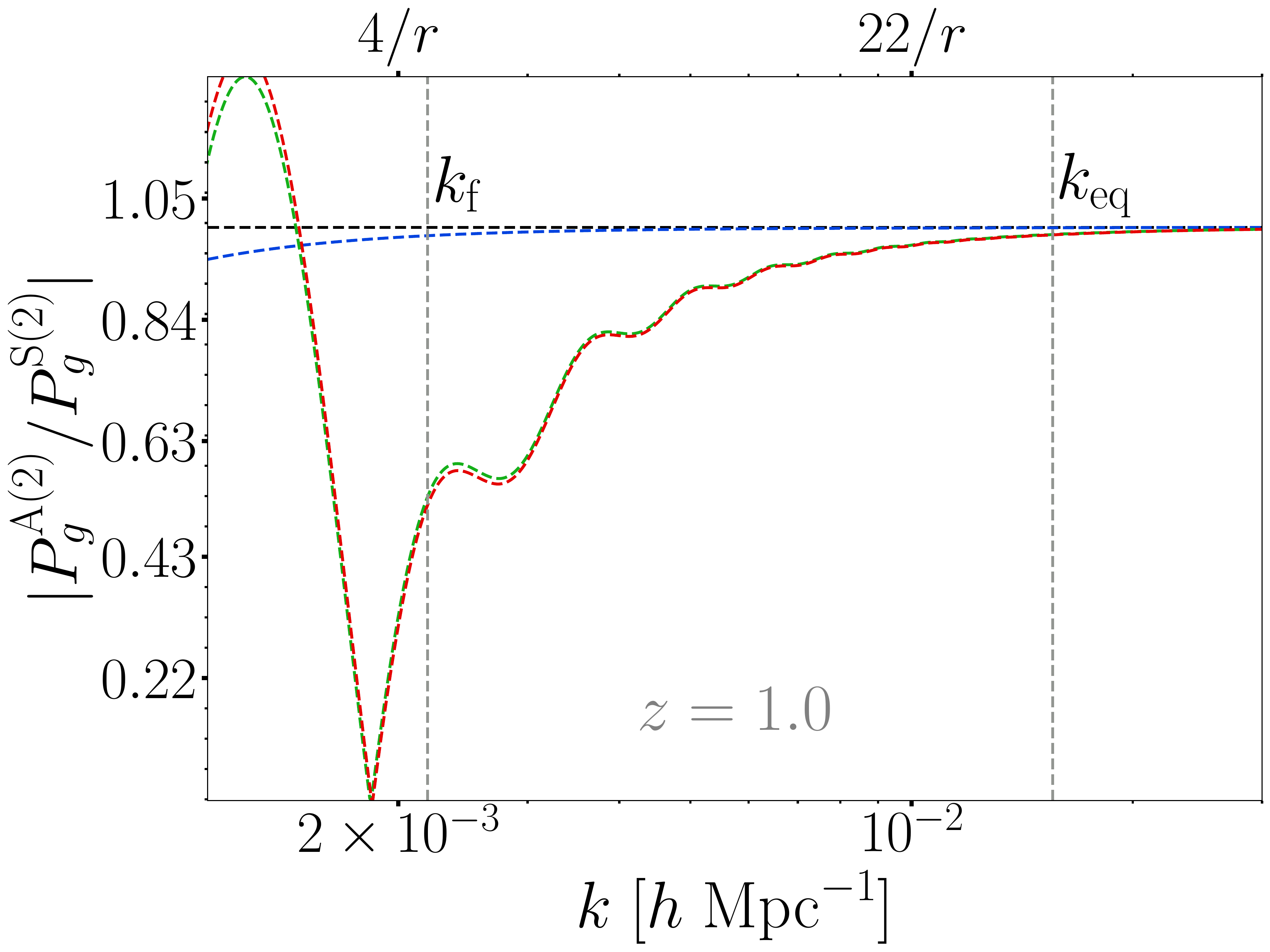}
\includegraphics[width=0.495\linewidth]{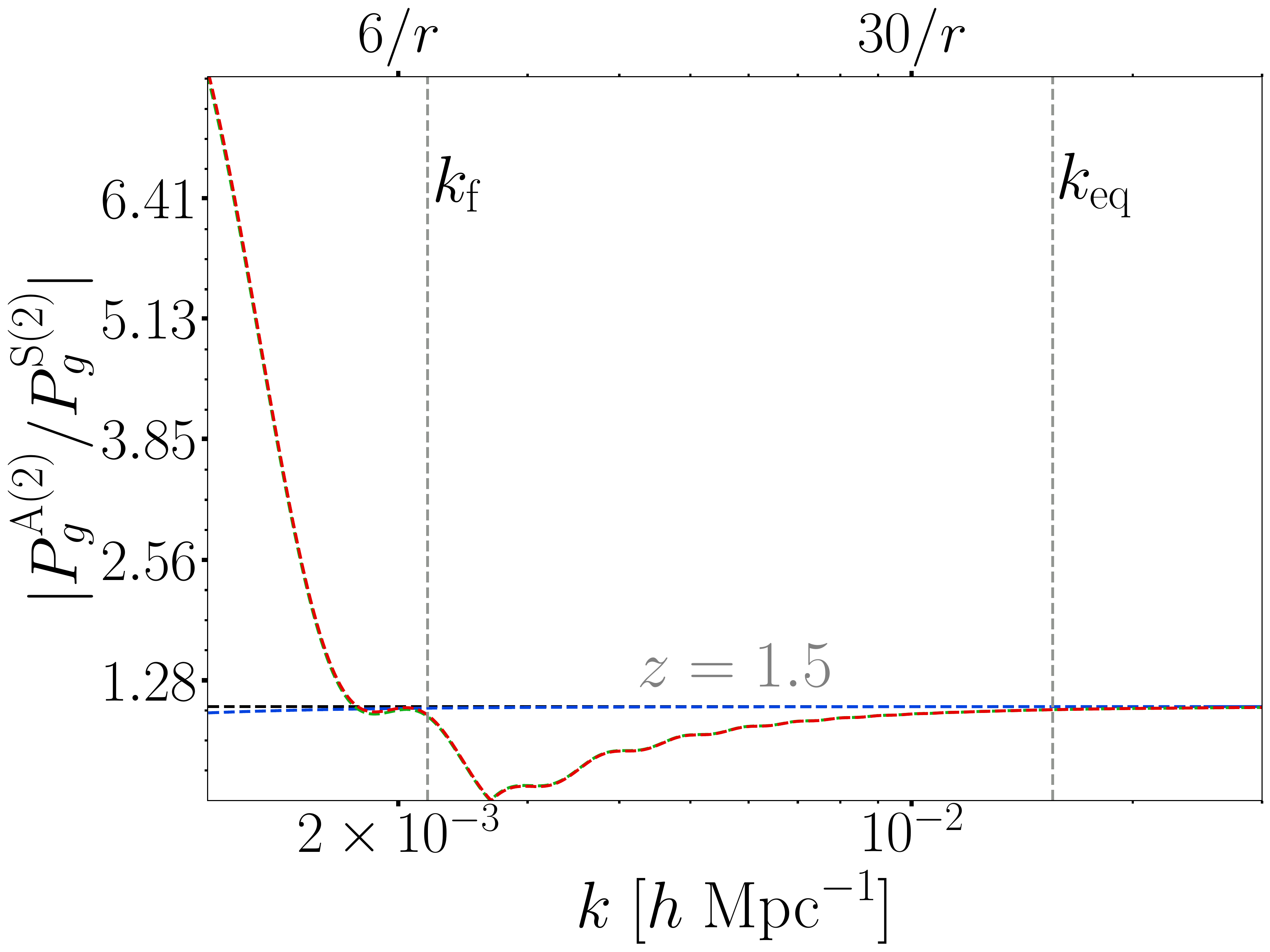}
\includegraphics[width=0.495\linewidth]{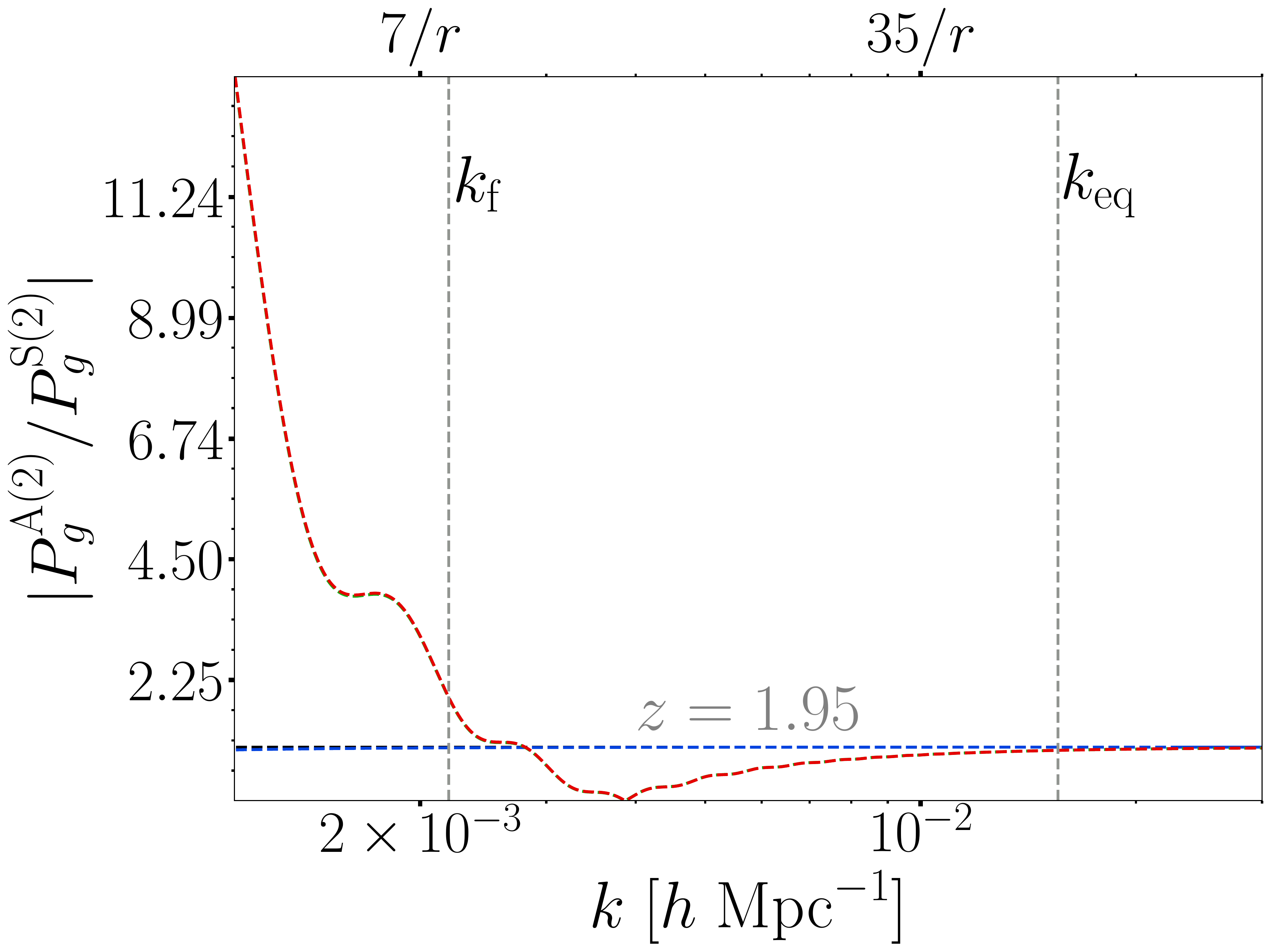}
\includegraphics[width=0.495\linewidth]{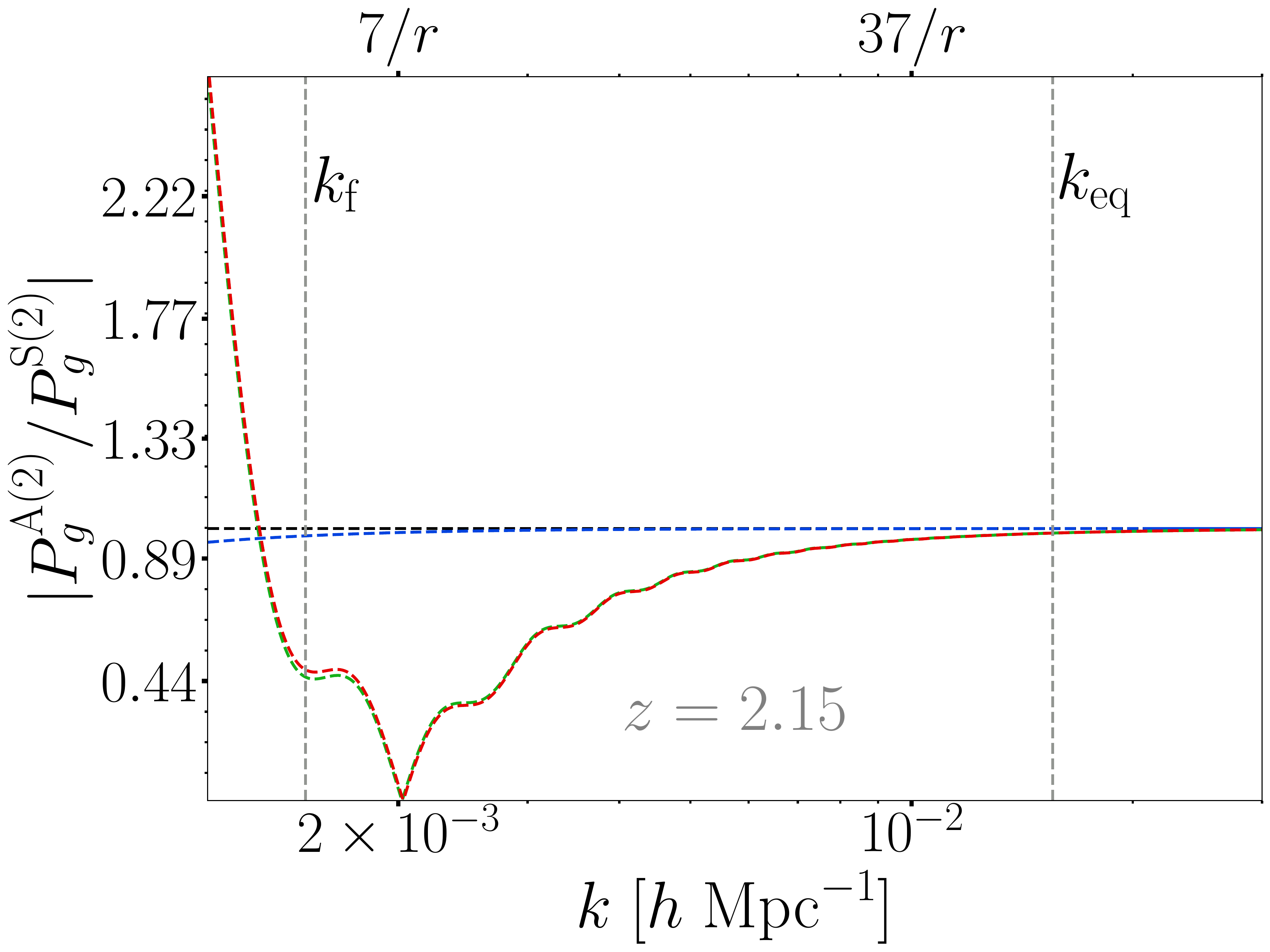}
\includegraphics[width=0.495\linewidth]{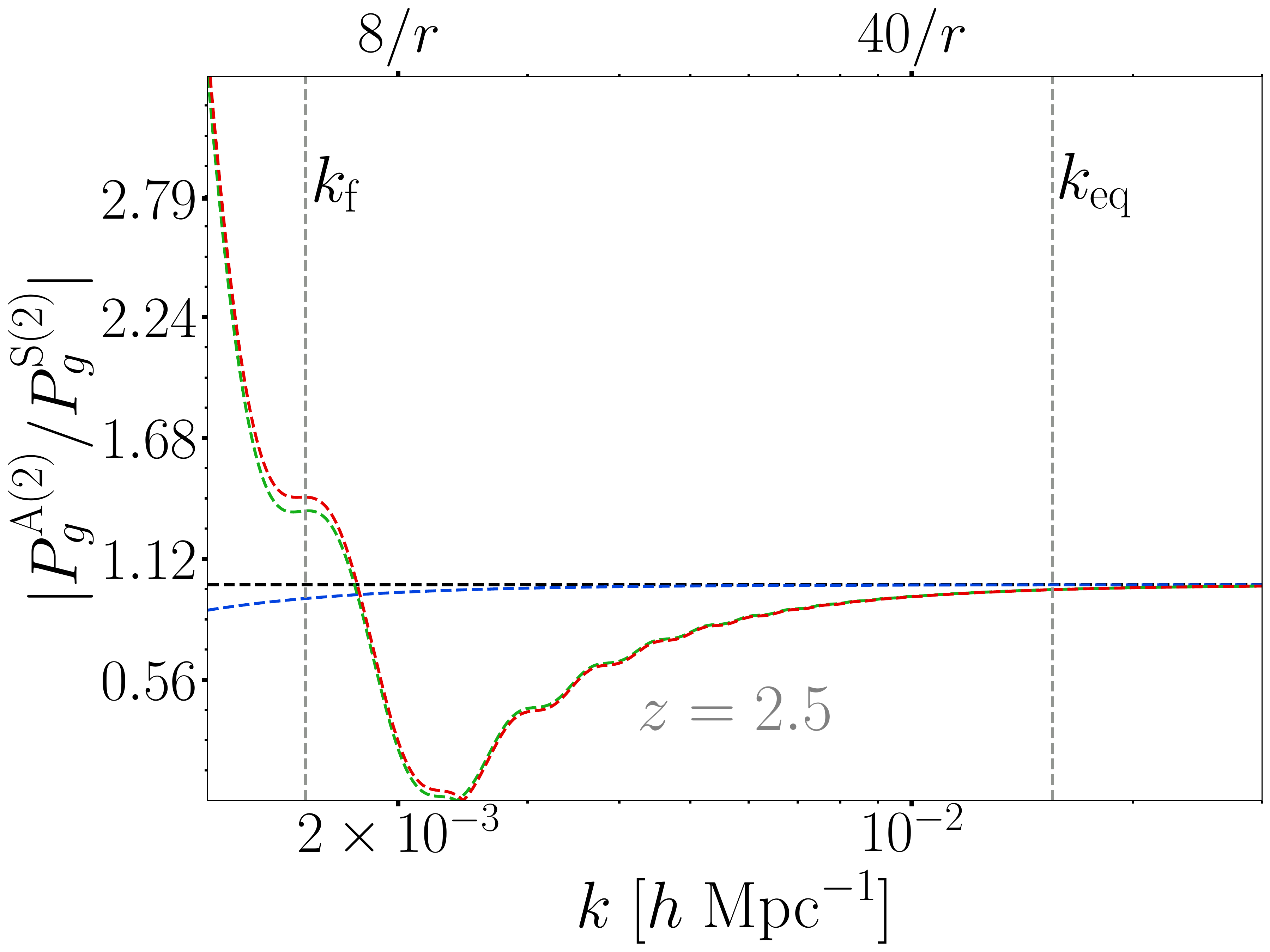}
\includegraphics[width=0.495\linewidth]{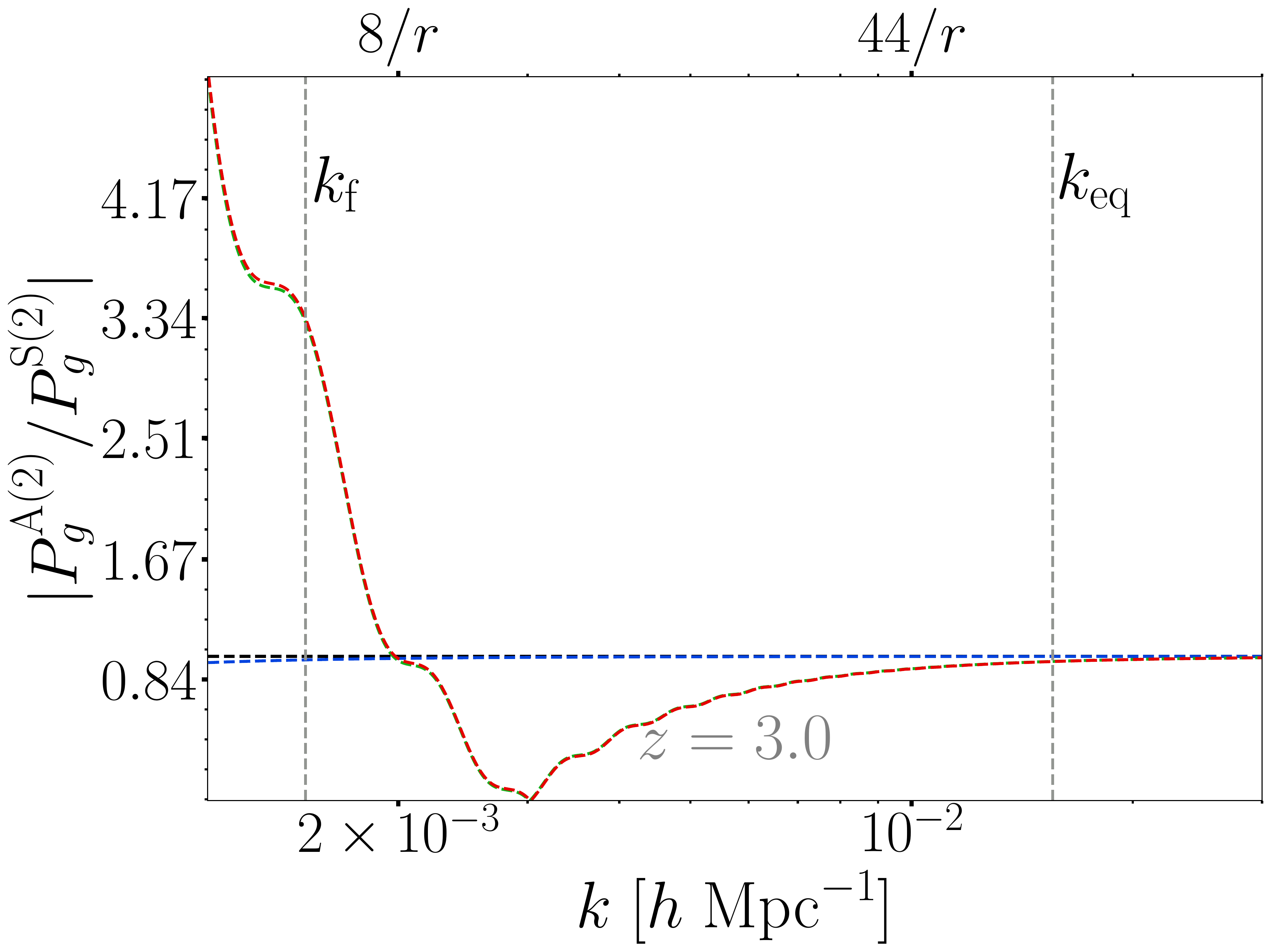}
\includegraphics[width=0.495\linewidth]{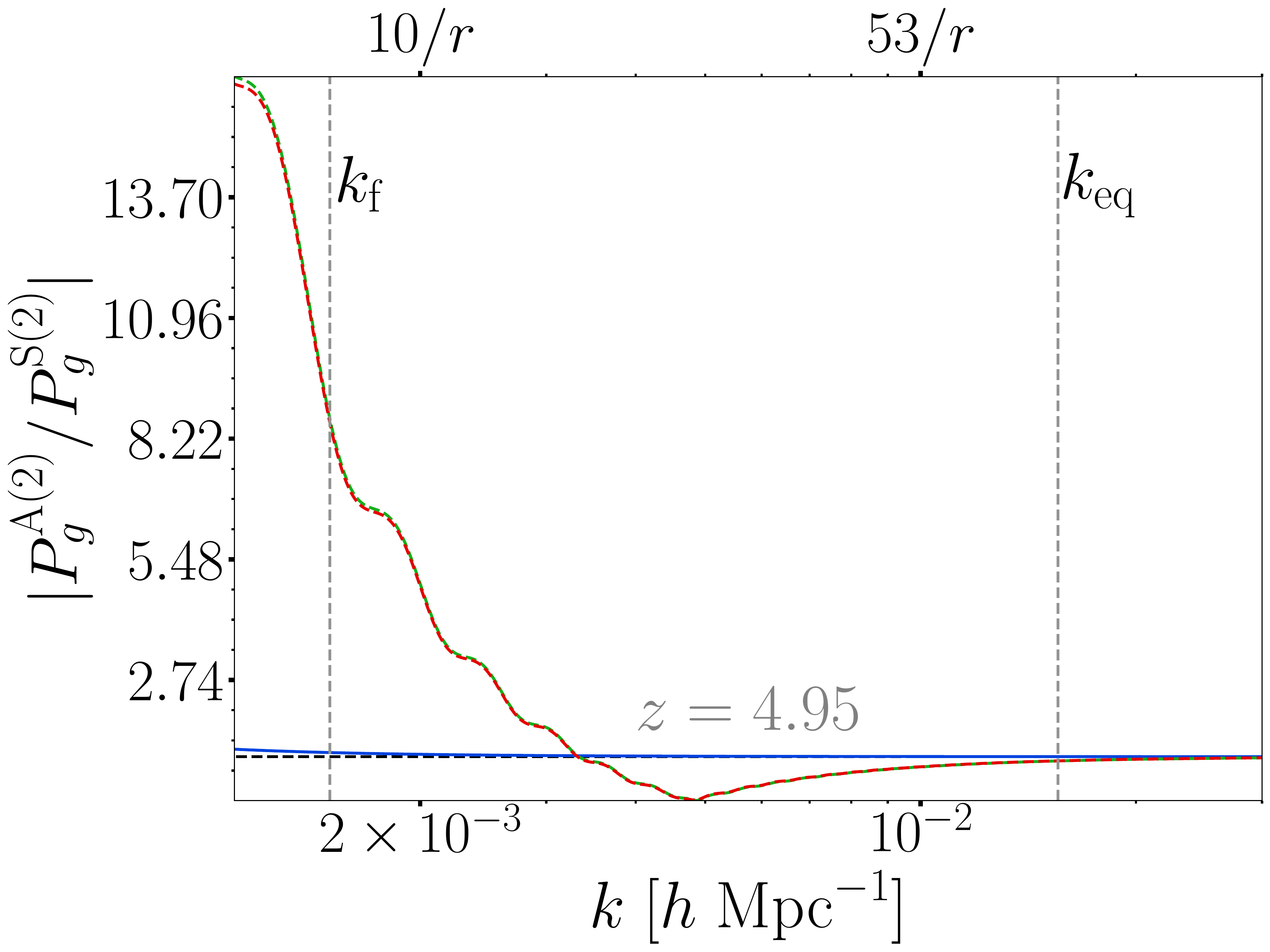}
\caption{Same as in  \autoref{fig:monopolesSKA2_LBG_SxI_SxNI}, but for the relative contributions
 of the non-integrated (NI), integrated (I) and total (NI + I) corrections to the standard quadrupole.}
\label{fig:quadrupolesSKA2_LBG_SxI_SxNI}
\end{figure}

In \autoref{fig:monopolesSKA2_LBG_SxI_SxNI} we show the relative contributions of the non-integrated (NI), integrated (I) and total (NI + I) corrections to the standard (S) power spectrum monopole. 
{The top horizontal axes show multiples of $1/r$ corresponding to $k$ on the bottom horizontal axes. The wide-angle expansions are made in terms of $1/(kr)$, see \autoref{ExpPtotal}, which must remain less than 1, so that $n>1$ for validity of the wide-angle approximation. In all cases, we see that $1/r<k_{\rm f}$, the longest wavelength mode in the redshift bin (see \autoref{sec3}).}
It is clear that the relativistic and wide-angle corrections increase in magnitude as $k$ decreases below the equality scale $k_{\rm eq}$. In addition, it is striking that in nearly all cases the non-integrated (wide-angle and local relativistic) corrections {\em decrease} the ultra-large scale power, while the integrated (lensing + time delay + integrated Sachs-Wolfe) corrections increase this power. As a result, the NI and I corrections partially cancel. {\autoref{fig:quadrupolesSKA2_LBG_SxI_SxNI} displays the quadrupole case. Note that there is no dipole (or higher odd multipoles), since we use the midpoint line of sight for each galaxy pair (see \cite{Jolicoeur:2024oij}).} 

\begin{figure}%[!htbp]
\centering
\vspace*{-1.0cm}
\includegraphics[width=0.495\linewidth]{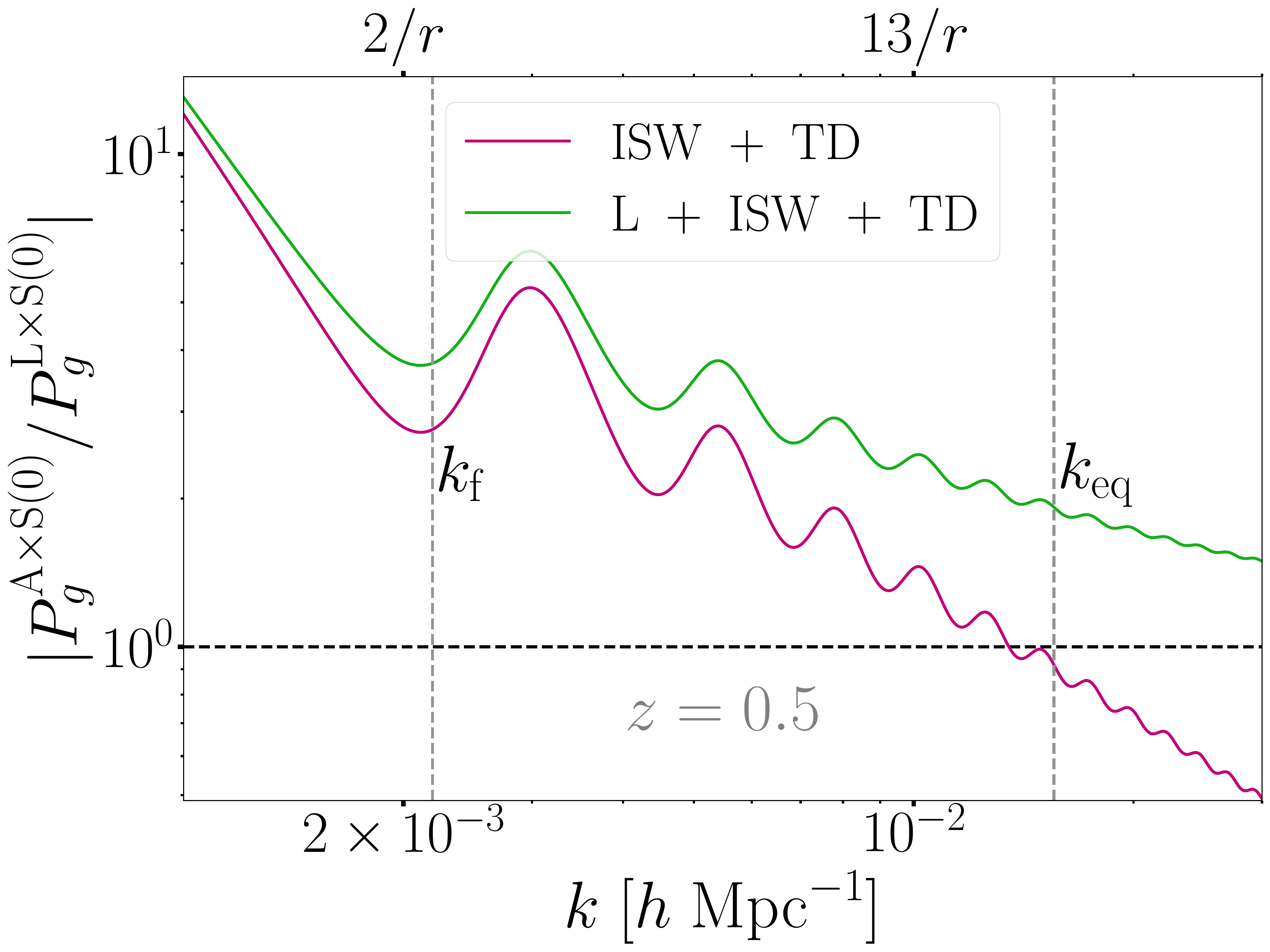}
\includegraphics[width=0.495\linewidth]{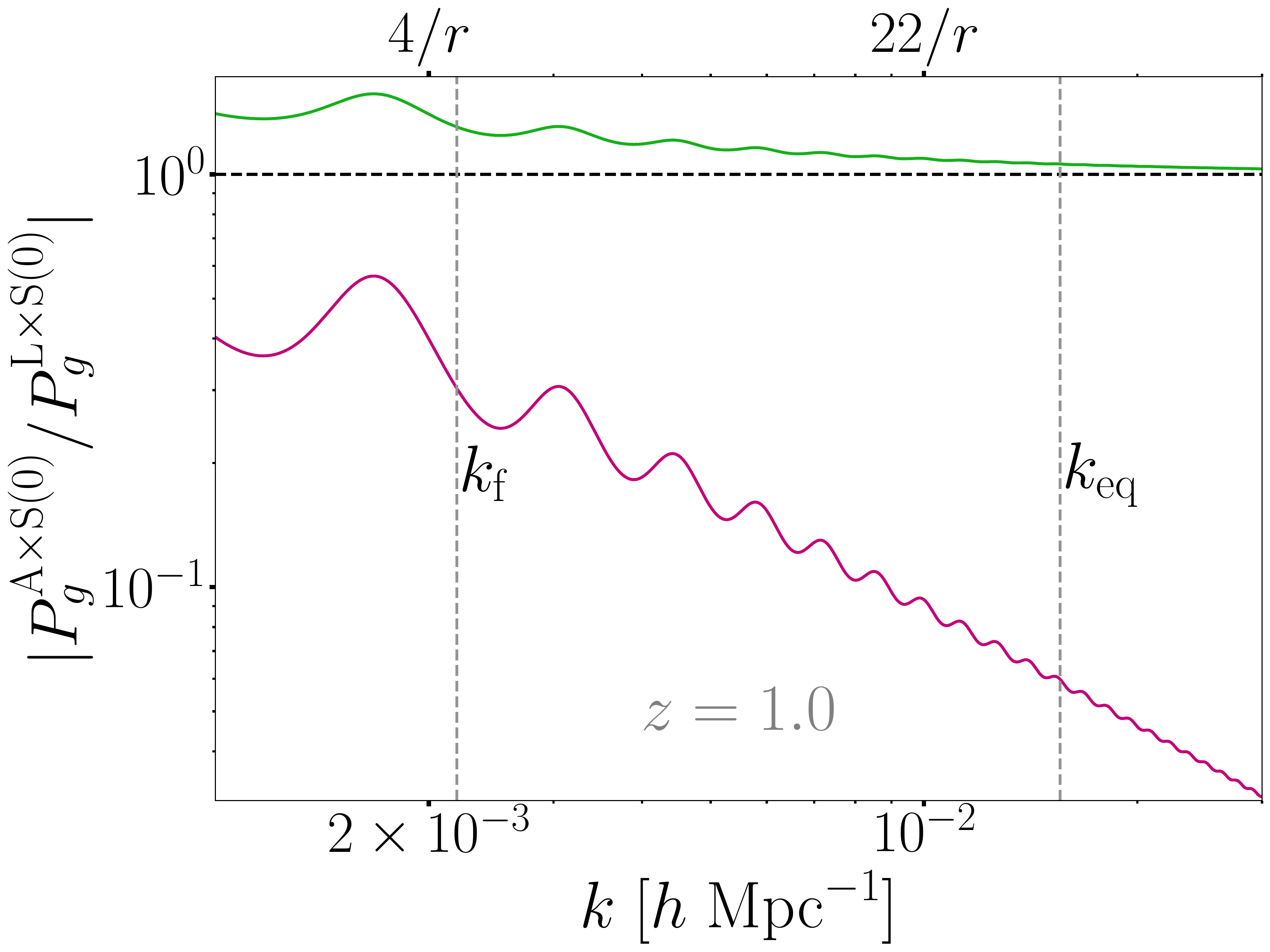}
\includegraphics[width=0.495\linewidth]{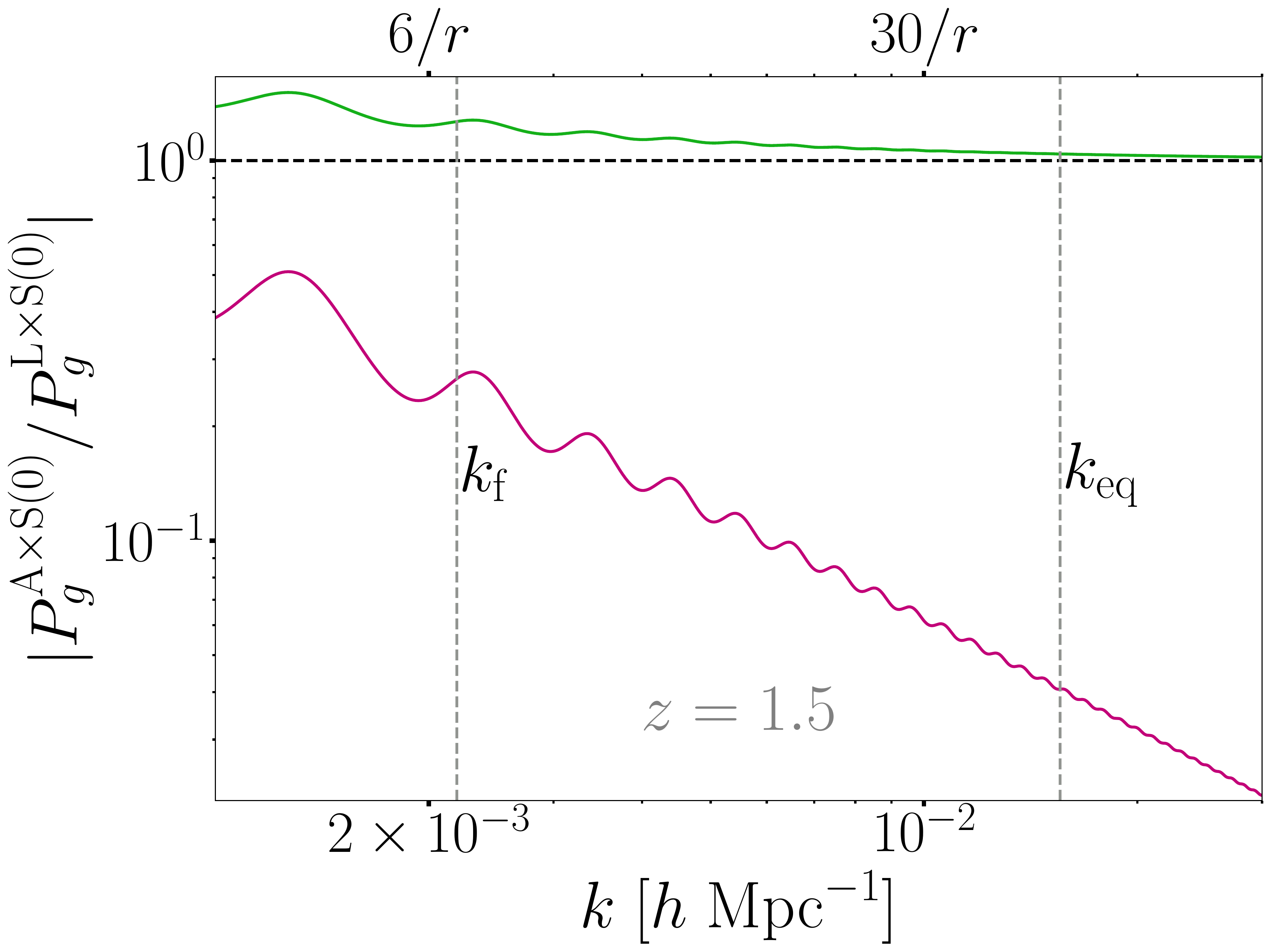}
\includegraphics[width=0.495\linewidth]{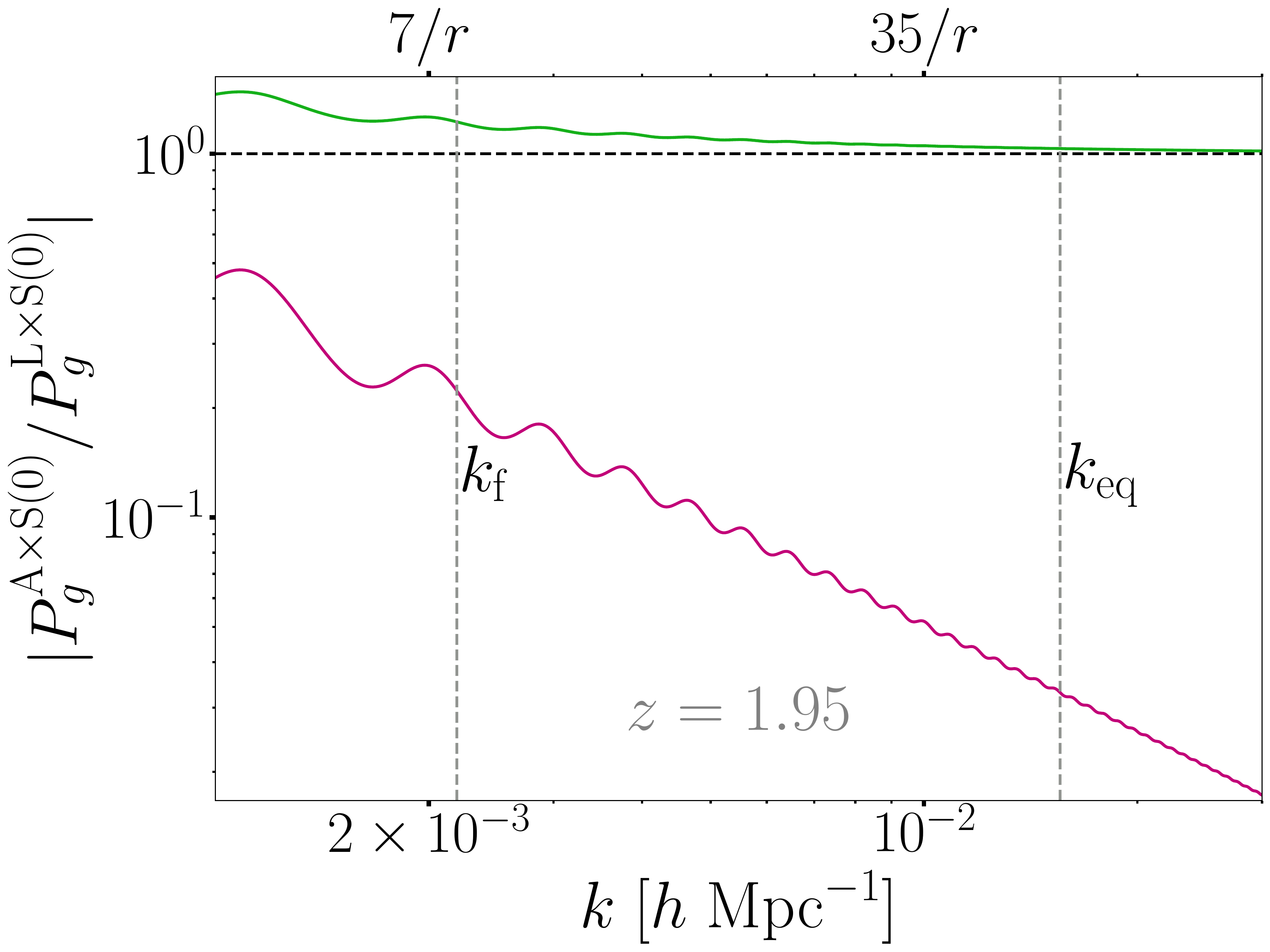}
\includegraphics[width=0.495\linewidth]{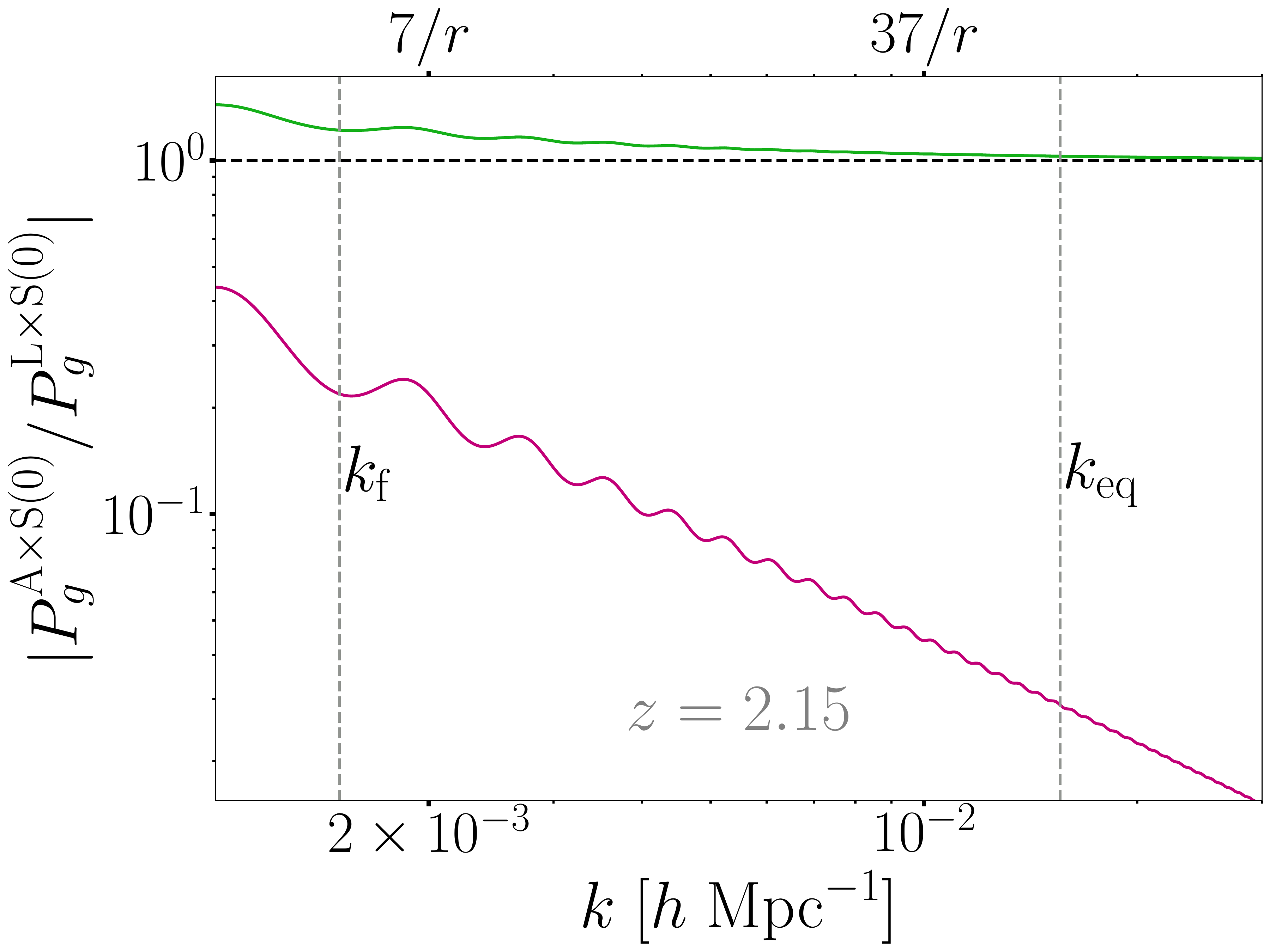}
\includegraphics[width=0.495\linewidth]{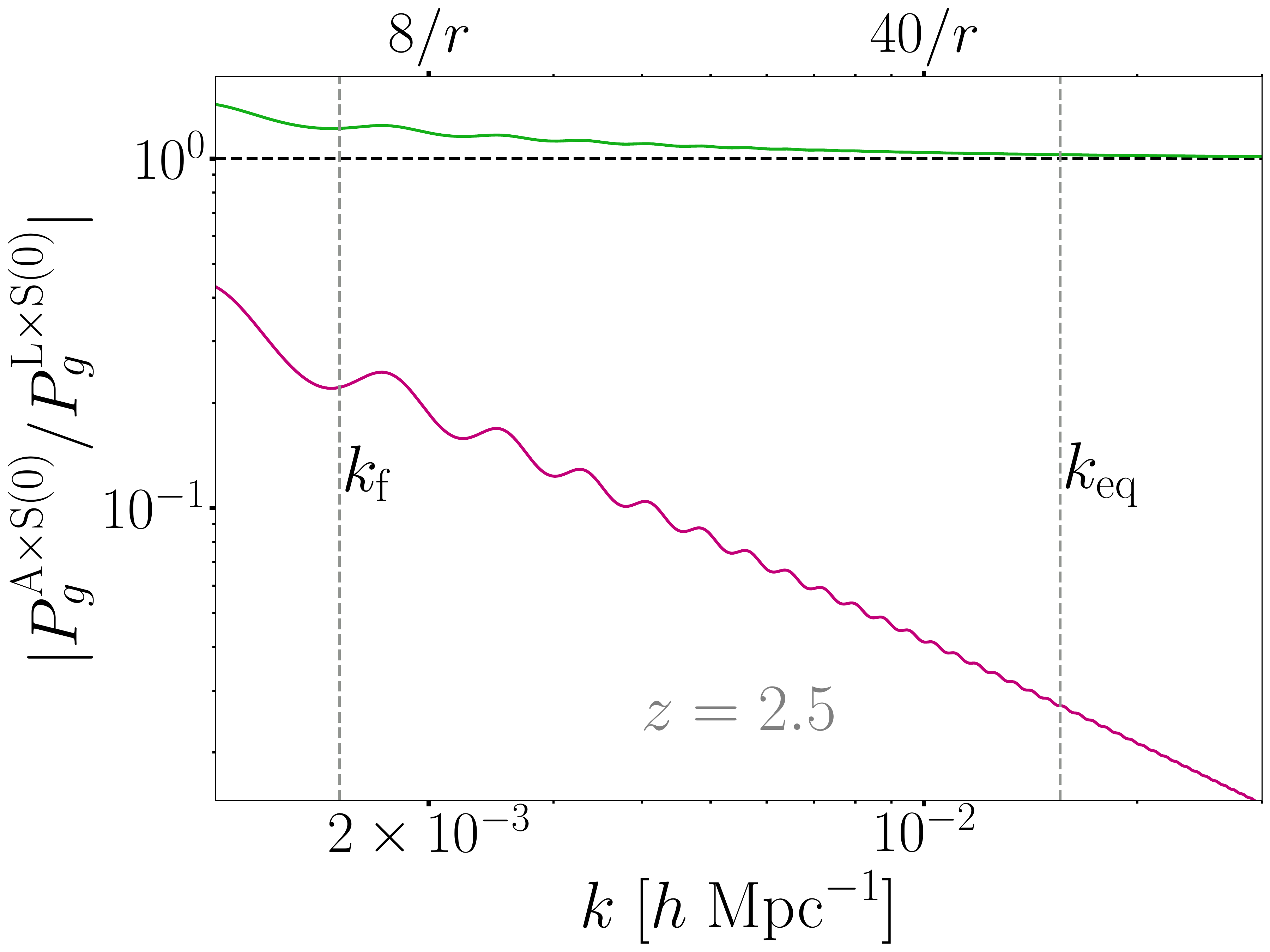}
\includegraphics[width=0.495\linewidth]{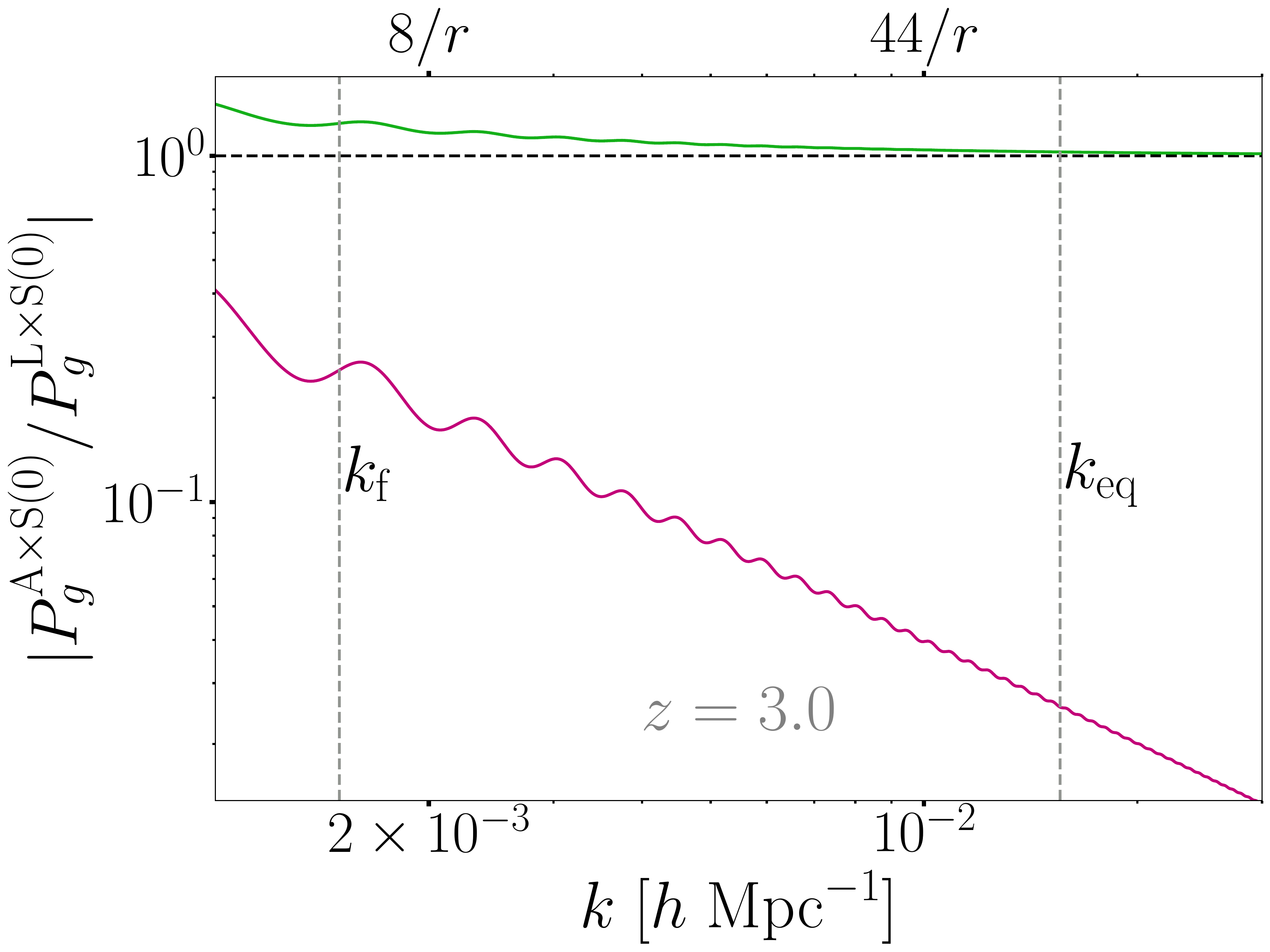}
\includegraphics[width=0.495\linewidth]{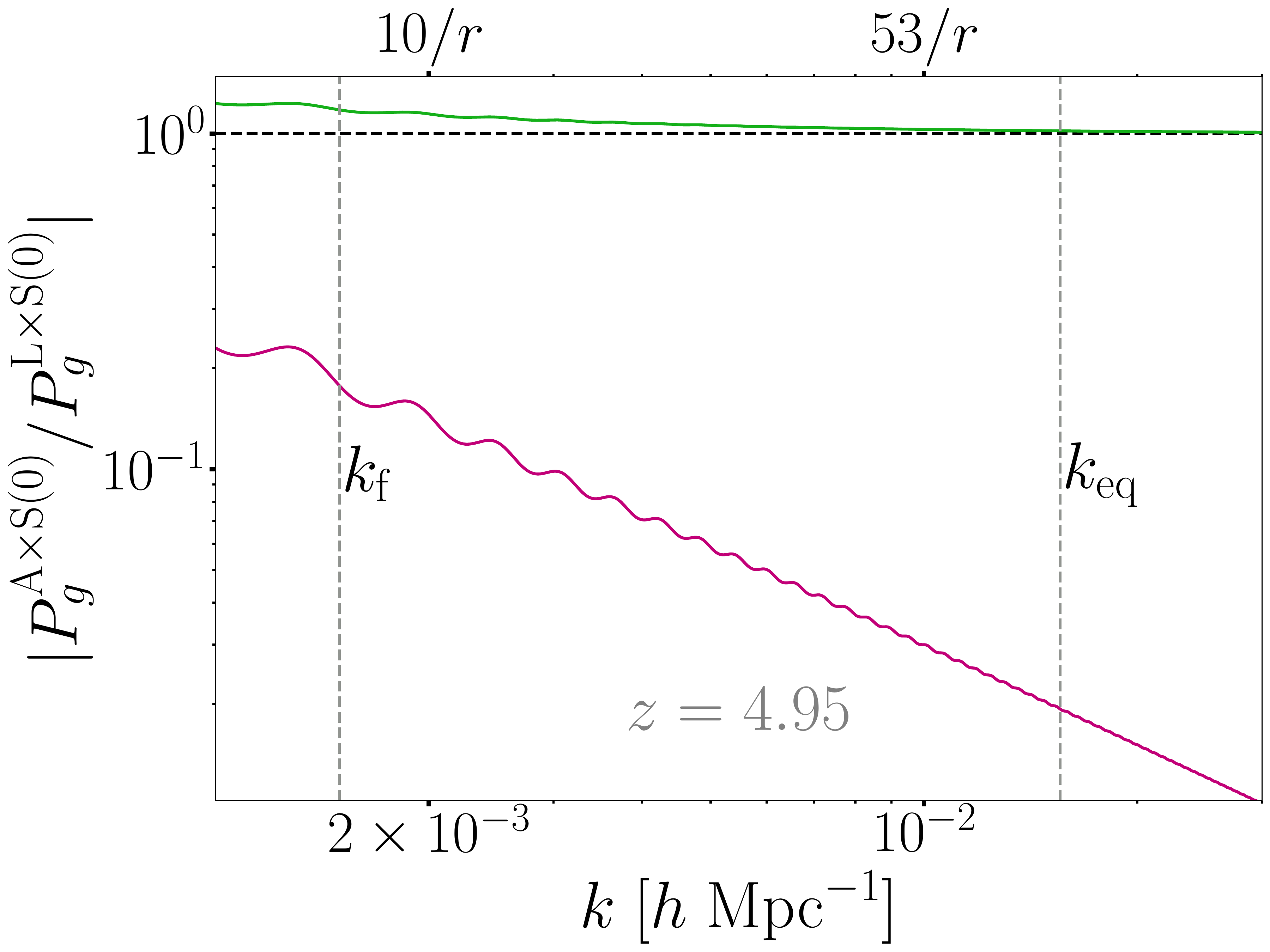}
\caption{For the same cases as in \autoref{fig:monopolesSKA2_LBG_SxI_SxNI}, but showing the magnitude of integrated Sachs-Wolfe + time-delay (ISW + TD) corrections relative to lensing (L)
-- i.e., {$P_g^{{\rm A}\times {\rm S}(0)}/P_g^{{\rm L}\times {\rm S}(0)}$}, where
 A = ISW+TD (magenta), L+ISW+TD (green). 
}
\label{fig:monopolesSKA2_LBG}
\end{figure}

\begin{figure}%[!htbp]
\centering
\includegraphics[width=0.495\linewidth]{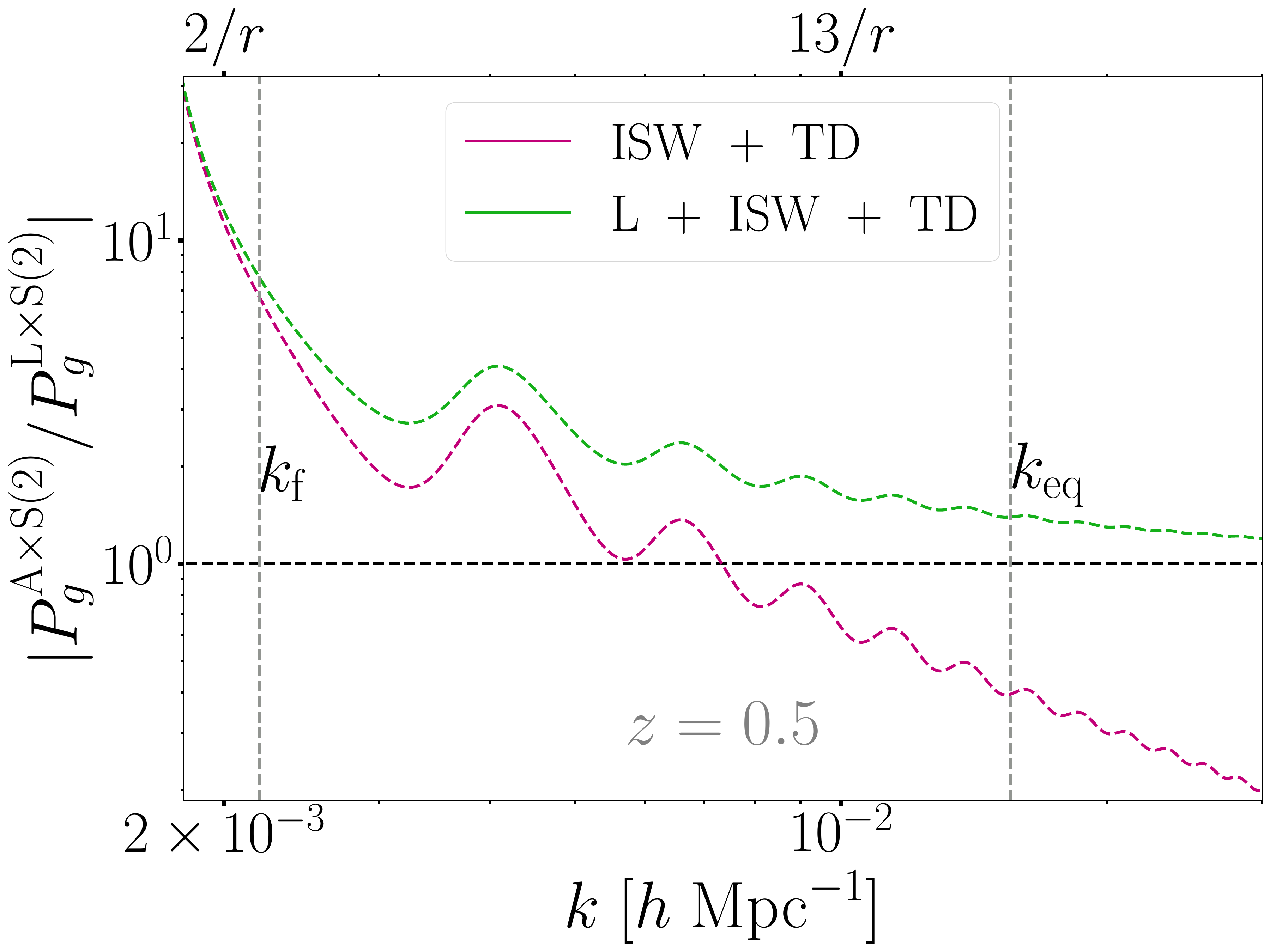}
\includegraphics[width=0.495\linewidth]{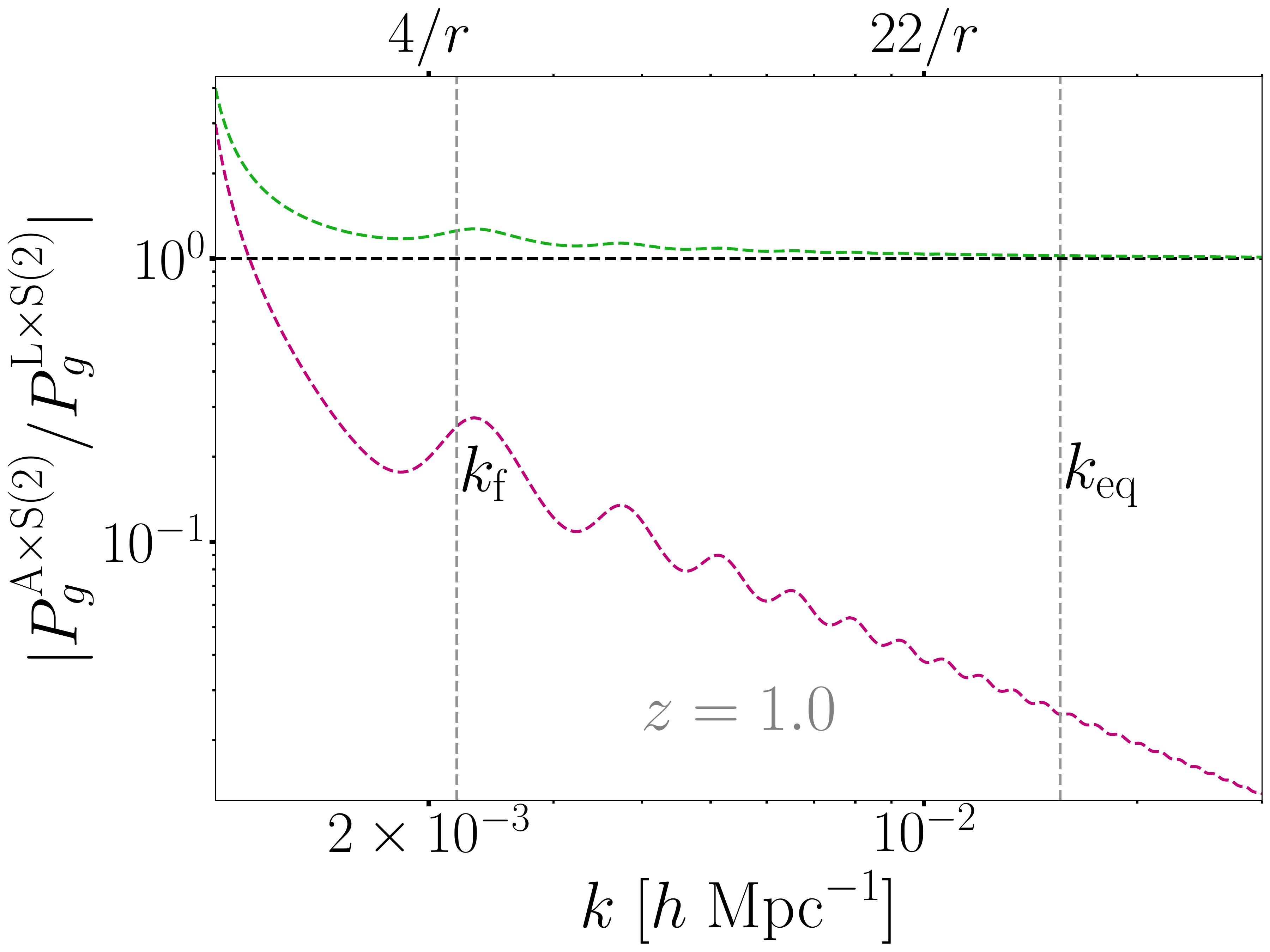}
\includegraphics[width=0.495\linewidth]{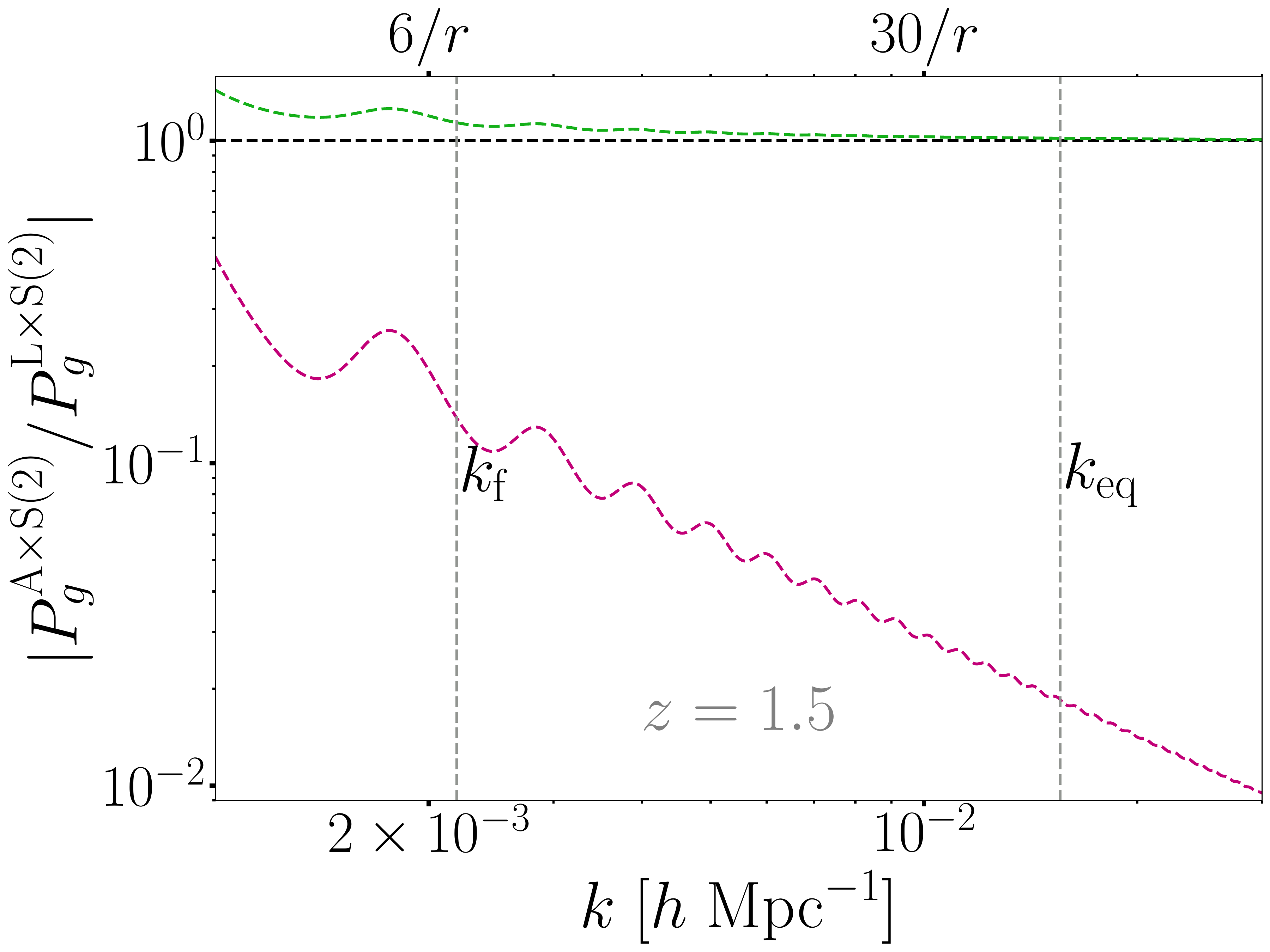}
\includegraphics[width=0.495\linewidth]{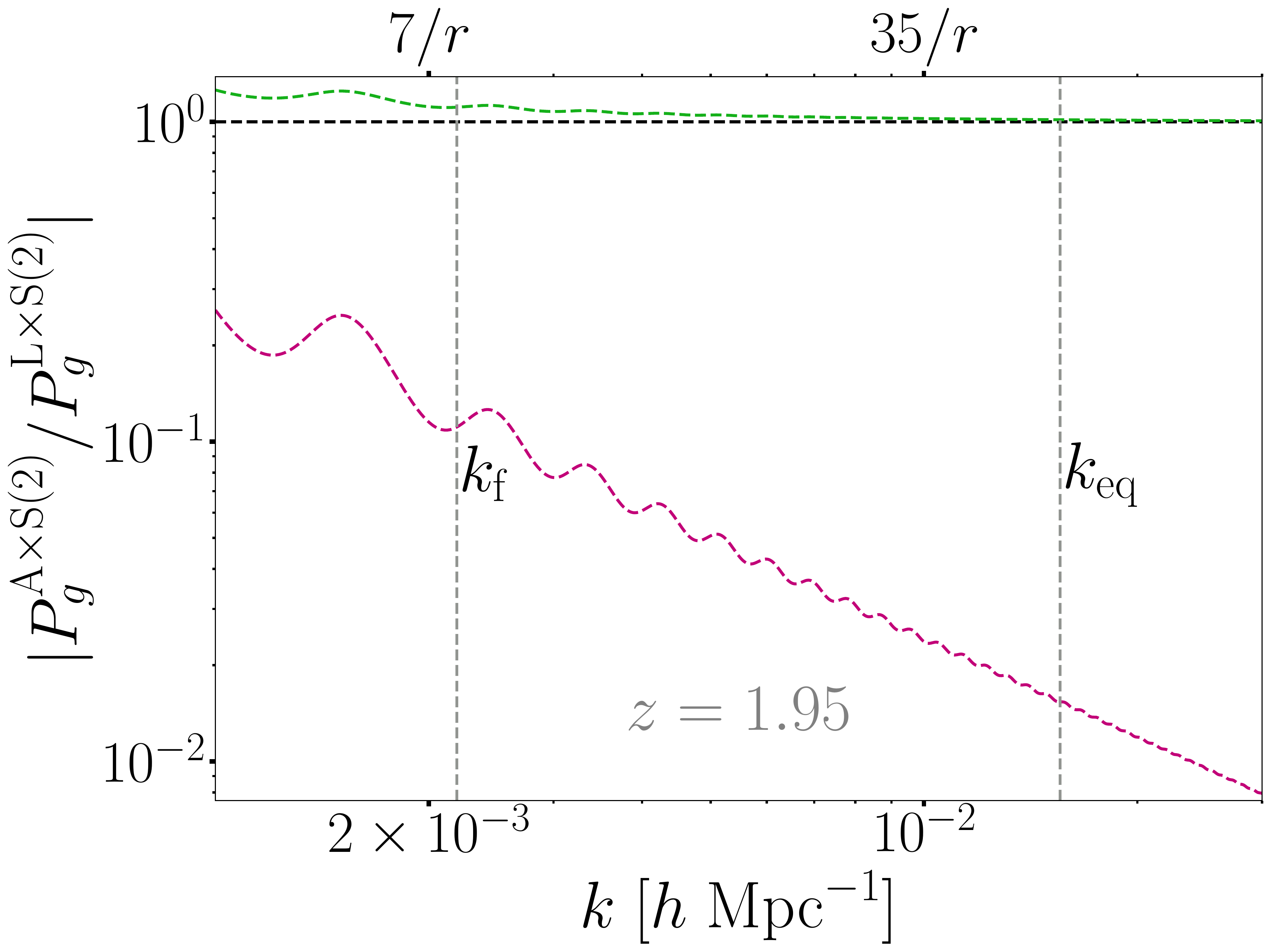}
\includegraphics[width=0.495\linewidth]{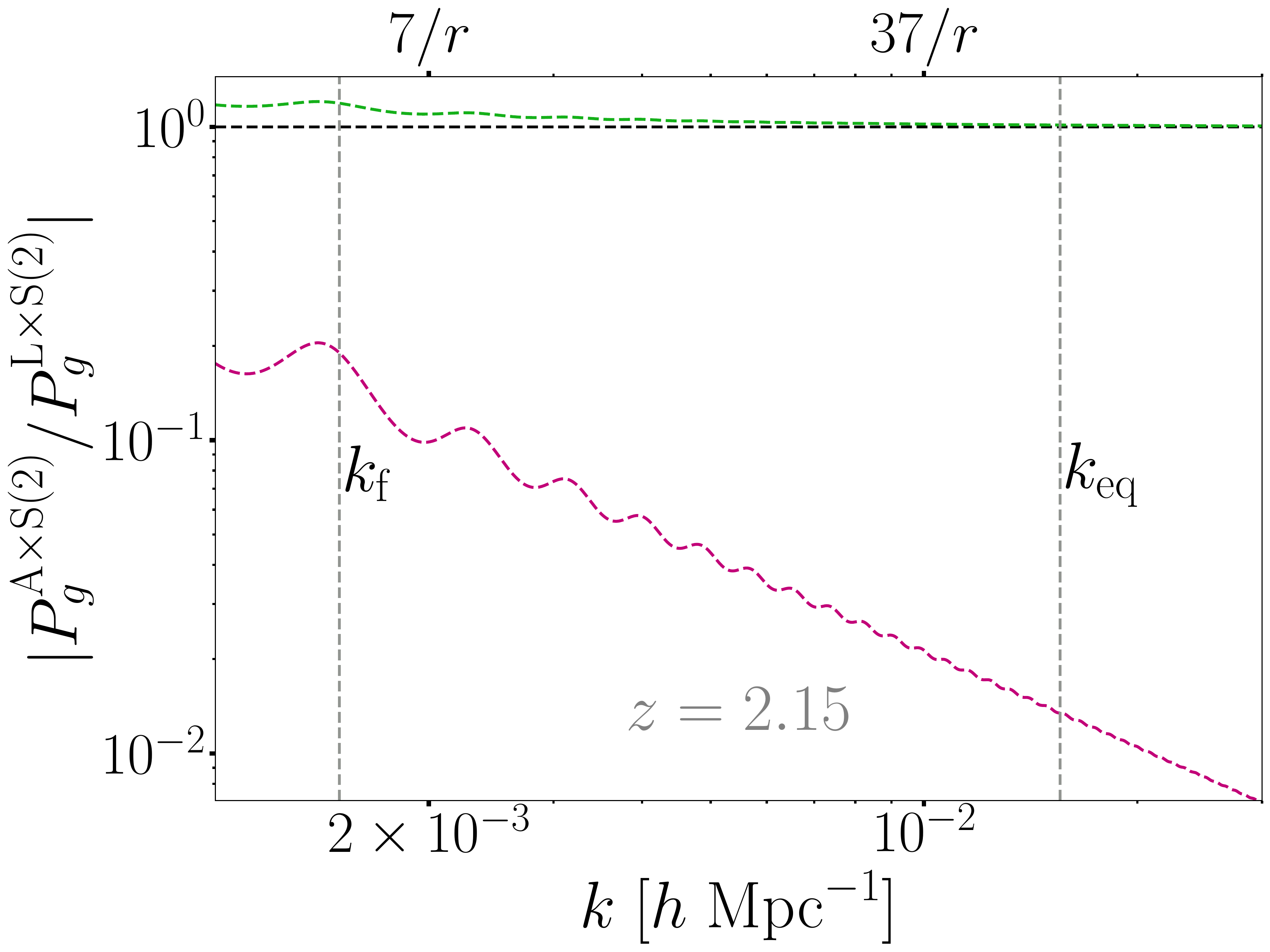}
\includegraphics[width=0.495\linewidth]{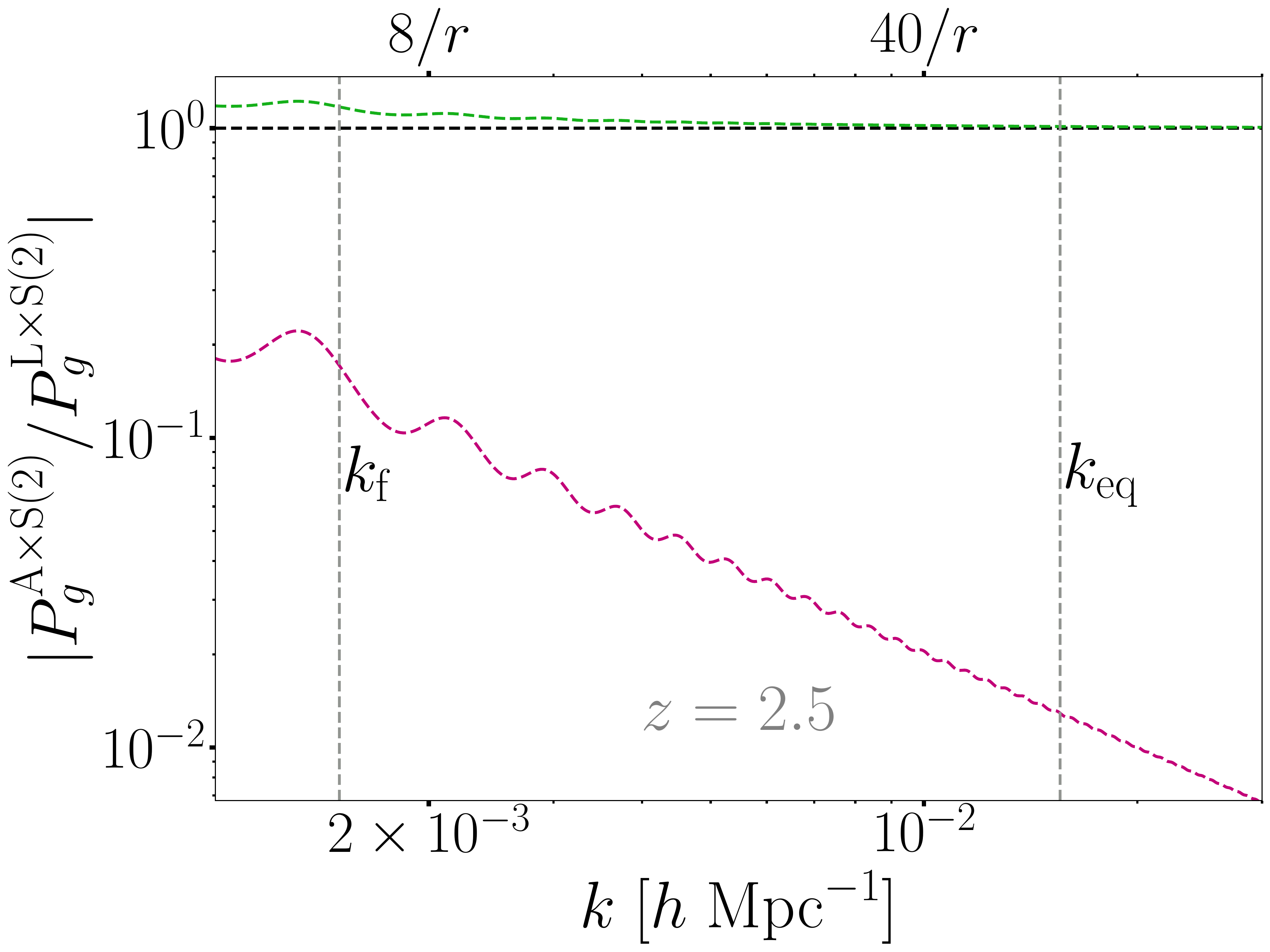}
\includegraphics[width=0.495\linewidth]{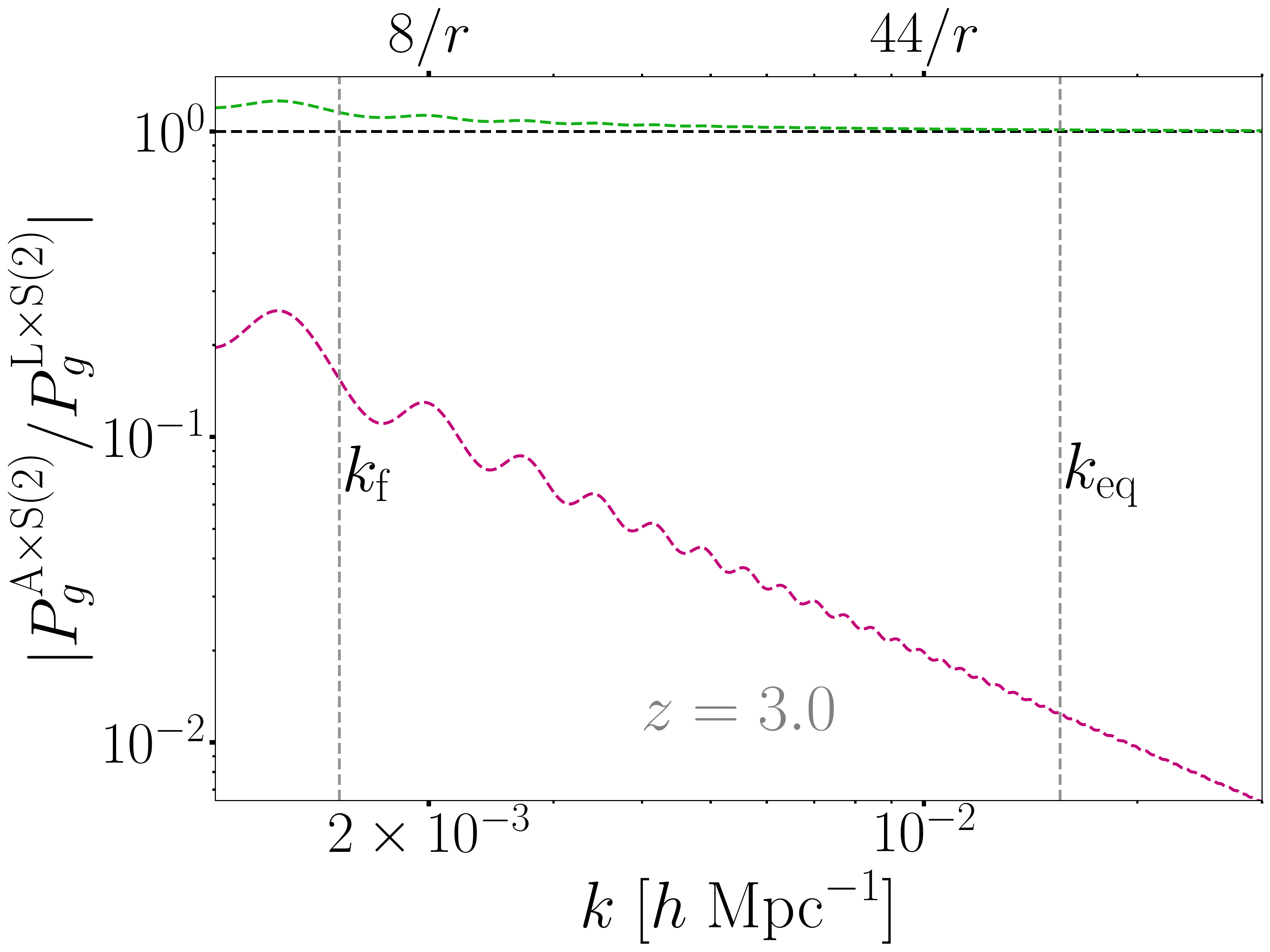}
\includegraphics[width=0.495\linewidth]{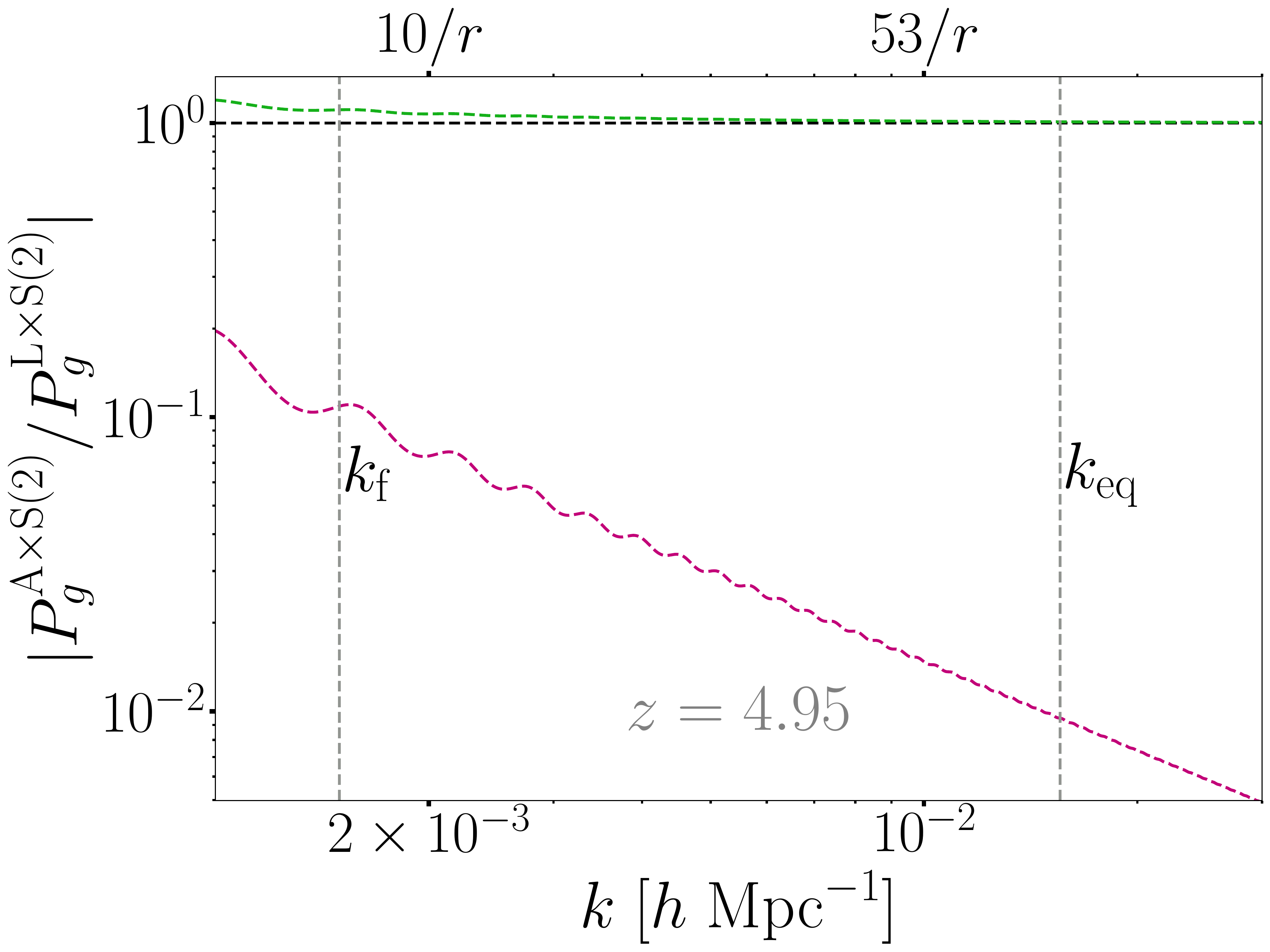}
\caption{{As in \autoref{fig:monopolesSKA2_LBG}, but for the quadrupoles}. \red{Top horizontal axes show multiples of $1/r$ corresponding to $k$ on the bottom axes (wide-angle expansions require $k>1/r$).} 
}
\label{fig:quadrupolesSKA2_LBG}
\end{figure}

It is also interesting to check whether the integrated Sachs-Wolfe and time delay (ISW + TD)  corrections are much smaller than the lensing  (L) correction, as is typically assumed.
\autoref{fig:monopolesSKA2_LBG} shows  
the magnitude of the ISW + TD corrections relative to the L power spectrum monopole, {while the quadrupole case is displayed in \autoref{fig:quadrupolesSKA2_LBG}.} It is apparent that ISW + TD is mainly much smaller than L, except at low redshifts.

%\allowdisplaybreaks

\begin{figure}%[!htbp]
\centering
\includegraphics[width=0.49\linewidth]{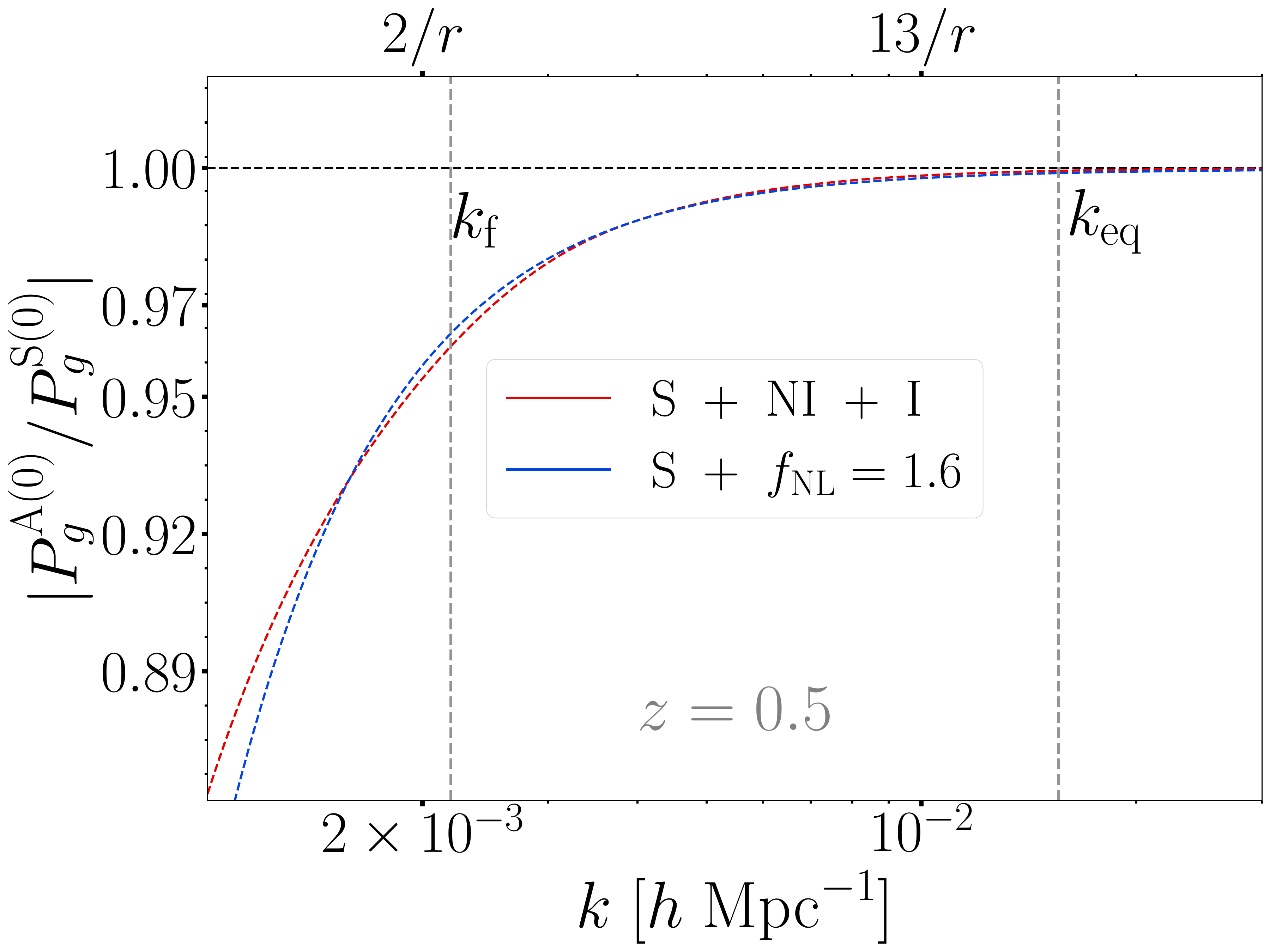}
\includegraphics[width=0.49\linewidth]{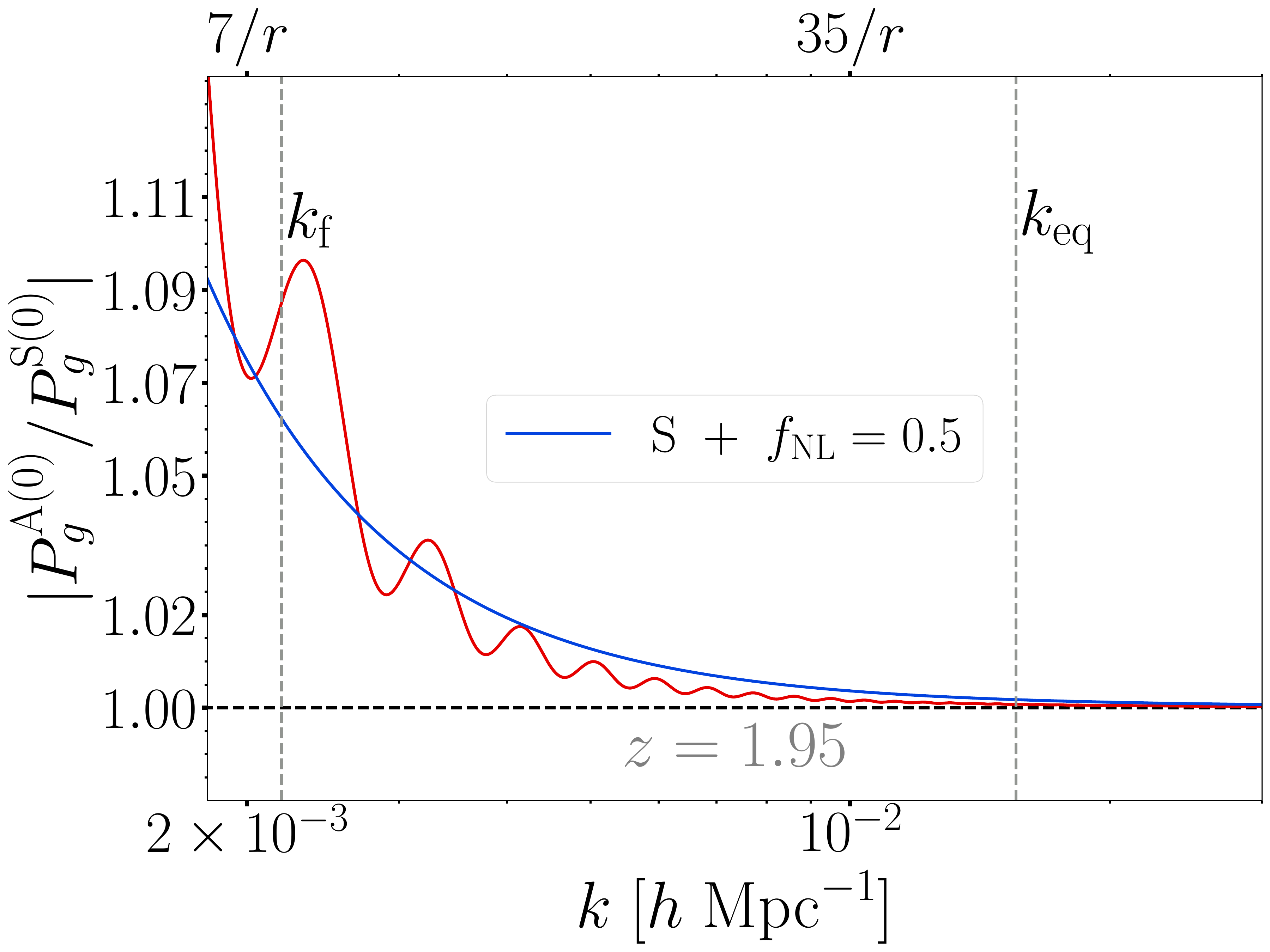}
\includegraphics[width=0.49\linewidth]{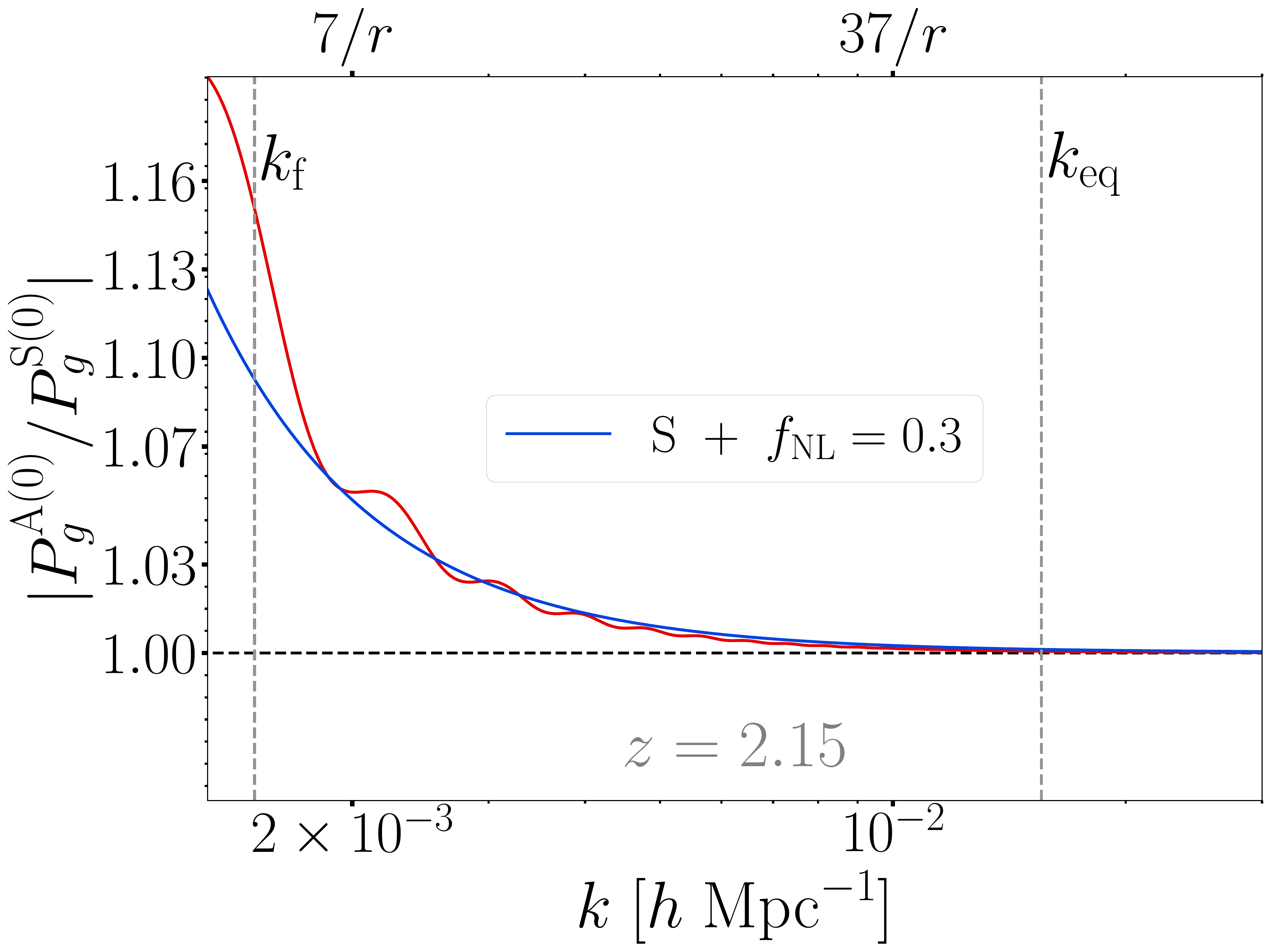}
\includegraphics[width=0.49\linewidth]{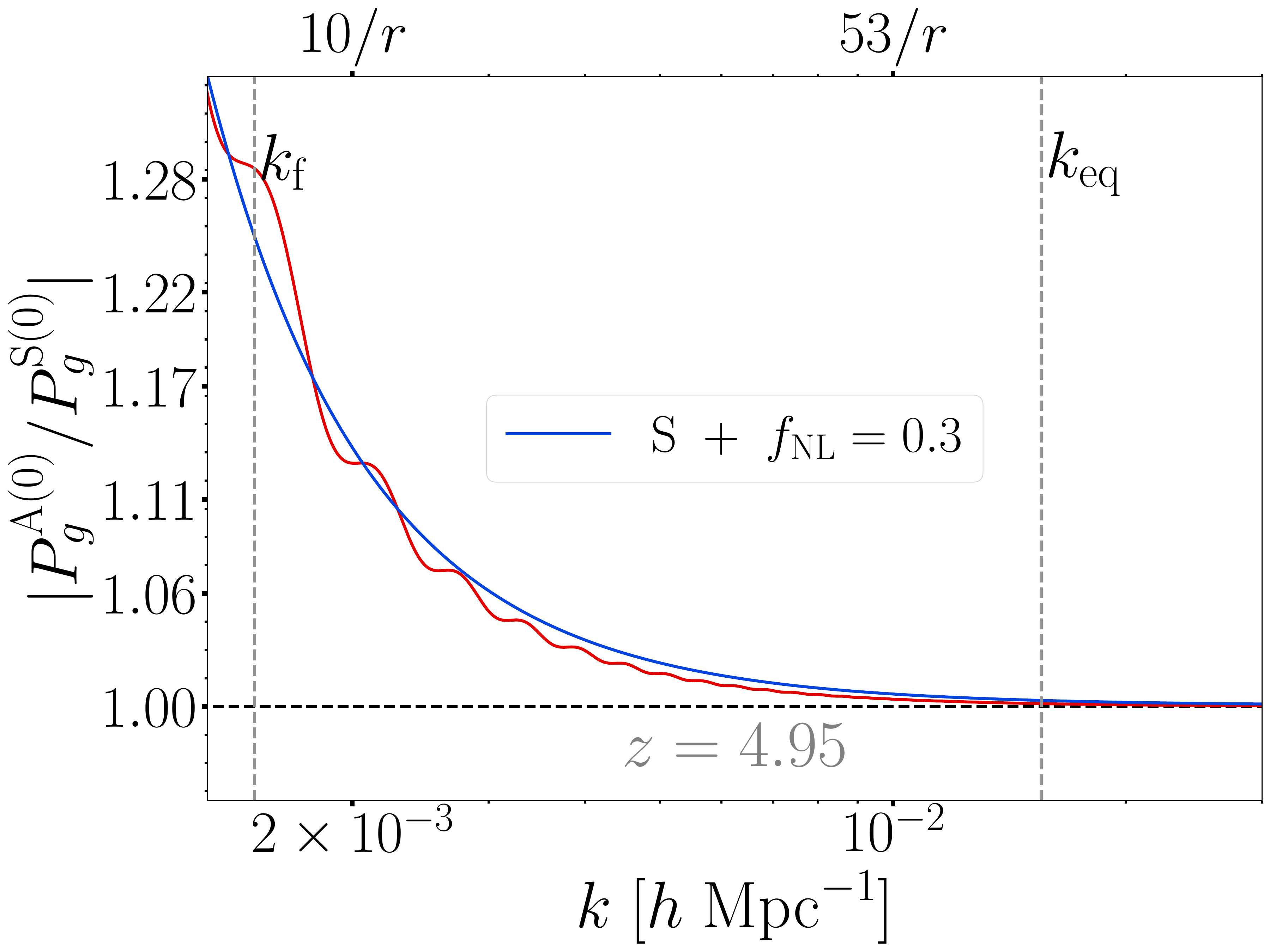}
\caption{Comparison between relativistic + wide-angle effects and local primordial non-Gaussianity -- using the ratios $P_g^{{\rm A}(0)}/P_g^{{\rm S}(0)}$, where
 A = S+NI+I (red) and  S+$\fnl$ (blue). 
 The  value of $\fnl$ indicated in the plots is chosen so that S+$\fnl$ approximately matches S+NI+I. {\em Top row} SKAO2, {\em bottom row} MegaMapper.
\red{Top horizontal axes show multiples of $1/r$ corresponding to $k$ on the bottom axes (wide-angle expansions require $k>1/r$).} } \label{fig:monopolesLBGSKA2fnl}
\end{figure}

\clearpage
It is clear from \autoref{fig:monopolesSKA2_LBG_SxI_SxNI} (S + NI + I curves) that the relativistic and wide-angle corrections become significant on the largest scales -- similar to the effect of scale-dependent bias from local PNG.
\autoref{fig:monopolesLBGSKA2fnl} confirms this expectation, showing how a suitable choice of $\fnl$ at each redshift leads to behaviour that is similar to the total correction (NI + I) in the absence of local PNG ($\fnl=0$).

\section{Constraining local primordial non-Gaussianity}
\label{sec3}

{In the Newtonian and flat-sky approximation,}
the covariance between the multipoles $\ell$ and $\ell^{\prime}$ is (e.g. \cite{Berti:2022ilk})
\begin{align}\label{cov}
    C_{\ell \ell^{\prime}}(z, k) = \frac{(2\ell+1)(2\ell^{\prime}+1)}{2} \, \int_{-1}^{1} \, \ud\mu \; \mathcal{L}_{\ell}(\mu)\, \mathcal{L}_{\ell^{\prime}}(\mu)\, \sigma(z, k, \mu)^{2},
\end{align}
where 
the variance  per $k$- and $\mu$-bin in each $z$-bin is 
\begin{align}\label{var}
\sigma^{2} = 
\frac{2}{N_{k}}\, \Big[P_{g}+ \frac{1}{\bar n_g}\Big]^2\,. 
\end{align}
Here
\begin{align}
N_{k} = \frac{4 \pi k^{2} \Delta k}{k_{\mathrm{f}}^{3}}\,,~~ k_{\rm f}=\frac{2\pi}{V_{\rm s}^{1/3}}\,,~~
V_{\rm s}(z)=\frac{\Omega_{\rm sky}}{3}\Big[r(z+\Delta z/2)^3- r(z-\Delta z/2)^3\Big] \;,
\label{Nmodes}
\end{align}
are the number of modes, the fundamental mode in each redshift bin and the bin volume.
{The covariance defined by \autoref{cov} neglects the mode coupling induced by wide-angle effects. Even in a Gaussian approximation, this mode-coupling is very difficult to model, typically requiring various approximations or simulations \cite{Castorina:2017inr, Blake:2018tou,Wadekar:2019rdu, Wadekar:2020hax,Castorina:2021xzs} (although there is recent progress based on the spherical Fourier-Bessel power spectrum \cite{Wen:2024hqj}). 
Incorporating mode-coupling in the covariance is beyond the scope of our paper.
Hence we neglect this mode-coupling, which will lead to over-optimistic precision. The same applies in general to the Fisher analysis that we use.}

\begin{figure}[!htbp]
\centering
\includegraphics[width=0.495\linewidth]{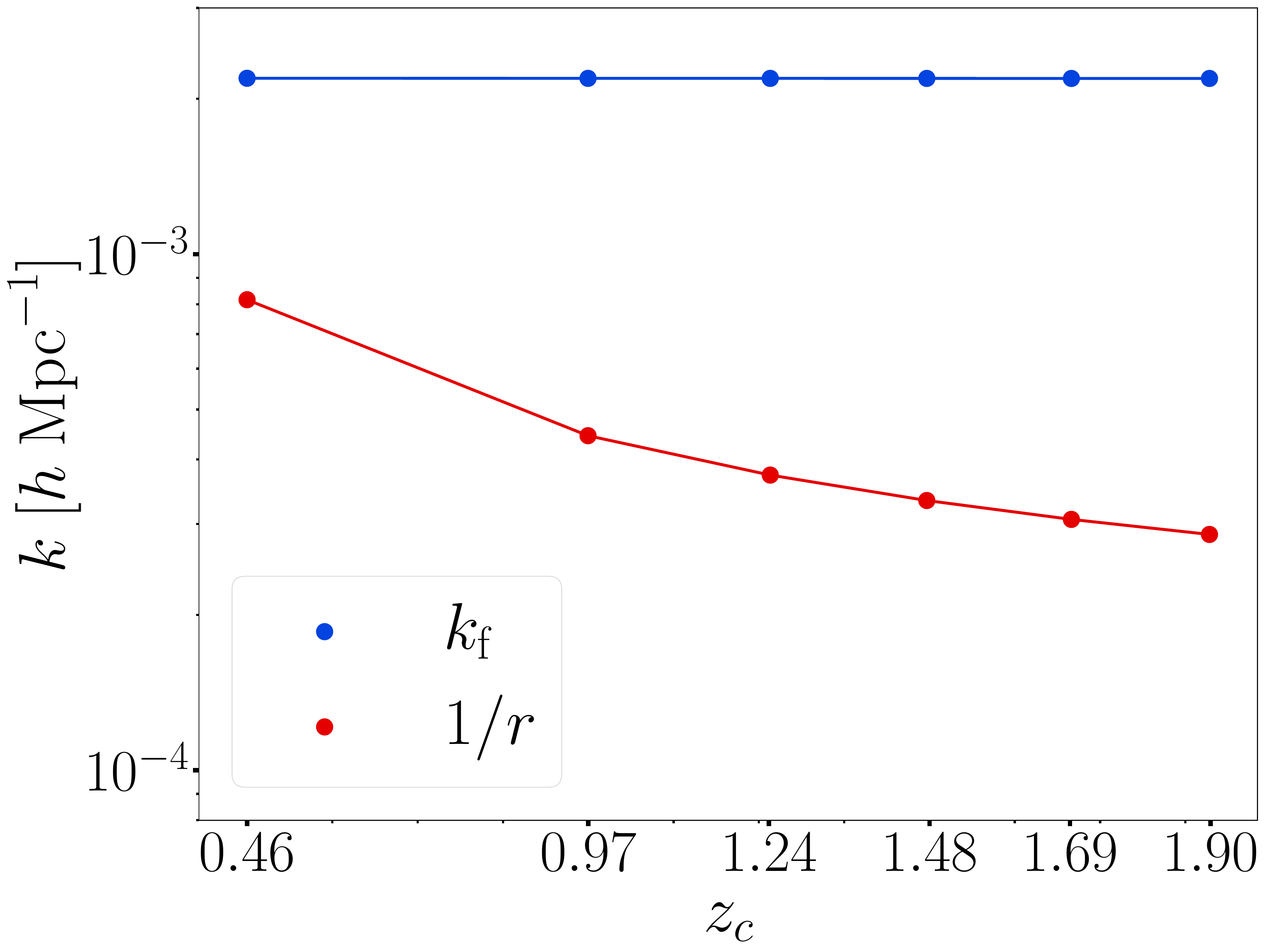}
\includegraphics[width=0.495\linewidth]{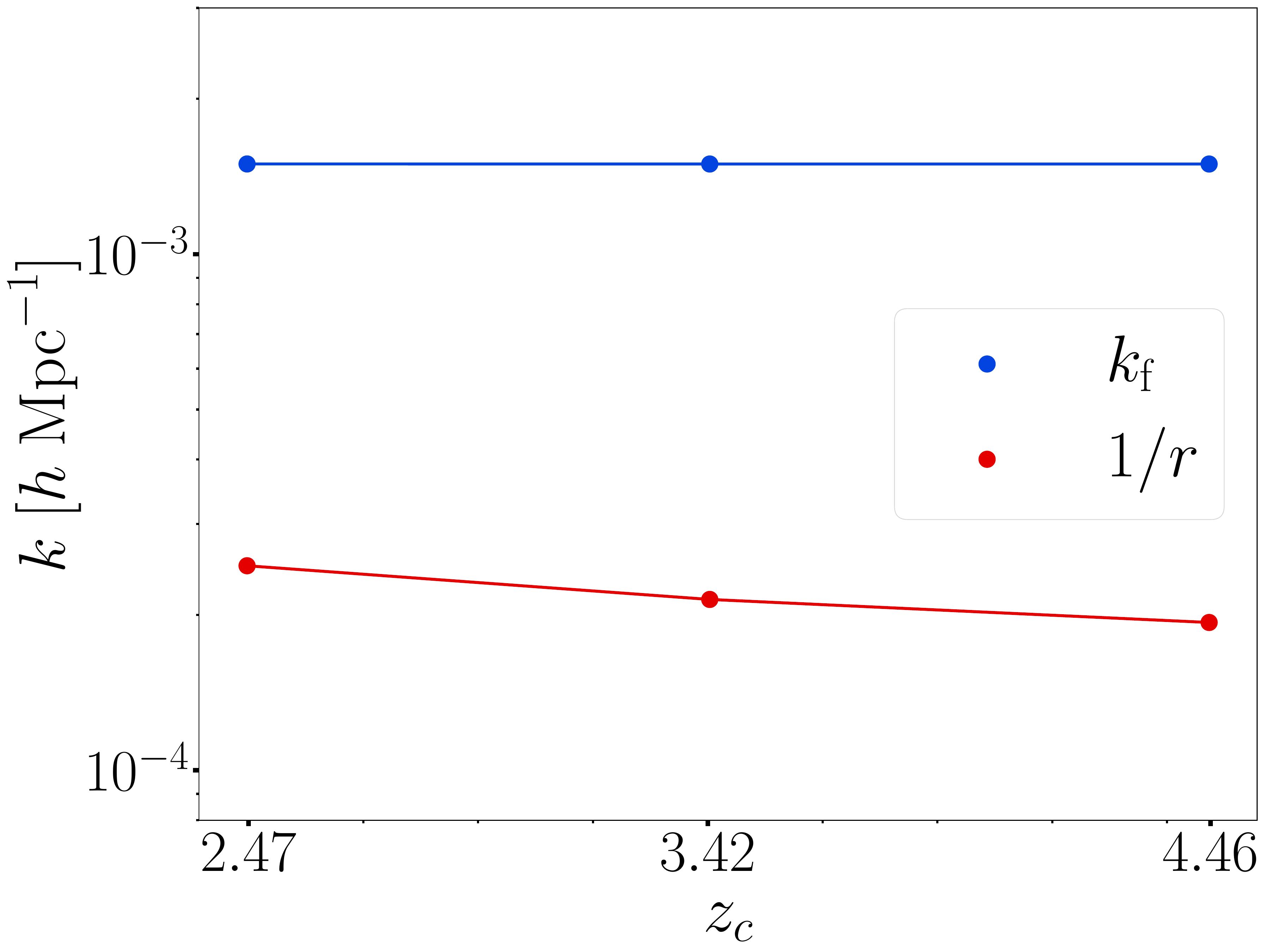}
\caption{{Modes $k_{\rm f}$ and $k=r^{-1}$ for SKAO2 (\emph{left}) and MegaMapper (\emph{right}), at the redshift bin centres of
\autoref{tab:zbinning}. Note that $k_{\rm f}$ is constant (equal-volume bins).}
} \label{figkmin}
\end{figure}

For the mode binning, we set $\Delta k = k_{\mathrm{f}}$. On small scales (large $k$), 
in order to avoid nonlinearity, we choose a conservative maximum mode {of $0.08\,h$/Mpc at $z=0$. For the redshift evolution we follow \cite{Smith:2002dz}, leading to}
\begin{align}
     k_{\rm max}(z) = 0.08\left(1+z\right)^{2/(2+n_{s})}\;h\,\mathrm{Mpc}^{-1}\;.
\end{align}
The minimum mode, corresponding to the largest wavelength accessible in each $z$-bin, is determined in each redshift bin by (a)~the longest-wavelength mode and (b)~the  perturbative approximation required for the wide-angle corrections \cite{Jolicoeur:2024oij}:
\begin{align}
 k_{\rm min}(z) = {{\rm max}\,\bigg[k_{\rm f}(z)\,, \,  \frac{1}{r(z)}\bigg].}
\end{align}
{By \autoref{Nmodes}, the fundamental mode $k_{\rm f}$ depends on the redshift binning, which we consider below (see \autoref{tab:zbinning}). For the equal-volume bins that we choose, $k_{\rm f}$ is constant and  $k_{\rm min}(z) = k_{\rm f}$ for both surveys,} as shown in \autoref{figkmin}. 

We use the monopole and quadrupole, neglecting the information from (and cross-correlation with) the $\ell \ge4$ multipoles. The general covariance matrix  per $k$-bin in a $z$-bin is
\begin{align}
    {\bm C}(z, k) = \begin{pmatrix}
                C_{00}(z, k) & ~~~ &  C_{02}(z, k) \\
                             & ~~~ &                \\
                C_{20}(z, k) & ~~~ &  C_{22}(z, k)
                \end{pmatrix} \; , \label{Eqcov_k}
\end{align}
and the Fisher matrix is 
\begin{align}
    F_{\alpha \beta}(z) = \sum_{k=k_{\rm min}}^{k_{\rm max}} \partial_{\alpha} {\bm D}(z,k) \, {\bm C}(z,k)^{-1} \, \partial_{\beta} {\bm D}(z,k)^{\sf T}  \; , \label{EqFisher}
\end{align}
where \red{${\bm D}=(P_g^{(0)},
P_g^{(2)})$} is the data vector of the multipoles of the power spectrum and $\partial_{\alpha} = \partial/\partial \vartheta_{\alpha}$, with $\vartheta_{\alpha}$ the parameters to be constrained. \red{Our forecasts are based on the full data vector but for comparison we also show some results using the monopole and quadrupole on their own.}

Since we are using optimistic forecasts, 
we include only the two cosmological parameters that directly affect the large-scale power spectrum, together 
with $\fnl$ and three nuisance parameters:
\begin{align}
    \vartheta_{\alpha} = \big( A_{s}, \; n_{s}, \; f_{\rm NL};\; b_{0}, \; \mathcal{E}_{0}, \; \mathcal{Q}_{0}\big) \,.     \label{EqConstrParams}
\end{align}
The fiducial values $\bar{A}_{s} = 2.105 \times 10^{-9}$, $\bar{n}_{s} = 0.9665$ for the $\Lambda$CDM cosmological parameters  are taken from Planck \cite{Planck:2018vyg}, and the remaining $\Lambda$CDM parameters are fixed at their Planck values. We take $\bar{f}_{\rm NL} = 0$ and the nuisance parameter fiducials are $\bar{b}_{0} = 1$, $\bar{\mathcal{E}}_{0} = 1$, $\bar{\mathcal{Q}}_{0} = 1$ {(we do not impose priors on the nuisance parameters)}.\\~\\

\begin{figure}[!htbp]
\centering
\includegraphics[width=0.45\linewidth]{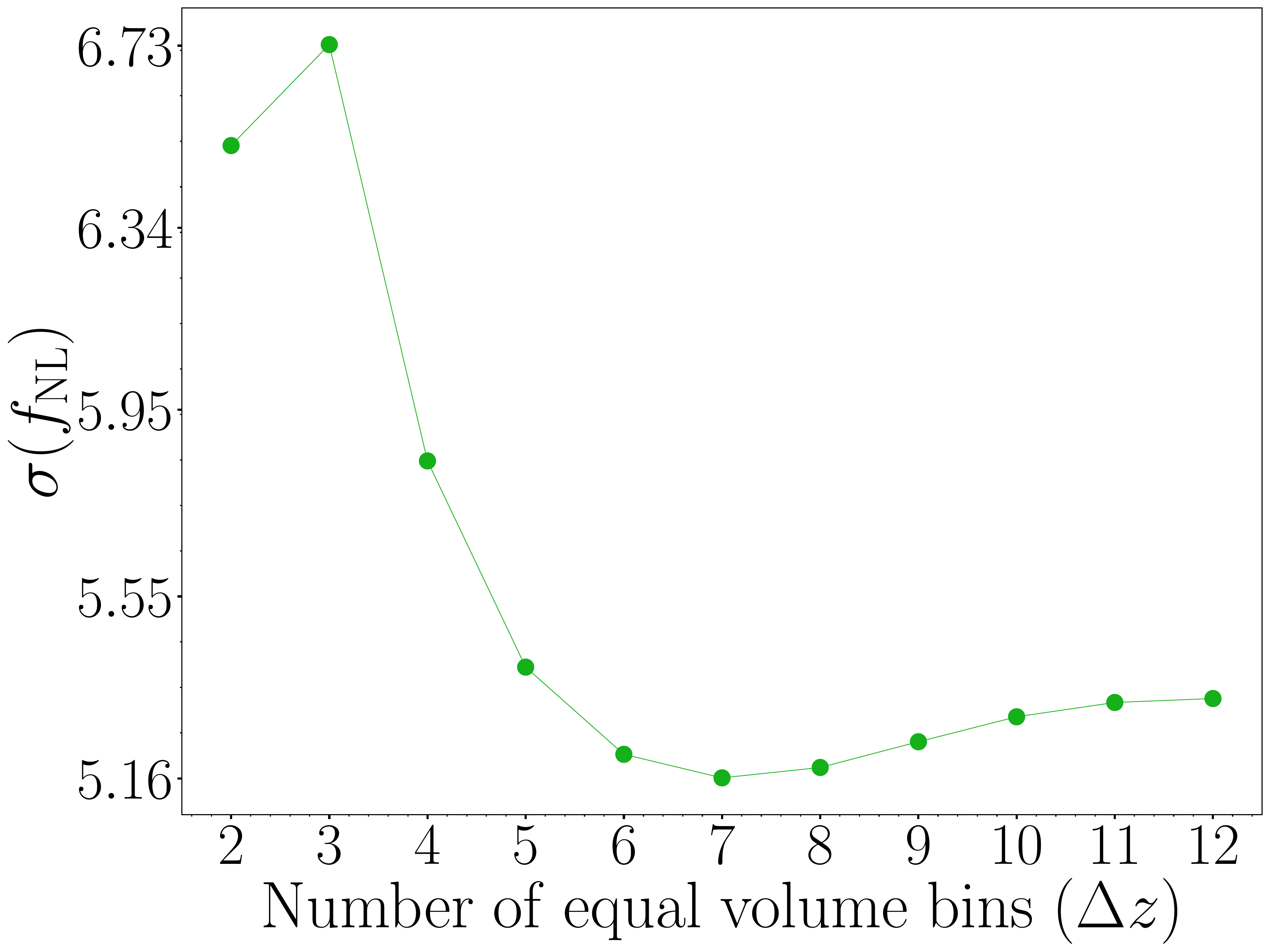}  
\includegraphics[width=0.45\linewidth]{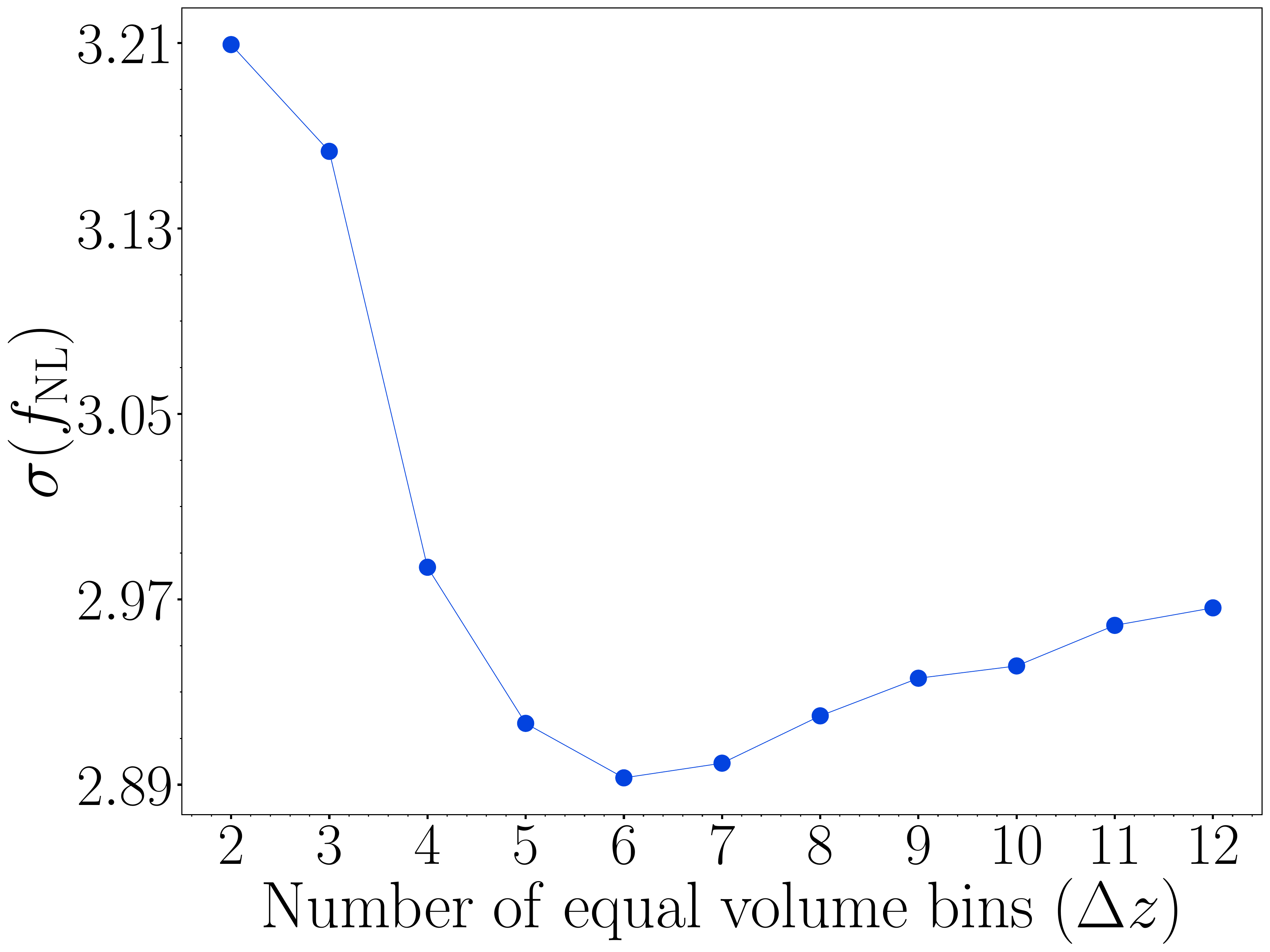} 
\includegraphics[width=0.45\linewidth]{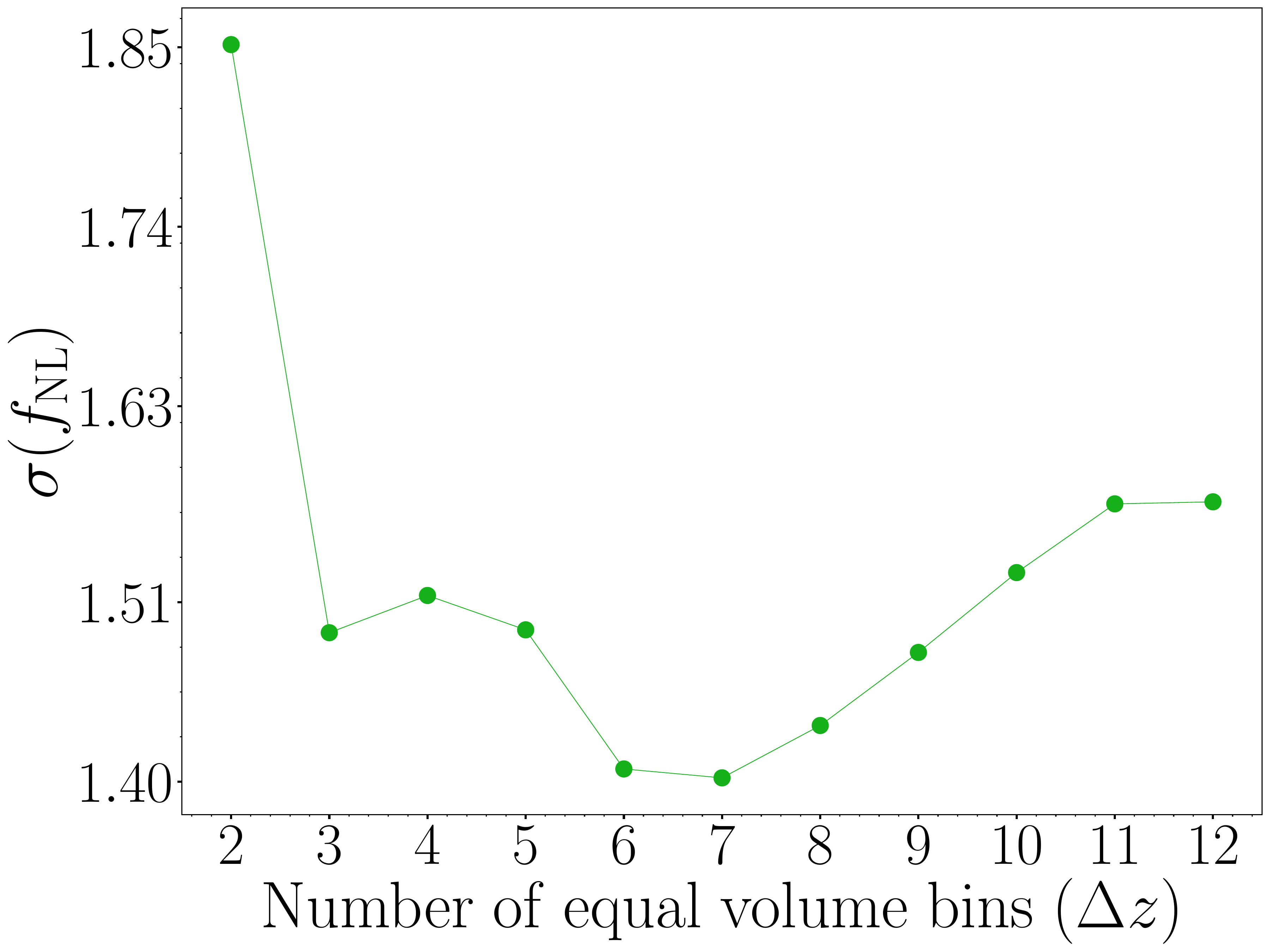}  
\includegraphics[width=0.45\linewidth]{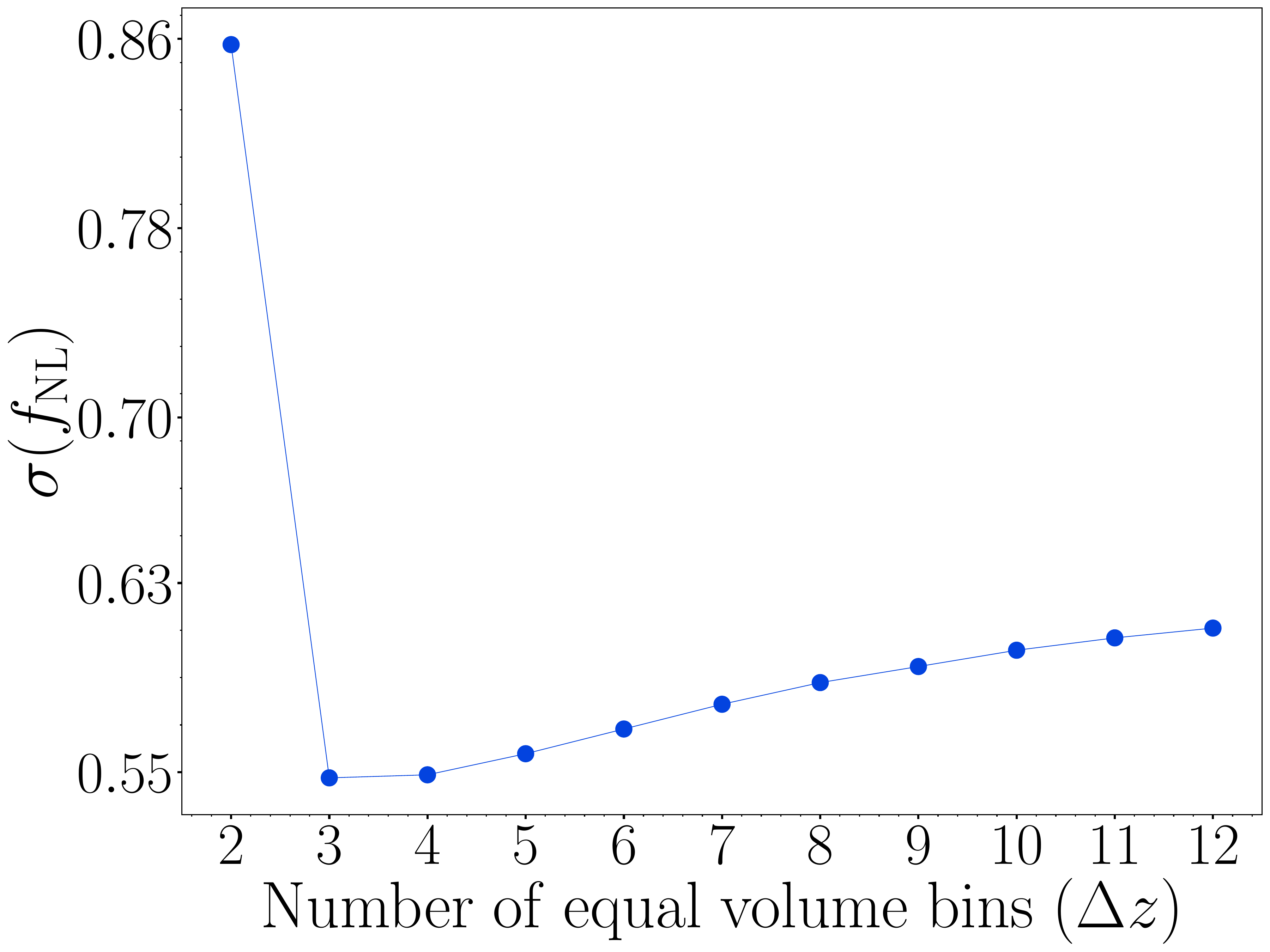}
\caption{{Effect on $\sigma(f_{\mathrm{NL}})$ of varying the number of equal-volume redshift bins for SKAO2 (\emph{top}) and MegaMapper (\emph{bottom}) -- using \red{$P_{g}^{(0)}$ ({green}), %$P_{g}^{(0)}+P_{g}^{(2)}$ ({red}) 
and $P_{g}^{(0)} \times P_{g}^{(2)}$ ({blue}).}}} \label{fig:zbinning}
\end{figure}
\begin{table}[h]
\centering
\caption{Redshift bin width, $\Delta z$, and bin centres, $z_c$, of  equal-volume  bins for SKAO2  and MegaMapper, following from \autoref{fig:zbinning}.}
\vspace{0.3cm}
\label{tab:zbinning}
\begin{tabular}{c|c|c|c|c|c|c||c|c|c|}
\cline{2-10}
 &   \multicolumn{6}{ c|| }{SKAO2} & \multicolumn{3}{ c| }{MegaMapper}  \\ 
\hline
\hline
\multicolumn{1}{ |l|| }{ $\Delta z$ ~~~} &  0.720 &  0.2991 &  0.2457 &  0.2223 &  0.2099 &  0.2029 & 0.9340 &  0.9803 &  1.0857  \\ 
\hline
\multicolumn{1}{ |l||  }{$ z_c $ } & 0.460 &  0.9696 &  1.2420 &  1.4760 &  1.6921 &  1.8985 & 2.4670 &  3.4241 &  4.4571 \\ 
\hline
\end{tabular}
\end{table}
%~\\

%We consider four cases:
%\begin{align}
%{\bm D} &= P_{g}^{(0)} ~~\mbox{and}~~{\bm C}= C_{00}:\quad   \mbox{monopole only}, \label{D1}\\
%{\bm D} &= P_{g}^{(2)}~~\mbox{and}~~{\bm C} = C_{22}: \quad \mbox{quadrupole only}, \label{D2}\\
%{\bm D} &= \big(P_{g}^{(0)}, \; P_{g}^{(2)}\big) ~~\mbox{and}~~C_{02} = C_{20} = 0: ~~ P_{g}^{(0)} + P_{g}^{(2)}\quad  \mbox{(uncorrelated)}
%\label{D3}\\
%{\bm D} &= \big(P_{g}^{(0)}, \; P_{g}^{(2)}\big)~~\mbox{and}~~ C_{02} = C_{20} \neq 0:~~  P_{g}^{(0)} \times P_{g}^{(2)}\quad \mbox{(correlated)}. \label{D4} 
%\end{align}

{For the redshift binning, 
we follow  \cite{D:2023oxn}, which recommends  equal-volume bins for improved $\fnl$ constraints, rather than equal-number bins. \autoref{fig:zbinning} shows that the smallest $\sigma(\fnl)$ is obtained using the correlated monopole and quadrupole, $P_{g}^{(0)} \times P_{g}^{(2)}$. The optimal number of equal-volume redshift bins is then 6 for SKAO2 and 3 for MegaMapper.
\autoref{tab:zbinning} gives  the bin centres and bin widths.}

\begin{figure}[!htbp]
\centering
\includegraphics[width=0.49\linewidth]{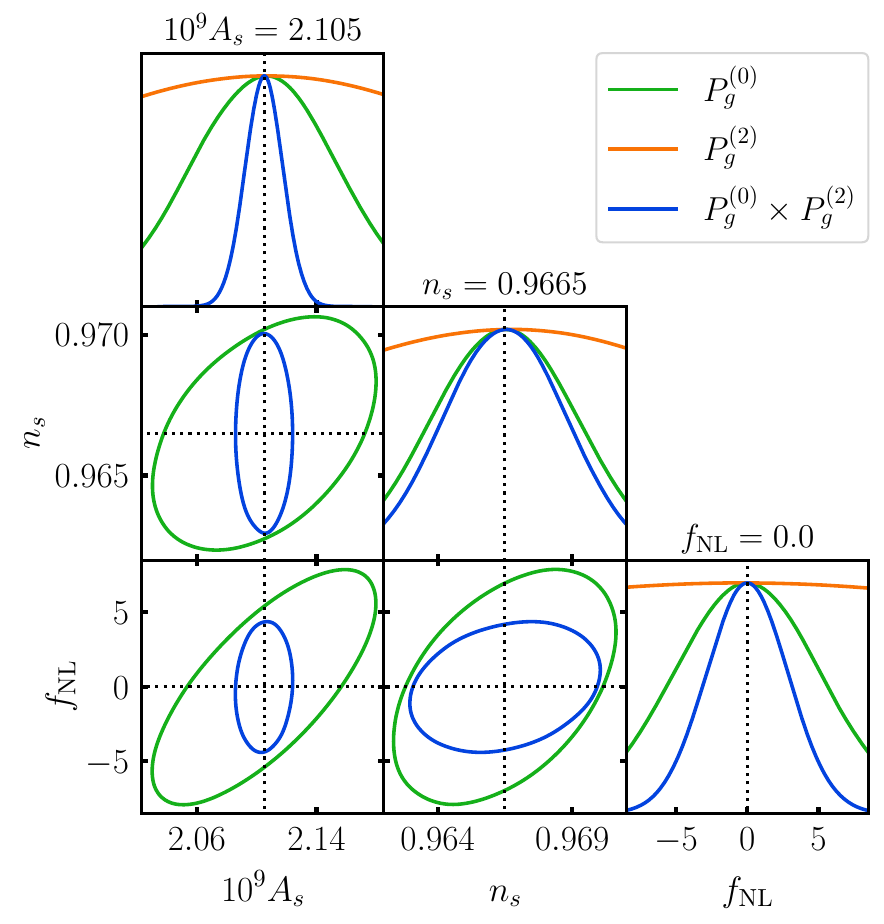}
\includegraphics[width=0.49\linewidth]{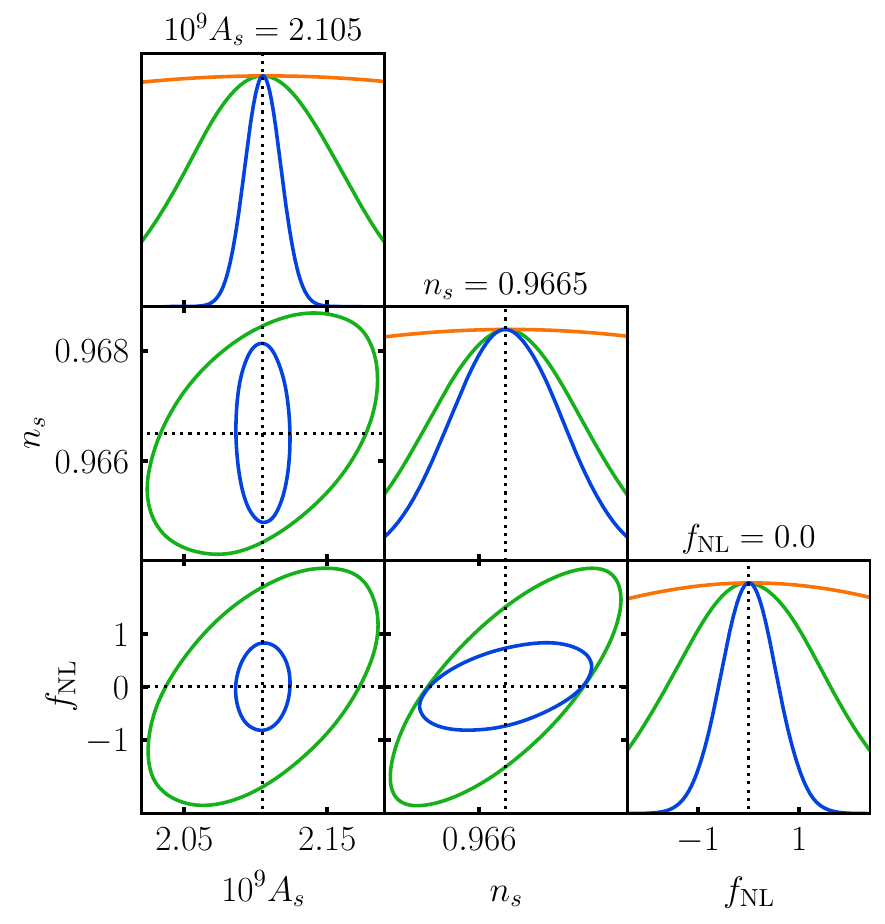}
\caption{{Contour plots of  $1\sigma$ marginal errors for different data vectors: SKAO2  (\emph{left}) and MegaMapper (\emph{right}).}
} \label{fig:fishercontourLBGSKA2}
\end{figure}

The results of the Fisher analysis are displayed in \autoref{fig:fishercontourLBGSKA2} and summarised in \autoref{tab:sigmaParameters2}. (See also \autoref{appd},  \autoref{fig:fishercontourLBGSKA2eachdata}.) 
%{The significant improvement in $\sigma(\fnl)$ for  the correlated multipole case $P_{g}^{(0)} \times P_{g}^{(2)}$ compared to the uncorrelated case $P_{g}^{(0)} + P_{g}^{(2)}$ is not due to an increased signal-to-noise. This is given for the 4 cases in \autoref{D1}--\autoref{D4} by 
%\begin{align}\label{snr}
% {\rm SNR}^2= {\bm D} \, {\bm C}^{-1} \, {\bm D}^{\sf T}   \,. 
%\end{align}
%\autoref{fig:snr_k} in \autoref{appg} shows that the SNR of $P_{g}^{(0)} + P_{g}^{(2)}$ and $P_{g}^{(0)} \times P_{g}^{(2)}$ are almost the same. The better constraints on $\fnl$ in the correlated case come from the breaking of parameter degeneracies.}

\allowdisplaybreaks

\begin{table}[hp!]
    \centering
    \caption{Marginalised $1\sigma$  errors on $\fnl, A_s, n_s$ from   SKAO2 and MegaMapper surveys.
    }
     \label{tab:sigmaParameters2}
        \vspace{0.3cm}
\begin{tabular}{llccc}
\toprule
Survey \hspace{1.50cm} & Case \hspace{3.0cm}&~~~ $\sigma(f_{\mathrm{NL}})$~~~ & ~~~$\sigma(A_s)/\bar{A}_s$~~~ & ~~~$\sigma(n_s)/\bar{n}_s$ \\
\midrule
\midrule
SKAO2 & $P_{g}^{(0)}$     & 5.21 & 0.024 & 0.003 \\
 & $P_{g}^{(2)}$     & 41.4 & 0.091 & 0.011 \\
 %& $P_{g}^{(0)}+P_{g}^{(2)}$  & 3.33 & 0.006 & 0.002 \\
 & $P_{g}^{(0)} \times P_{g}^{(2)}$ & \bf{2.89} & 0.006 & 0.002 \\
\midrule
MegaMapper & $P_{g}^{(0)}$     & 1.49 & 0.025 & 0.002 \\
 & $P_{g}^{(2)}$     & 6.44 & 0.174 & 0.001 \\
 %& $P_{g}^{(0)}+P_{g}^{(2)}$  & 0.629 & 0.006 & 0.001 \\
 & $P_{g}^{(0)} \times P_{g}^{(2)}$ & \bf{0.548} & 0.006 & 0.001 \\
\bottomrule
\end{tabular}
\end{table}
{In \autoref{appe} we show the effects on $\sigma(\fnl)$ of {increasing
$k_{\rm max}(0)$ (\autoref{figkmax}) and $k_{\rm min}(z)$ (\autoref{figkmin2}).} 
Increasing  $k_{\rm max}(0)$ from our conservative value of $0.08h$/Mpc to the more commonly used $0.1h$/Mpc, leads to only a small improvement in precision, as expected. There is a larger effect of increasing $k_{\rm min}(z)$ by 50\% and then 100\%, as expected -- since the strongest PNG signal is on the largest scales. It is noticeable that there is a large increase in errors on the quadrupole {($\sim100\%$ for SKAO2, $>200\%$ for MegaMapper) when $k_{\rm min}(z)$ is doubled.} However, this is significantly mitigated by combining with the correlated monopole.}

%\newpage
\section{Shift in the estimate of $f_{\rm NL}$}
\label{sec4}

As shown in \autoref{fig:monopolesLBGSKA2fnl}, relativistic and wide-angle corrections can approximately mimic the effect of  $f_{\rm NL}$. We thus expect that measurements of $f_{\rm NL}$ {\em when using the standard power spectrum}, i.e., without relativistic + wide-angle corrections, could be shifted away from the true value of $f_{\rm NL}$. In order to quantify this shift (or bias), we introduce a `theory parameter' $\varepsilon$ which distinguishes between the true power spectrum ($\varepsilon=1$) and the approximate standard power spectrum ($\varepsilon=0$):
\begin{align}
  P_{g}= P_{g}^{\rm S} + \varepsilon  P_{g}^{\rm corr} \,.
\end{align}

\begin{figure}%[!htbp]
\centering
\includegraphics[width=0.49\linewidth]{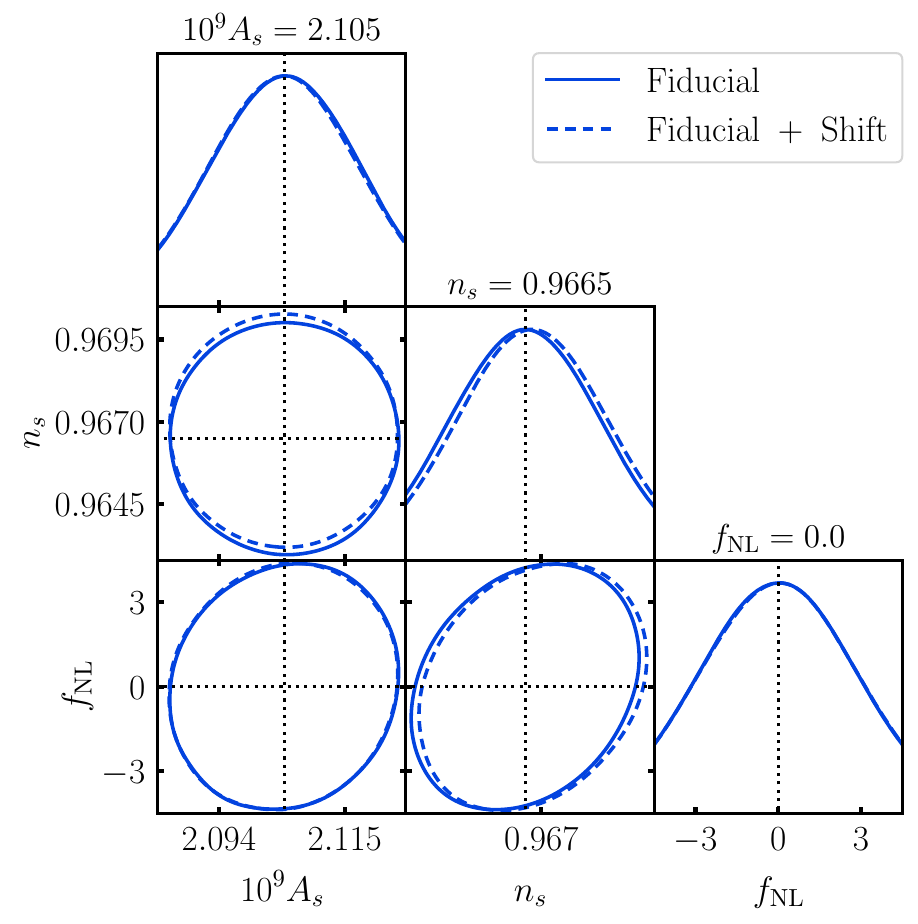}
\includegraphics[width=0.49\linewidth]{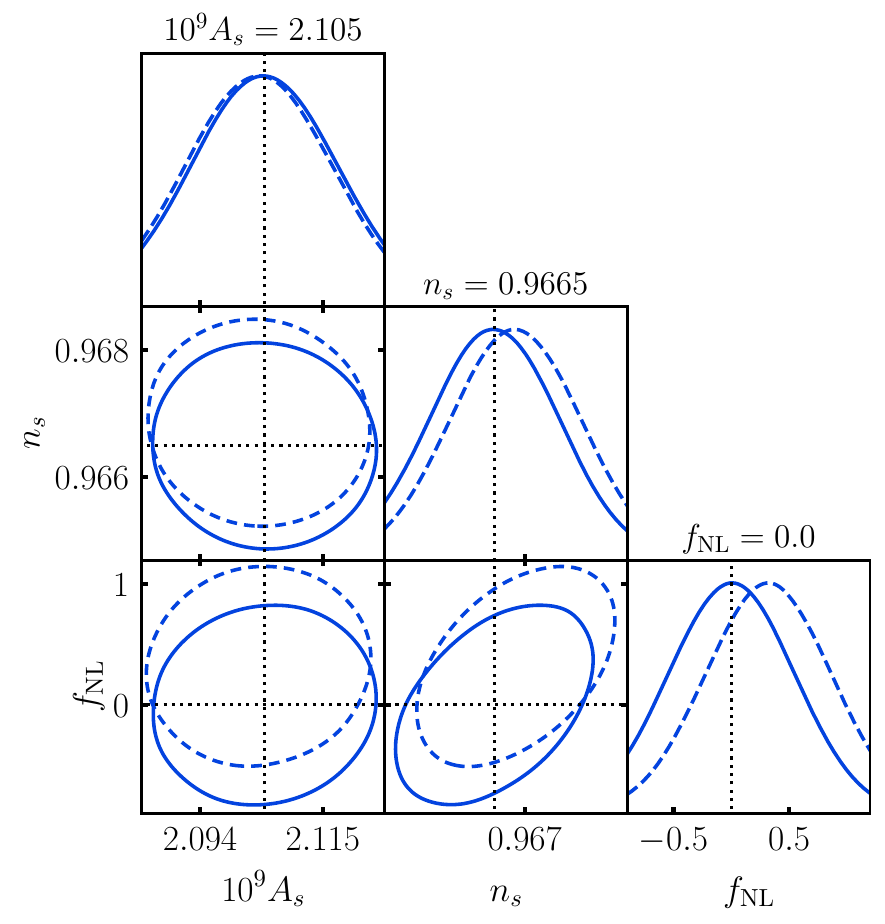}
\caption{{Contour plots showing shifts in parameters from $P_{g}^{(0)} \times P_{g}^{(2)}$, for SKAO2  (\emph{left}) and MegaMapper   (\emph{right}).}
} \label{fig:fishercontourfidshiftLBGSKA2}
\end{figure}

Since the true and incorrect models are competing and nested models, they have the same number 6 of common parameters, and the correct model has 1 extra parameter. 
The value of $\fnl$ in the incorrect model shifts to compensate for the fact that $\varepsilon$ is being kept fixed at the incorrect fiducial value $\varepsilon = 0$. 
Within a Gaussian Fisher formalism, the shift  in the value of $f_{\rm NL}$ may be computed as (see e.g. \cite{Camera:2014dia} and references therein)
\begin{align}\label{bfbias}
\delta f_{\rm NL}= \fnl^{\rm true} - \fnl^{\rm wrong}=-\big({\overset{{}_0}{F}}{}^{-1} \big)_{\fnl\alpha}\,\overset{{}_1}{ F}_{\alpha\varepsilon}\,\delta\varepsilon\,.
\end{align}
Here there is a sum over $\alpha$, i.e. over the parameters in $\overset{{}_0}{\bm F}$, which is the Fisher matrix in the incorrect model ($\varepsilon=0$). The Fisher matrix in the true model, $\overset{{}_1}{\bm F}$, has one extra row and column due to the extra parameter $\varepsilon$. By definition of $\varepsilon$, its shift is $\delta\varepsilon= \varepsilon^{\rm true} - \varepsilon^{\rm wrong} =1$.

The results of the shift $\delta(\vartheta_i)$ in the value of $\vartheta_i=\big(f_{\rm NL}, \, A_s,\, n_s\big)$ are presented in \autoref{tab:deltaParameters_fNL} --\autoref{tab:deltaParameters_ns}. Since the shift must be compared with the precision of measurement, we give the normalised shift $\delta(\vartheta_i)/\sigma(\vartheta_i)$. Note that $\sigma(\vartheta_i)$ is the marginalised error from the correct model.  

\begin{table}[h!]
    \centering
    \caption{Normalised shift, $\delta(f_{\rm NL})/\sigma(f_{\rm NL})$, in the value of $f_{\rm NL}$ for the correlated case $P_{g}^{(0)} \times P_{g}^{(2)}$, from SKAO2 and MegaMapper,  with different $P_{g}^{\rm corr}$.
    }
     \label{tab:deltaParameters_fNL}
        \vspace{0.3cm}
\begin{tabular}{lccc}
\toprule
Survey \hspace{2.0cm} & ~~~~~~~~~~ NI ~~~~~~~~~~ & ~~~~~~~~~~ I ~~~~~~~~~~ & ~~~~~~~~~~ NI + I ~~~~~~~~~~ \\
\midrule
\midrule
SKAO2   & 0.513 & $-0.498$ & $\bf{-0.001}$ \\
\midrule
MegaMapper  & 0.410 & $-0.177$ & \bf{~~0.614} \\ 
\bottomrule
\end{tabular}
\end{table}

\begin{table}[h!]
    \centering
    \caption{As in \autoref{tab:deltaParameters_fNL}, for $\delta(A_{s})/\sigma(A_{s})$.
    }
     \label{tab:deltaParameters_As}
        \vspace{0.3cm}
\begin{tabular}{lccc}
\toprule
Survey \hspace{2.0cm} & ~~~~~~~~~~ NI ~~~~~~~~~~ & ~~~~~~~~~~ I ~~~~~~~~~~ & ~~~~~~~~~~ NI + I ~~~~~~~~~~ \\
\midrule
\midrule
SKAO2   & $-0.216$ & ~~0.014 & $\bf{-0.107}$ \\ 
\midrule
MegaMapper  & $-0.365$ & $-0.091$ & $\bf{-0.471}$ \\ 
\bottomrule
\end{tabular}
\end{table}

\begin{table}[h!]
    \centering
    \caption{As in \autoref{tab:deltaParameters_fNL}, for $\delta(n_{s})/\sigma(n_{s})$.
    }
     \label{tab:deltaParameters_ns}
        \vspace{0.3cm}
\begin{tabular}{lccc}
\toprule
Survey \hspace{2.0cm} & ~~~~~~~~~~ NI ~~~~~~~~~~ & ~~~~~~~~~~ I ~~~~~~~~~~ & ~~~~~~~~~~ NI + I ~~~~~~~~~~ \\
\midrule
\midrule
SKAO2   & $-0.023$ & 0.152 & \bf{0.095} \\ 
\midrule
MegaMapper  & ~~0.257 & 0.146 & \bf{0.324} \\ 
\bottomrule
\end{tabular}
\end{table}

\newpage
\section{Conclusion} \label{sec5}

In this work, we investigate the impact on the galaxy power spectrum of the non-integrated (local) and integrated corrections and their implications for constraining the local primordial non-Gaussianity parameter, $f_{\rm NL}$. Relativistic and wide-angle corrections significantly influence the monopole {(see \autoref{fig:monopolesSKA2_LBG_SxI_SxNI})} and especially the quadrupole {(\autoref{fig:quadrupolesSKA2_LBG_SxI_SxNI})} of the power spectrum at scales \(k < k_{\rm eq}\). Non-integrated corrections, such as wide-angle and Doppler terms, and integrated corrections, including lensing,  exhibit opposing behaviours for the surveys considered, leading to a partial cancellation of their contributions. We find that  integrated Sachs-Wolfe and time-delay effects are generally much smaller than lensing corrections for $k>k_{\rm f}$, at redshifts $\lesssim1$
{(\autoref{fig:monopolesSKA2_LBG} and \autoref{fig:quadrupolesSKA2_LBG})},
but we include them in our numerical computations.

Our results demonstrate that incorporating {the quadrupole} provides significant gains in the precision on $f_{\rm NL}$. For SKAO2, the marginalised error on $f_{\rm NL}$ is reduced from {$\sigma(f_{\rm NL}) = 5.21$} when only the monopole is used, to {$\sigma(f_{\rm NL}) = 2.89$} when the correlated monopole and quadrupole are combined, an improvement of {$\sim40\%$}. MegaMapper achieves tighter constraints due to its higher redshift coverage, with {$\sigma(f_{\rm NL}) = 0.548$} for the correlated monopole and quadrupole combination, an improvement of {$\sim60\%$} compared to the monopole-only case. {This reflects the following features:
\begin{itemize}
    \item 
the quadrupole's sensitivity to  relativistic, wide-angle and PNG effects;
    \item
the additional signal in the quadrupole, when \red{correlated with the monopole},  leads to improved precision.
%    \item 
%the correlation between the quadrupole and monopole, $P_{g}^{(0)}\times P_{g}^{(2)}$, brings further improvement, due to the  breaking of degeneracies.
\end{itemize}
In addition, the constraints were significantly further improved by investigating the redshift binning:
\begin{itemize}
    \item 
we choose equal-volume redshift bins \cite{D:2023oxn};
    \item 
we compute the $P_{g}^{(0)}\times P_{g}^{(2)}$ constraints on $\fnl$ for $N=2, 3, \cdots$ bins and identify the minimum $\sigma(\fnl)$ (\autoref{fig:zbinning});
    \item
we find that 6 bins for SKAO2 and 3 for MegaMapper are optimal in the equal-volume case.
\end{itemize}}

Relativistic and wide-angle corrections also influence the measurement or estimation of $f_{\rm NL}$, as these effects can partly mimic the scale-dependent bias introduced by primordial non-Gaussianity (\autoref{fig:monopolesLBGSKA2fnl}). If these corrections are neglected, non-negligible shifts (or biases) may be introduced in $f_{\rm NL}$ estimates \cite{Namikawa:2011yr,Bruni:2011ta,Camera:2014sba,Lorenz:2017iez,Viljoen:2021ocx}. (Note that the effect of lensing is stronger in photometric than in spectroscopic surveys \cite{Namikawa:2011yr,Fonseca:2015laa,Jelic-Cizmek:2020pkh}.)
For SKAO2, the shift in $\fnl$ due to non-integrated corrections alone is {$\sim0.5\sigma$}, while for MegaMapper it is {$\sim0.4\sigma$}. When both integrated and non-integrated corrections are included, the shift
{for SKAO2 becomes negligible, effectively cancelled by a negative shift due to integrated terms. Although integrated terms also cause a negative shift for MegaMapper, in combination with the non-integrated terms the result is a larger positive shift of $\sim0.6\sigma\,$.}

Compared to previous studies, such as  \cite{Camera:2014sba} for SKAO2 and \cite{Viljoen:2021ocx} for Euclid-like spectroscopic surveys, the shifts we find in $f_{\rm NL}$ are smaller. This difference arises from the methodologies -- these earlier works employed angular power spectrum analyses, which naturally include all relativistic and wide-angle effects without approximation, as well as cross-bin correlations. In contrast, our Fourier-space analysis neglects such correlations, which likely contributes to the smaller shifts observed here. Indeed, there are no cross-bin correlations of any type in a standard Fourier analysis (see
    \cite{Bailoni:2016ezz} for a discussion).

Furthermore, our analysis, following the approximation introduced by \cite{Noorikuhani:2022bwc}, 
{shows that there are significant partial cancellations in the correction terms. This means  that the omitted correction terms can have a larger than expected influence.}
In particular, we omit several potentially important contributions:
\begin{itemize}
    \item 
    The lensing-lensing  correlation, which is not suppressed by any positive power of $k/\cH$, is neglected.
    \item 
    The off-diagonal (cross-bin) lensing-lensing correlations  are largest for the most widely separated redshift bins \cite{Viljoen:2021ocx}, and  are not incorporated. 
    \item
{The lensing-wide-angle correlations are neglected.}  
    \item
The standard perturbative approximation to wide-angle effects that we use here, following e.g. \cite{Reimberg:2015jma,Castorina:2017inr,Noorikuhani:2022bwc,Paul:2022xfx,Jolicoeur:2024oij}, cannot capture all wide-angle effects -- and therefore can also shift (bias) the estimate of $\fnl$, as shown in \cite{Benabou:2024tmn}.
    \item 
{Finally, we omitted mode-coupling from wide-angle effects in the covariance and this could have a significant effect in reducing the precision found in a Fisher analysis.}
\end{itemize}

\noindent We intend to address some of these limitations in a follow-up investigation. 

\[\]

\noindent{\bf Acknowledgements:}
We thank Stefano Camera, Chris Clarkson, Jessie Hammond and Pritha Paul for useful discussions, {and an anonymous reviewer for critical comments that helped us to improve the paper.}
SG and RM were supported by the South African Radio Astronomy Observatory and the National Research Foundation (grant no. 75415). SJ was supported by the Stellenbosch University Astrophysics Research Group fund.
%-------------------------------------

\clearpage
\appendix

\section{Multipoles of the non-integrated power spectrum} \label{AppB}

\allowdisplaybreaks
The monopoles  of the standard power spectrum and the non-integrated correction from \cite{Jolicoeur:2024oij} (in the midpoint case, i.e, $t = 1/2$) are: 
\begin{align}
\dfrac{P^{\mathrm{S}(0)}_{g}}{P} &= (b + b_{\mathrm{ng}})^{2} + \frac{2}{3} (b + b_{\mathrm{ng}}) f + \frac{1}{5} f^{2}
\\
\dfrac{P^{\mathrm{NI}(0)}_{g}}{P} &= \frac{1}{3} \frac{1}{k^{2}} \Big[(\gamma^{\rm D})^{2} + 2 (f +  3 b + 3 b_{\mathrm{ng}}) \gamma^{\Phi} \Big] 
\notag
\\
 & {} \quad + \frac{2}{15} \frac{f}{k r} \Bigg[- \frac{2}{k} \gamma^{\rm D} + \frac{\mathcal{H}}{k}  \left(\mathcal{E} - 2 \mathcal{Q} - \frac{{\mathcal{H}'}}{\mathcal{H}^2}\right) \Big(f  + 5 b + 5 b_{\mathrm{ng}} + 5 k \, \partial_{k} b_{\mathrm{ng}} \Big) \Bigg] 
\notag
\\
 & {} \quad + \frac{2}{15} \frac{f}{(k r)^{2}} \Bigg[\frac{11}{14} f + (1 - \mathcal{Q}) \Big[f - 5 b - 5 b_{\mathrm{ng}} - 5 k \, \partial_{k} b_{\mathrm{ng}}\Big] 
 \notag
 \\
 & {} \hspace{3.0cm}  - \frac{1}{2} \Big(3 b + 3 b_{\mathrm{ng}} + 5 k \, \partial_{k} b_{\mathrm{ng}} + k^{2} \, \partial_{k}^{2} b_{\mathrm{ng}} \Big) \Bigg] 
 \notag
\\
& {} \quad + \frac{2}{15} f \Bigg\{\frac{1}{k r} \Bigg[- \frac{2}{k} \gamma^{\rm D} + \frac{\mathcal{H}}{k} \left(\mathcal{E} - 2 \mathcal{Q} - \frac{{\mathcal{H}'}}{\mathcal{H}^2}\right) (f + 5 b + 5 b_{\mathrm{ng}}) \Bigg]
\notag
\\
& {} \hspace{2.0cm} + \frac{1}{(k r)^{2}} \Bigg[\frac{37}{14} f + (1 - \mathcal{Q}) (f - 5 b - 5 b_{\mathrm{ng}})  
\notag
\\
& {} \hspace{4.0cm} - \frac{1}{2} (5 b + 5 b_{\mathrm{ng}} + 2 k \, \partial_{k} b_{\mathrm{ng}}) \Bigg] \Bigg\} \frac{k \p_k P}{P}
\notag
\\
& {} \quad + \frac{1}{15} \frac{f}{(k r)^{2}} \Bigg\{ \frac{13}{7} f  - \Bigg[b + b_{\mathrm{ng}} + \frac{1}{k^{2}} \gamma^{\Phi} - \frac{3}{2} \frac{\mathcal{H}}{k} \left(\mathcal{E} - 2 \mathcal{Q} -  \frac{{\mathcal{H}'}}{\mathcal{H}^{2}}\right) \frac{1}{k} \gamma^{\rm D} \Bigg] 
\notag 
\\
& \hspace{2.50cm} + f \frac{\mathcal{H}^{2}}{k^{2}} \left(\mathcal{E} - 2 \mathcal{Q} - \frac{{\mathcal{H}'}}{\mathcal{H}^{2}}\right)^{2} \Bigg\} \frac{k^2 \p^2_k P}{P} 
\;.\label{ExpP0_NI}
\end{align}

\begin{figure}%[!htbp]
\centering
\includegraphics[width=0.5\linewidth]{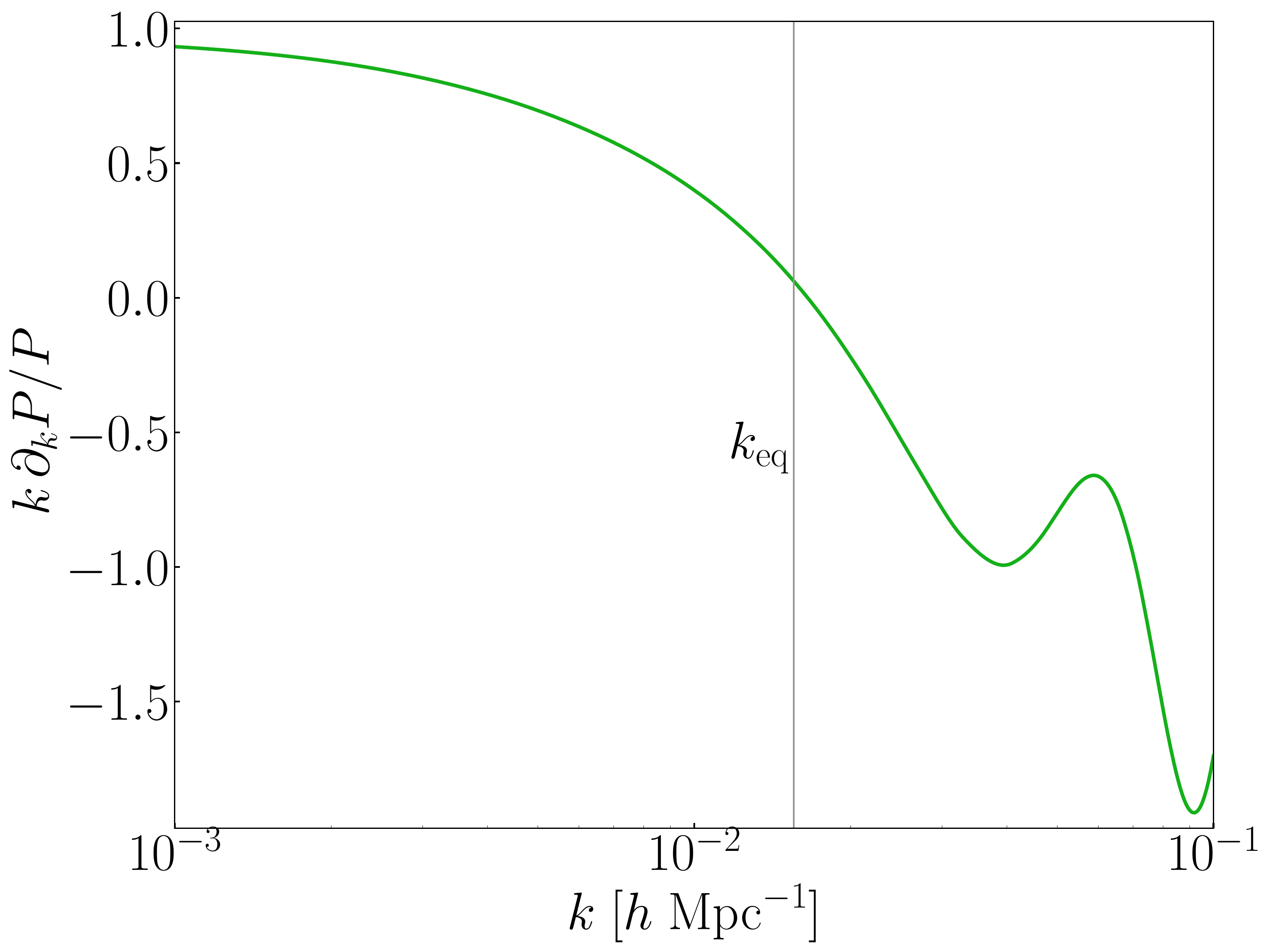}
\includegraphics[width=0.49\linewidth]{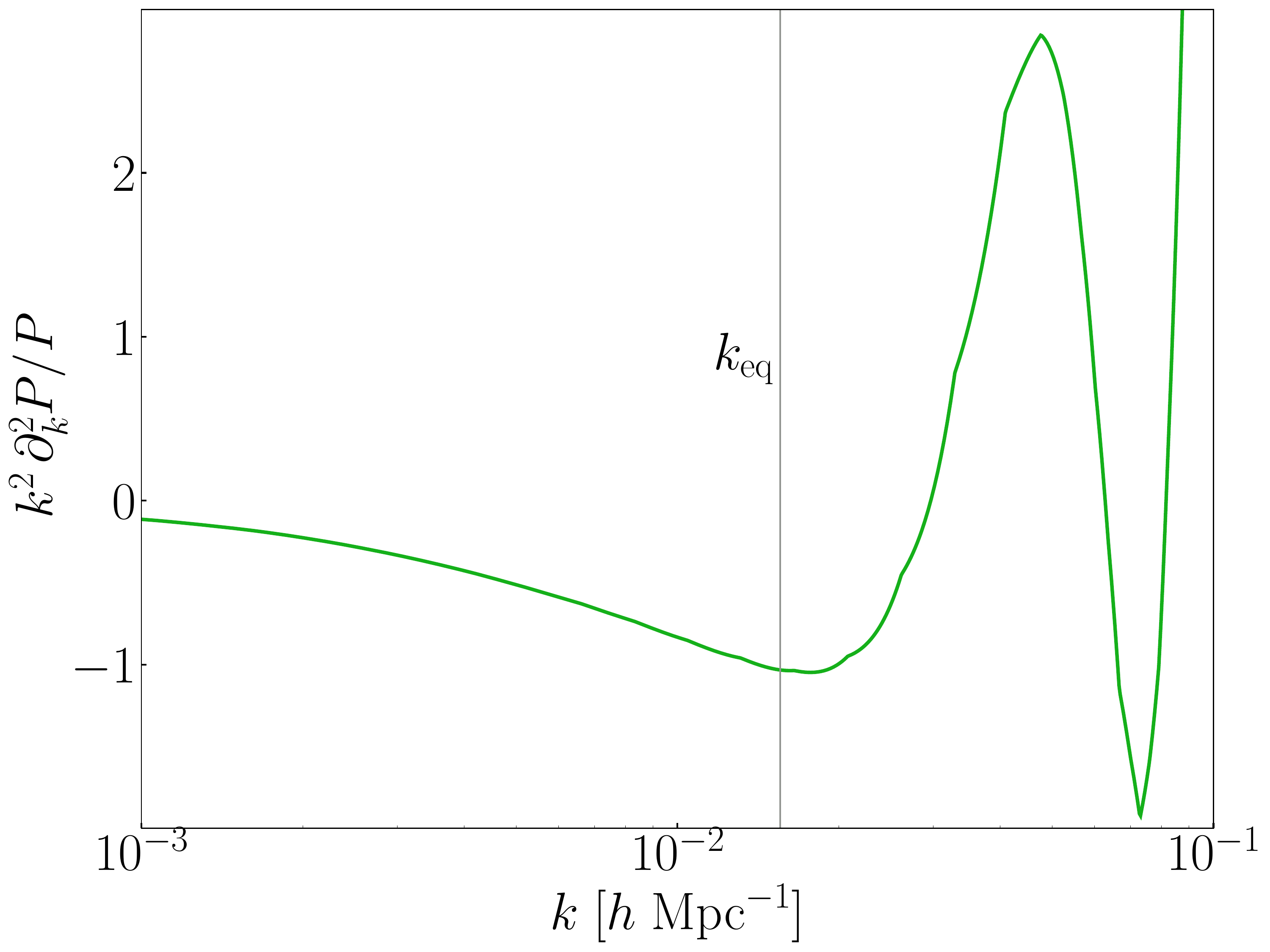}
\caption{Derivatives of the linear matter power spectrum $P$. } \label{fig:PmDerivatives}
\end{figure}

The quadrupoles are
\begin{align}
\dfrac{P^{\mathrm{S}(2)}_{g}}{P} &= \frac{4}{3} (b + b_{\mathrm{ng}}) f + \frac{4}{7} f^{2} 
\\
\dfrac{P^{\mathrm{NI}(2)}_{g}}{P} &= \frac{2}{3} \frac{1}{k^{2}} \Big[ (\gamma^{\rm D})^{2} + 2 f \, \gamma^{\Phi} \Big] 
\notag
\\
& {} \quad + \frac{2}{21} \frac{f}{k r} \Bigg[ \frac{\mathcal{H}}{k} \left(\mathcal{E} - 2 \mathcal{Q} - \frac{{\mathcal{H}'}}{\mathcal{H}^{2}}\right) \Big(10 f + 14 b + 14 b_{\mathrm{ng}} - 7 k \, \partial_{k} b_{\mathrm{ng}}\Big) - \frac{20}{k} \, \gamma^{\rm D}\Bigg]
\notag
\\
& {} \quad - \frac{2}{21} \frac{f}{(k r)^{2}} \Bigg[- f  + 2 (1 - \mathcal{Q}) \Big(13 f + 28 b + 28 b_{\mathrm{ng}} - 14 k \, \partial_{k} b_{\mathrm{ng}}\Big) 
\notag
\\
& {} \hspace{2.50cm}  + \frac{11}{2} \Big(6 b + 6 b_{\mathrm{ng}} - 2  k \, \partial_{k} b_{\mathrm{ng}} - k^{2} \, \partial_{k}^{2} b_{\mathrm{ng}}\Big) \Bigg] 
\notag
\\
& {} \quad + \frac{2}{21} f \Bigg\{\frac{1}{k r} \Bigg[\frac{\mathcal{H}}{k} \left(\mathcal{E} - 2  \mathcal{Q} - \frac{{\mathcal{H}'}}{\mathcal{H}^{2}}\right) (f - 7 b - 7 b_{\mathrm{ng}}) - \frac{2}{k} \, \gamma^{\rm D}\Bigg]
\notag
\\
& {} \hspace{2.0cm} + \frac{1}{(k r)^{2}} \Bigg[ 2 (1 - \mathcal{Q}) (5 f + 14 b + 14 b_{\mathrm{ng}}) 
\notag
\\
& {} \hspace{4.0cm} + 11 (f + b + b_{\mathrm{ng}} +  k \, \partial_{k} b_{\mathrm{ng}}) \Bigg] \Bigg\} \frac{k \p_k P}{P} 
\notag
\\
& {} \quad + \frac{1}{21} \frac{f}{(k r)^{2}} \Bigg\{  3 f + 11 \big(b + b_{\mathrm{ng}} + \frac{1}{k^{2}} \, \gamma^{\Phi} \big) + \frac{3}{2} \frac{\mathcal{H}}{k} \left(\mathcal{E} - 2 \mathcal{Q} - \frac{{\mathcal{H}'}}{\mathcal{H}^{2}} \right) \frac{1}{k} \, \gamma^{\rm D} 
\notag
\\
& {} \hspace{2.50cm} - 4 f \, \frac{\mathcal{H}^{2}}{k^{2}} \left(\mathcal{E} - 2 \mathcal{Q} - \frac{{\mathcal{H}'}}{\mathcal{H}^{2}}\right)^{2} \Bigg\} \frac{k^2 \p^2_k P}{P} 
\; .
\label{ExpP2_NI}
\end{align}
where 
\begin{align}
\gamma^{\mathrm{D}} &= \mathcal{H}f\bigg[\mathcal{E} - 2\mathcal{Q} + \frac{2\big(\mathcal{Q} - 1\big)}{r \mathcal{H}} - \frac{\mathcal{H}^{\prime}}{\mathcal{H}^{2}}\bigg]\;,\label{gamma1} 
\\
\gamma^{\Phi} &= \frac{3}{2}\Omega_{m}\mathcal{H}^2\bigg[2 + \mathcal{E} - f - 4\mathcal{Q}  + \frac{2\big(\mathcal{Q} - 1\big)}{r \mathcal{H}} - \frac{\mathcal{H}^{\prime}}{\mathcal{H}^{2}}\bigg] +\mathcal{H}^2 f(3-\mathcal{E})  \;.\label{gamma2}
\end{align}
The derivatives of $P$ are shown in \autoref{fig:PmDerivatives}.

%\newpage
\section{Derivation of the integrated correction to the power spectrum} \label{AppA}

{Here we show how to derive the equations in \autoref{sec2} and thus recover the results of \cite{Noorikuhani:2022bwc}.}
The standard (S) contribution to the number count density contrast is the galaxy density contrast and linear Kaiser redshift-space distortions. It is given by 
\begin{equation}
\Delta_{g}^{\rm S}(\bm{x}_a) = \int \frac{\ud^{3}\bm{k}_a}{(2\pi)^{3}}\,\exp{\big(\mathrm{i}\,\bm{k}_a \cdot \bm{x}_a\big)}\,
\mathcal{K}^{\rm S}{(x_a,k_a,\mu_a)}
\,{D(x_a)}\,\delta_{0}(\bm{k}_a)\;,
\label{E1}
\end{equation}
where $\delta_{0}$ is the matter density contrast at $z=0$ and the standard Fourier kernel is given by \autoref{pp}.
The integrated (I) relativistic correction to \eqref{E1} is   
\begin{equation}
\Delta_{g}^{\rm {int}}(\bm{x}_a) = \int \frac{\ud^{3}\bm{k}_a}{(2\pi)^{3}}\,\int^{x_a}_{0} \ud \tilde{r}\,\exp{\big(\mathrm{i}\,\bm{k}_a \cdot \tilde{\bm{r}}\big)}\,\mathcal{K}^{\rm {int}} {(x_a,\tilde{r}, k_a, \mu_a)} 
{D(\tilde{r})} \,\delta_{0}(\bm{k}_{a})\;,\label{E3}
\end{equation}
where $\mathcal{K}^{\rm {int}}$ is  
given by \autoref{eKI}.  
Then the cross-correlation between $\Delta_{g}^{\rm S}$ and $\Delta_{g}^{\rm {int}}$ is given by 
\begin{align} \label{E5}
\big \langle \Delta_{g}^{\rm S}(\bm{x}_{1}) \, \Delta_{g}^{\rm {int}}(\bm{x}_{2}) \big \rangle &= 
\int\! \frac{\ud^3 \bm{k}_{1}}{(2\pi)^{3}}\!\int \!\frac{\ud^3 \bm{k}_{2}}{(2\pi)^{3}}\!
\int^{x_{2}}_{0} \! \ud \tilde{r}\,\exp{\bigg\{\mathrm{i}\bigg[\bm{k}_{1} \cdot \bm{x}_{1} + \Big(\frac{\tilde{r}}{x_{2}}\Big)\bm{k}_{2} \cdot \bm{x}_{2} \bigg] \bigg\} } 
 \\ \nonumber
& \qquad \qquad \times \mathcal{K}^{\rm S}
{(x_1,k_1,\mu_1)}
\,\mathcal{K}^{\rm {int}} {(x_2,\tilde{r}, k_2, \mu_2)} {D(x_1)\,D(\tilde{r})} \, 
\big \langle \delta_{0}(\bm{k}_{1}) \delta_{0}(\bm{k}_{2})\big \rangle\;.
\end{align}
Here we {follow \cite{Noorikuhani:2022bwc} and}  make the flat-sky approximation \autoref{pp2}, so that {$\hat{\tilde{\bm r}} = \bm{\hat{x}}_{1} = \bm{\hat{x}}_{2}$ at zero order}. Thus we neglect mixing with wide-angle corrections, as in \cite{Noorikuhani:2022bwc}.
For the mid-point configuration (\autoref{geom}), 
$\bm{x}_{1} = \bm{r} - \bm{x}/2 \quad \text{and} \quad \bm{x}_{2} = \bm{r} + \bm{x}/2\,.$
Using
$\langle \delta_{0}(\bm{k}_{1}) \delta_{0}(\bm{k}_{2})\big \rangle = (2\pi)^{3}P_{0}(k_{1})\delta^{\rm D}(\bm{k}_{1}+\bm{k}_{2})$
and integrating over $\bm{k}_{2}$, \autoref{E5} becomes
\begin{align}
\big \langle \Delta_{g}^{\rm S}(\bm{x}_{1}) \, \Delta_{g}^{\rm {int}}(\bm{x}_{2}) \big \rangle &= 
\int\! \frac{\ud \bm{k}_{1}}{(2\pi)^{3}}\!
\int^{x_{2}}_{0} \!\ud^3 \tilde{\bm r}\,\exp{\bigg\{\mathrm{i}\bigg[\Big(\frac{x_{2}-\tilde{r}}{x_{2}}\Big)\bm{k}_{1} \cdot \bm{r} -\Big(\frac{x_{2}+\tilde{r}}{2x_{2}}\Big) \bm{k}_{1} \cdot \bm{x} \bigg] \bigg\} } \nonumber \\
& \qquad \qquad  ~~~~ \times \mathcal{K}^{\rm S} 
{(x_1,k_1,\mu_1)}
\,\mathcal{K}^{\rm {int}\,*}
{(x_2,\tilde{r}, k_1, \mu_1)} {D(x_1)\,D(\tilde{r})} 
\,P_{0}(k_{1})\;.\label{E8}
\end{align}
At leading order, {$ x_{1} =  x_{2} =  r$} (since we  neglect mixing with wide-angle corrections)
and then \autoref{E8} becomes
\begin{align} \label{E11}
\big \langle \Delta_{g}^{\rm S}({\bm{r} - \bm{x}/2}) \, \Delta_{g}^{\rm {int}}({\bm{r} + \bm{x}/2}) \big \rangle &= 
\int\! \frac{\ud^3 \bm{k}_{1}}{(2\pi)^{3}}\!
\int^{r}_{0} \!\ud \tilde{r}\,\exp\bigg\{\! \mathrm{i}\bigg[\!\Big(\!\frac{r-\tilde{r}}{r} \Big)\bm{k}_{1} \cdot \bm{r} -\Big(\frac{r+\tilde{r}}{2r}\Big) \bm{k}_{1} \cdot \bm{x} \bigg]\! \bigg\}\! 
\\ 
& \qquad \qquad  ~~~~ \times \mathcal{K}^{\rm S}
{(r,k_1,\mu_1)}
\,\mathcal{K}^{\rm {int}\,*}
{(r,\tilde{r}, k_1, \mu_1)}
 {D(x_1)\,D(\tilde{r})} \, P_{0}(k_{1})\;. \notag
\end{align}

The Fourier transform of \autoref{E11} gives the cross-power spectrum:
\begin{align}
P_{g}^{\rm S \times {int}}(\bm{r}, \bm{k}) &= \int\! \ud^{3}\bm{x}\, \exp\big(-\mathrm{i}\,\bm{k} \cdot \bm{x}\big)\,\big \langle \Delta_{g}^{\rm S}({\bm{r} - \bm{x}/2}) \, \Delta_{g}^{\rm {int}}({\bm{r} + \bm{x}/2}) \big \rangle 
\notag\\
 &= \int^{r}_{0} \! \ud \tilde{r} \int\! \ud^{3}\bm{k}_{1}\,\delta^{\rm D}\left[\bm{k} + \Big(\frac{r+\tilde{r}}{2r}\Big)\bm{k}_{1}\right]\,\exp{\left[\mathrm{i}\Big(\frac{r-\tilde{r}}{r}\Big)\bm{k}_{1} \cdot \bm{r}\right]} \nonumber \\
& \qquad \qquad  \qquad ~~~ \times \mathcal{K}^{\rm S}
{(r,k_1,\mu_1)}
\,\mathcal{K}^{\rm {int}\,*}
{(r,\tilde{r}, k_1, \mu_1)}
{D(r)\,D(\tilde{r})} \, P_{0}(k_{1})\;. \label{E13}
\end{align}
The Dirac-delta function can be simplified:
\begin{equation}
\delta^{\rm D}\big[\bm{k}_{1} + G(r,\tilde{r})\,\bm{k}\big] = G(r,\tilde{r})^{-3}\delta^{\rm D}\left[\bm{k}_{1} + G(r,\tilde{r})^{-1}\bm{k}\right]\;,\label{E14}
\end{equation}
where we used \autoref{eG}.
Then \eqref{E13} becomes
\begin{align}
P_{g}^{\rm S \times {int}}({r, k, \mu}) &= \int^{r}_{0} \!\ud \tilde{r} \,G(r,\tilde{r})^{-3}\,\exp\!{\left[-\mathrm{i}\, \mu \, G(r,\tilde{r})^{-1} k \, (r-\tilde{r})\right]} \nonumber \\
& \qquad \quad ~\times 
\mathcal{K}^{\rm S}
{(r,k,\mu)}
\,\mathcal{K}^{\rm {int}\,*}
{(r,\tilde{r}, k, \mu)}
D(r)\,D(\tilde{r}) \, P_{0}\left(k/G(r,\tilde{r})\right)\nonumber \\
&\equiv \int^{r}_{0} \! \ud \tilde{r} \, \mathcal{J}({r,\tilde r, k, \mu}) \;. \label{E15}
\end{align}
The total cross-power spectrum for the integrated relativistic effects at leading order is
\begin{align}
P_{g}^{\rm I}({r, k, \mu}) &= P_{g}^{\rm S \times {int}}({r, k, \mu}) + P_{g}^{\rm {int} \times S}({r, k, \mu})
=\int^{r}_{0} \ud \tilde{r} \,\big[\mathcal{J}({r,\tilde r, k, \mu}) + \mathcal{J}^{*}({r,\tilde r, k, \mu})\big]\;. \label{E17}
\end{align} 
%---------------------------------------------------------------------------------

\subsection*{Multipoles of the integrated power spectrum} \label{AppB2}

Gauss-Legendre quadrature is an efficient method for numerically approximating definite integrals by transforming the integration interval into a standard domain. For a 2D numerical integration over a rectangular domain $[x_{\rm min}, x_{\rm max}] \times [y_{\rm min}, y_{\rm max}]$, the integral
\begin{align}
F = \int_{x_{\rm min}}^{x_{\rm max}} \int_{y_{\rm min}}^{y_{\rm max}} \ud x \, \ud y\, f(x, y)   \; , \label{Eq2dI}
\end{align}
must first be transformed into the reference interval $[-1, 1] \times [-1, 1]$ before applying the Gauss-Legendre quadrature rule:
\begin{align}
x = \frac{x_{\rm max} - x_{\rm min}}{2} \xi + \frac{x_{\rm max} + x_{\rm min}}{2}, \quad y = \frac{y_{\rm max} - y_{\rm min}}{2} \eta + \frac{y_{\rm max} + y_{\rm min}}{2},
\end{align}
where $\xi$ and $\eta$ are the new variables of integration in the interval $[-1, 1]$. The integral \autoref{Eq2dI} becomes:
\begin{align}
F = A\! \int_{-1}^{1}\! \int_{-1}^{1}\! \ud\xi \, \ud\eta\, f\!\left(\!\frac{x_{\rm max} - x_{\rm min}}{2} \xi + \frac{x_{\rm max} + x_{\rm min}}{2},  \frac{y_{\rm max} - y_{\rm min}}{2} \eta + \frac{y_{\rm max} + y_{\rm min}}{2}\!\right)\!  ,
\end{align}
where $A=(x_{\rm max} - x_{\rm min}) \, (y_{\rm max} - y_{\rm min})/4$. Using Gauss-Legendre quadrature, the integral is approximated by a weighted sum over the quadrature points:
\begin{align}
F \approx A \sum_{i=1}^n \sum_{j=1}^n w_i \, w_j \, f\!\left(\!\frac{x_{\rm max} - x_{\rm min}}{2} \xi_{i} + \frac{x_{\rm max} + x_{\rm min}}{2},  \frac{y_{\rm max} - y_{\rm min}}{2} \eta_{j} + \frac{y_{\rm max} + y_{\rm min}}{2}\!\right)\!,
\end{align}
where $\xi_i$ and $\eta_j$ are the Gauss-Legendre quadrature points on $[-1, 1]$, and $w_i$, $w_j$ are the corresponding weights.

The integrated contribution \autoref{EqPSxI} is of the form:
\begin{align}
    P^{\rm I}_{g} = \int_{0}^{r} \ud\tilde r\,\mathcal{
    I}(\tilde r, \mu)  \; ,
\end{align}
and its multipoles can be written as:
\begin{align}
    P^{{\rm I}(\ell)}_{g} &= \frac{2\ell+1}{2} \, \int_{-1}^{1} \int_{0}^{r}\ud\tilde r \ud\mu \,\mathcal{L}_{\ell}(\mu) \, \mathcal{I}(\tilde r, \mu) 
     \approx \frac{2\ell+1}{2} \, \frac{r}{2} \sum_{i=1}^n \sum_{j=1}^n w_i \, w_j \,  \mathcal{L}_{\ell}(\xi_{i}) \, \mathcal{I}\left(   \frac{r}{2} \eta_{j} + \frac{r}{2},\xi_{i}\right) \; .
\end{align}

\clearpage
\section{Additional survey information}
\label{appc}

\begin{table}[ht]
    \centering
    \caption{MegaMapper LBG: number density, evolution bias and magnification bias (from \cite{Rossiter:2024tvi}).}
    \vspace{0.3cm}
    \begin{tabular}{|c|c|c|c|}
        \hline
        ~~~~~ $z$ ~~~~~ & ~~~~~ $\bar n_{g}$ ~~~~~ & ~~~~~ $\mathcal{E}$ ~~~~~ & ~~~~~ $\mathcal{Q}$ ~~~~~  \\
        & $(10^{-3} \; h^3 \, \mathrm{Mpc}^{-3})$  &    &    \\
        \hline  \hline
        2.1 &  2.26 & 2.99 & 1.92 \\
        2.2 &  1.79 & 3.61 & 1.97 \\
        2.3 &  1.41 & 3.78 & 2.02 \\
        2.4 &  1.12 & 3.37 & 2.08 \\
        2.5 &  0.91 & 2.41 & 2.14 \\
        2.6 &  0.76 & 1.12 & 2.21 \\
        2.7 &  0.66 & -0.13 & 2.28 \\
        2.8 &  0.60 & -1.04 & 2.36 \\
        2.9 &  0.55 & -1.50 & 2.45 \\
        3.0 &  0.50 & -1.56 & 2.55 \\
        3.1 &  0.46 & -1.34 & 2.65 \\
        3.2 &  0.41 & -0.99 & 2.76 \\
        3.3 &  0.37 & -0.61 & 2.88 \\
        3.4 &  0.33 & -0.31 & 3.00 \\
        3.5 &  0.29 & -0.14 & 3.12 \\
        3.6 &  0.25 & -0.16 & 3.24 \\
         3.7 &  0.22 & -0.39 & 3.36 \\
        3.8 &  0.19 & -0.81 & 3.46 \\
        3.9 &  0.17 & -1.34 & 3.54 \\
        4.0 &  0.15 & -1.82 & 3.61 \\
        4.1 &  0.14 & -2.17 & 3.65 \\
        4.2 &  0.12 & -2.32 & 3.68 \\
        4.3 &  0.11 & -2.20 & 3.70 \\
        4.4 &  0.10 & -1.78 & 3.71 \\
        4.5 &  0.09 & -0.98 & 3.71 \\
        4.6 &  0.08 & -0.24 & 3.71 \\
        4.7 &  0.07 &  1.97 & 3.70 \\
        4.8 &  0.06 &  4.29 & 3.70 \\
        4.9 &  0.05 &  7.37 & 3.71 \\
        5.0 &  0.03 & 11.40 & 3.72 \\
        \hline
    \end{tabular}
    \label{tab:LBGParameters}
\end{table}

\clearpage
\section{Additional contour plots}
\label{appd}

\begin{figure}[!htbp]
\centering
\includegraphics[width=0.49\linewidth]{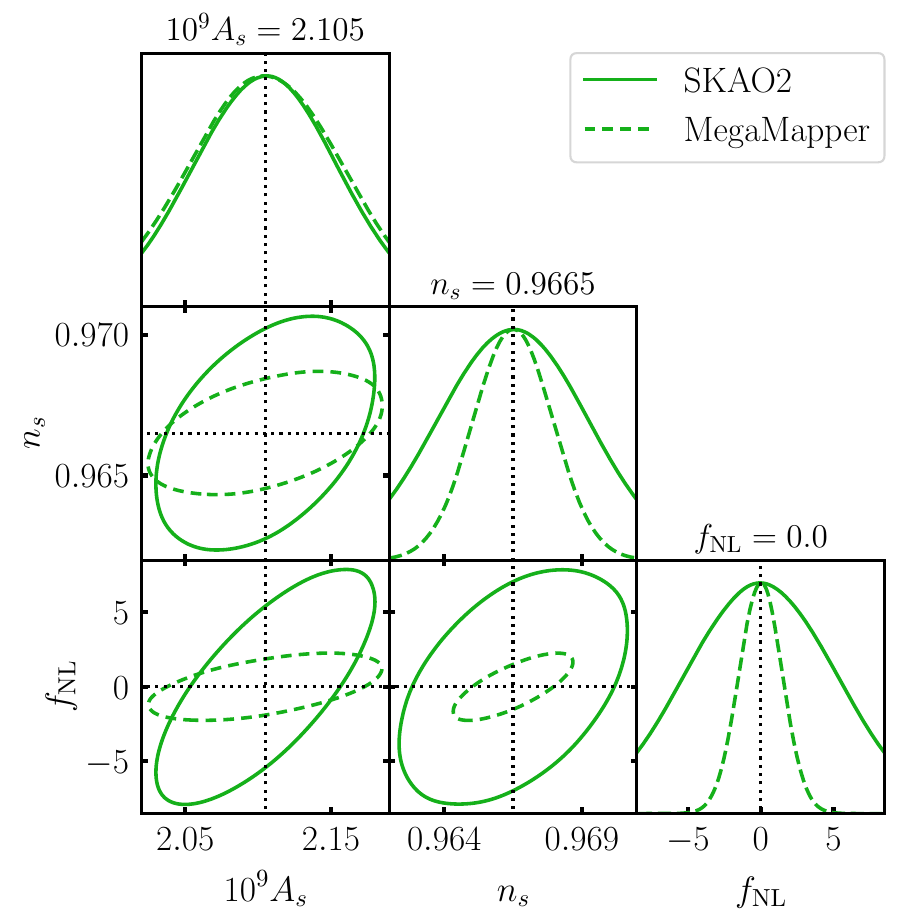}
\includegraphics[width=0.49\linewidth]{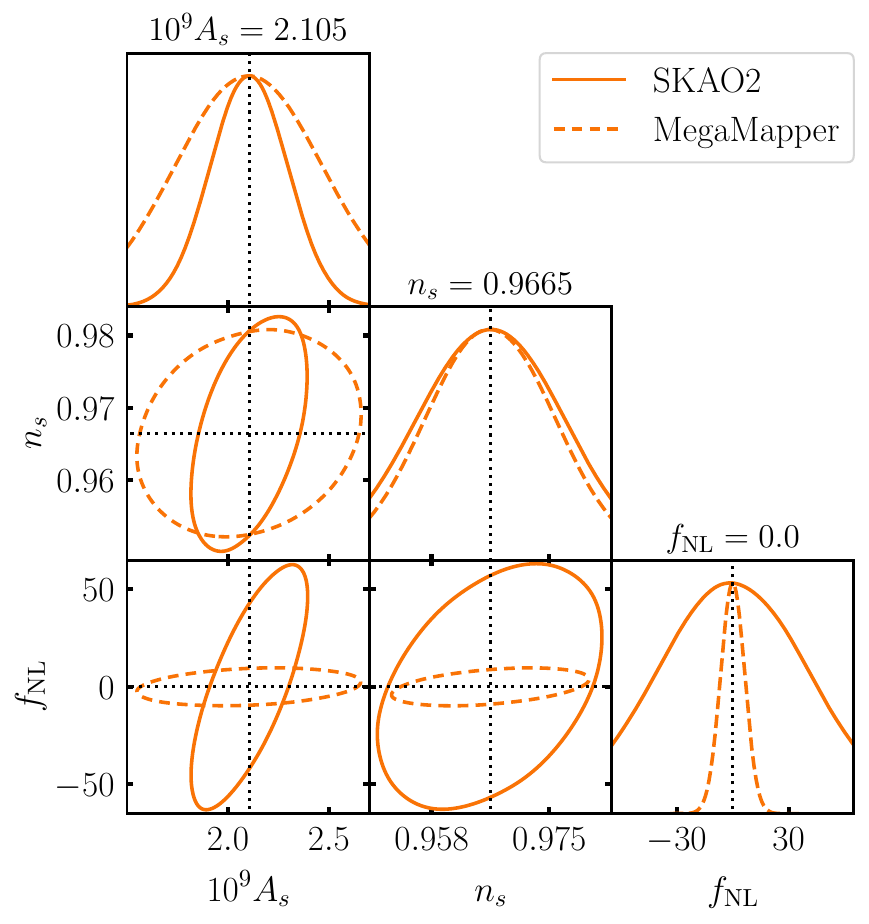}
\includegraphics[width=0.6\linewidth]{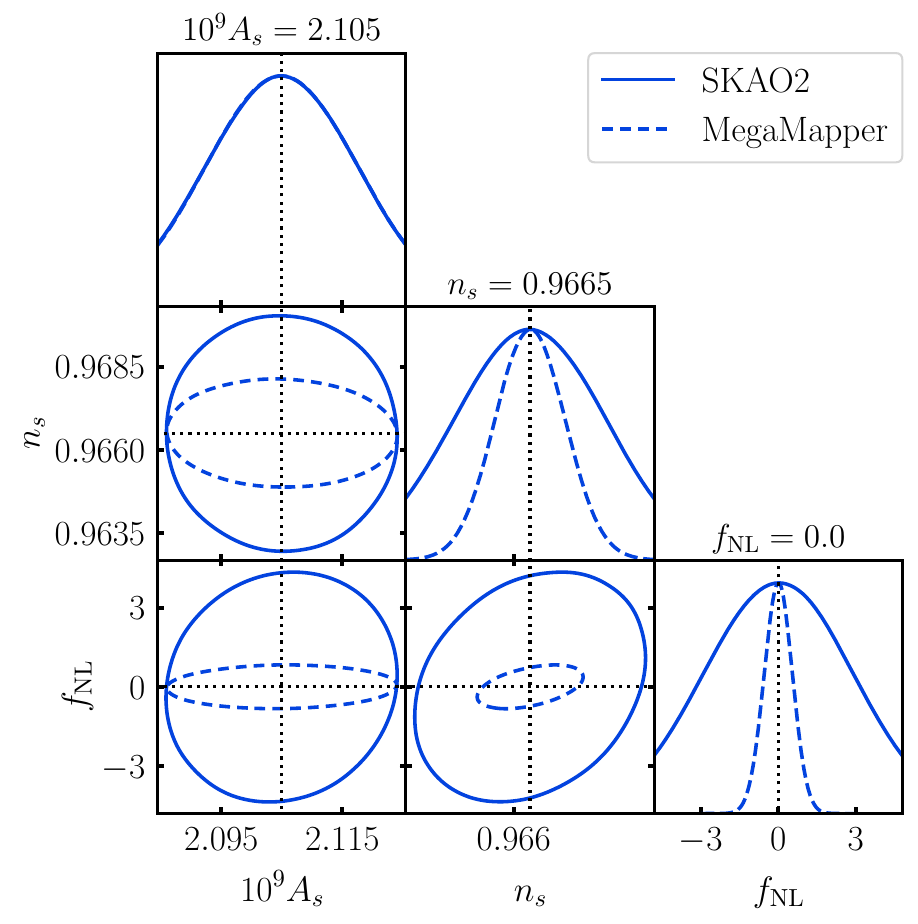}
\caption{{Contour plots of  $1\sigma$ marginal errors from SKAO2  and MegaMapper, for the cases:  $P_{g}^{(0)}$ (\emph{top left}), $P_{g}^{(2)}$ (\emph{top right}), %$P_{g}^{(0)} + P_{g}^{(2)}$ (\emph{bottom left}), 
\red{and   $P_{g}^{(0)}\times P_{g}^{(2)}$ (\emph{bottom}).}}
} \label{fig:fishercontourLBGSKA2eachdata}
\end{figure}

\iffalse
%%%%%%%%
\clearpage
\section{\teal{Signal-to-noise ratios}}
\label{appg}

\begin{figure}[!htbp]
\centering
\includegraphics[width=0.325\linewidth]{snr_z0.46_SKA2.pdf}  
\includegraphics[width=0.325\linewidth]{snr_z0.9696_SKA2.pdf}   
\includegraphics[width=0.325\linewidth]{snr_z1.8985_SKA2.pdf} 
\includegraphics[width=0.325\linewidth]{snr_z2.467_MegaMapper.pdf}  
\includegraphics[width=0.325\linewidth]{snr_z3.4241_MegaMapper.pdf}   
\includegraphics[width=0.325\linewidth]{snr_z4.4571_MegaMapper.pdf}
\caption{\teal{Signal-to-noise ratio, given by \autoref{snr}, at selected redshift bin centres, corresponding to the 4 different combinations in \autoref{D1}--\autoref{D4}: $P_{g}^{(0)}$ ({green}), 
$P_{g}^{(2)}$ ({orange}),
$P_{g}^{(0)}+P_{g}^{(2)}$ ({red}) and $P_{g}^{(0)} \times P_{g}^{(2)}$ ({blue}) for SKAO2 (\emph{upper panels}) and MegaMapper (\emph{lower panels}).}} \label{fig:snr_k}
\end{figure}
%%%%%
\fi

\clearpage
\section{{Effect of changing  ${k}$ limits}}
\label{appe}

\begin{figure}[!htbp]
\centering
\includegraphics[width=0.49\linewidth]{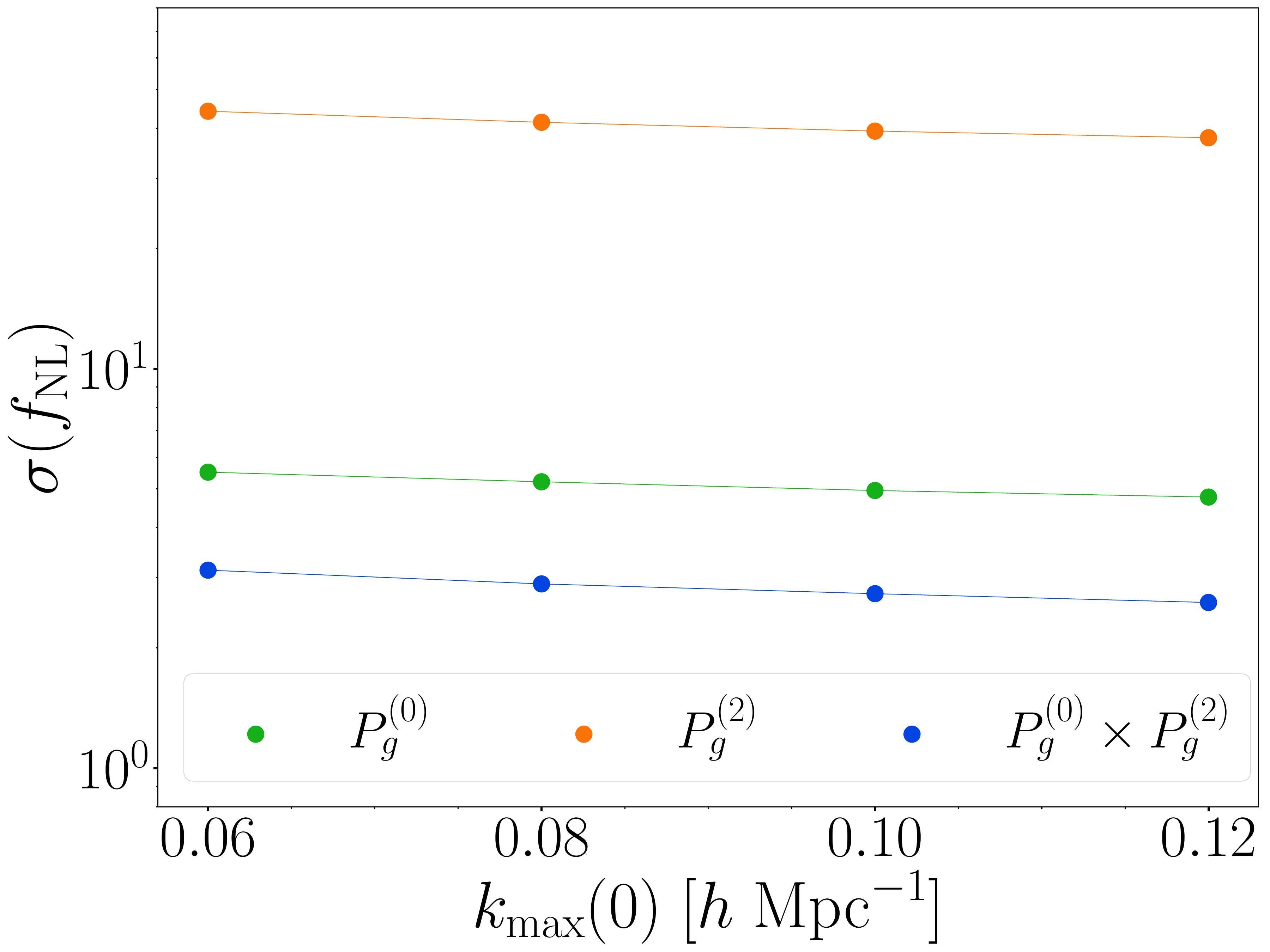}
\includegraphics[width=0.49\linewidth]{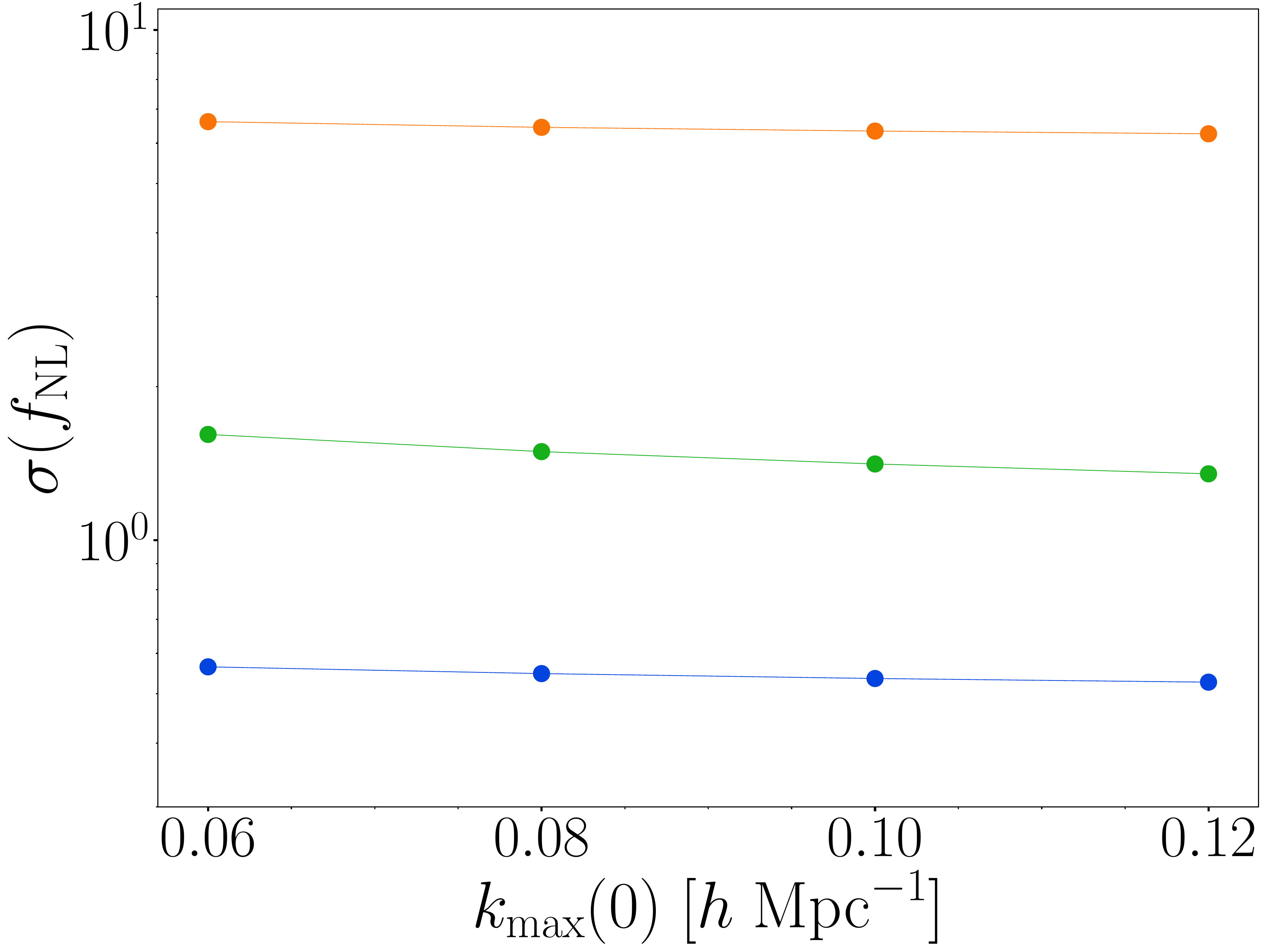}
\caption{{Effect on $\sigma(\fnl)$ of %\autoref{figdelz}, for 
increasing $k_{\rm max}$
for SKAO2 (\emph{left}) and MegaMapper (\emph{right}).}} \label{figkmax}
\end{figure}

\begin{figure}[!htbp]
\centering
\includegraphics[width=0.49\linewidth]{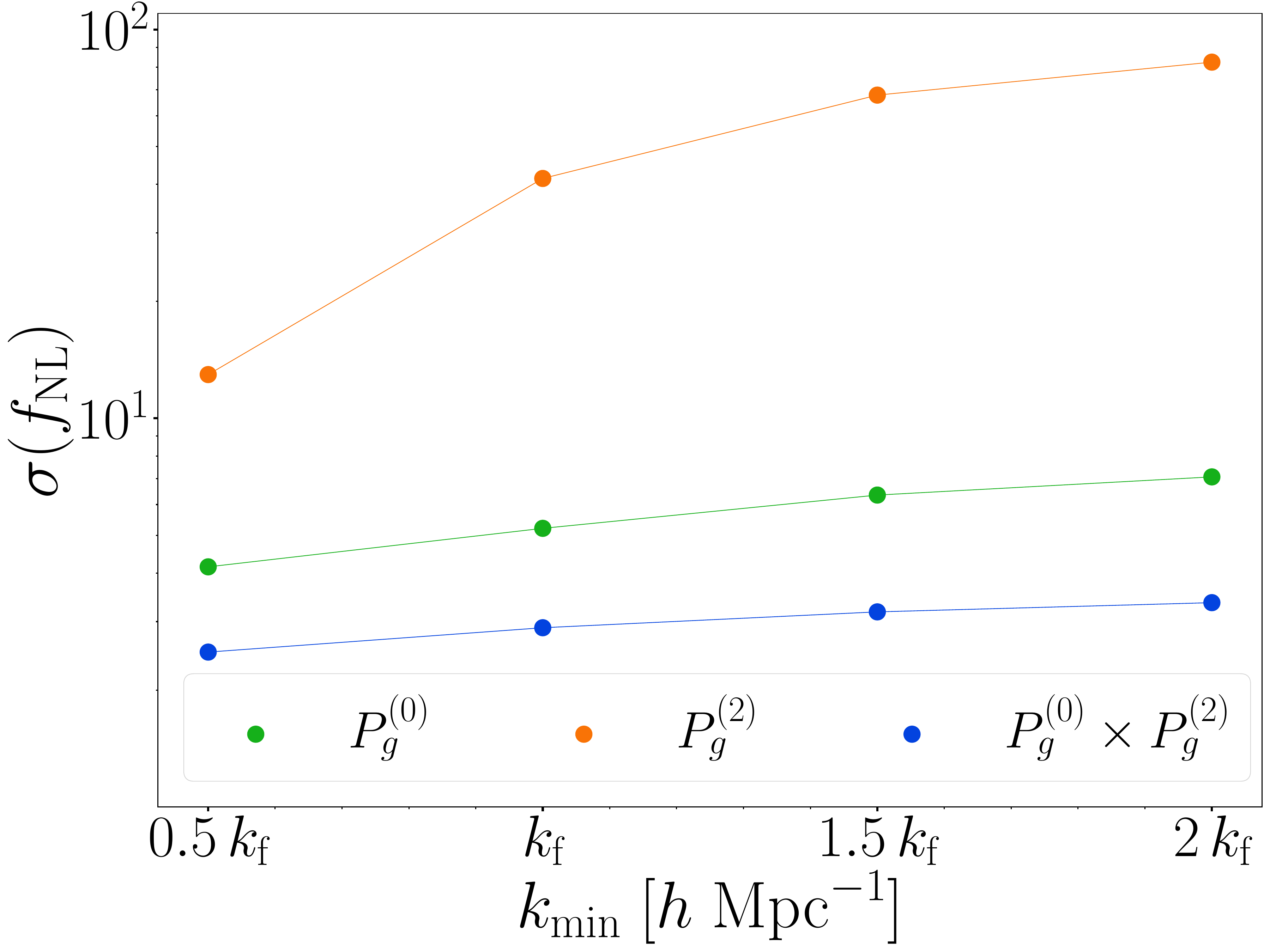}
\includegraphics[width=0.49\linewidth]{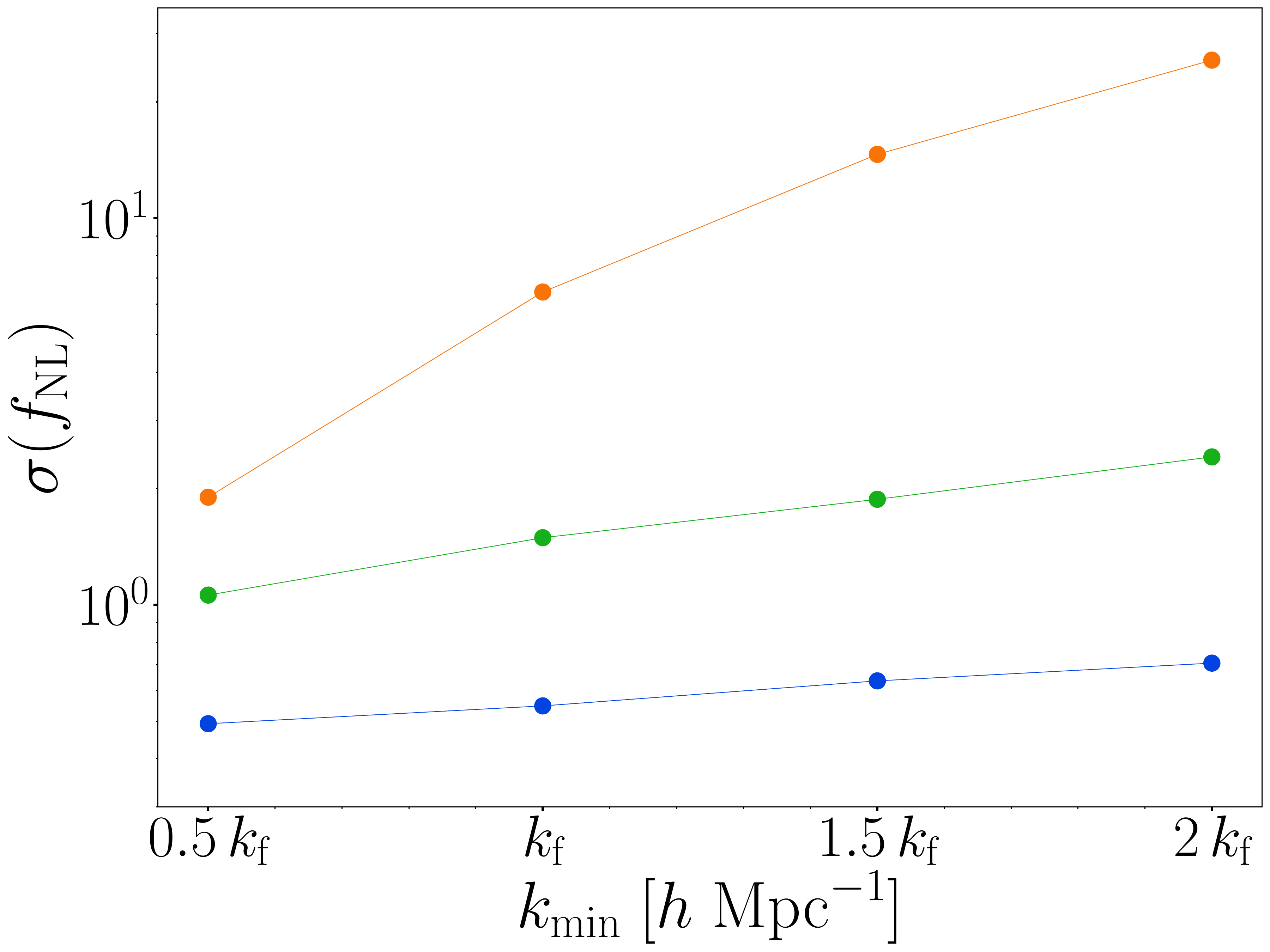}
\caption{{As in \autoref{figkmax}, for increasing $k_{\rm min}$.}
} \label{figkmin2}
\end{figure}

\clearpage
\bibliographystyle{JHEP}
\bibliography{reference_library}

\end{document}